\documentclass[a4paper]{article}
\usepackage{amsfonts}
\usepackage{amssymb}
\usepackage{amsmath}
\usepackage{graphicx}
\usepackage{graphics}
\usepackage{geometry}
\usepackage{caption} 
\usepackage{float, times}
\usepackage[hyperindex=true, colorlinks=true] {hyperref}
\usepackage{mathptmx}
\usepackage{newcent}
\usepackage{setspace}
\usepackage{enumerate}
\usepackage{epstopdf}
\newcommand{\R}{\mathbb{R}}

\newcommand{\edem}{\hfill $\Box$ }

\newcommand{\dsum}{\displaystyle\sum}
\newcommand{\dlim}{\displaystyle\lim}

\newcommand{\diag}{\mathrm{diag}}
\newcommand{\mb}{\mathbf}
\newcommand{\mc}{\mathcal}

\newcommand{\defn}[1]{\emph{#1}}

\newcommand{\bftmp}{ }

\newcommand{\ciVar }{\xi_{i} }
\newcommand{\fiVar }{\phi_{i} }

\newcommand{\kVar }{\kappa }
\newcommand{\kiVar }{\kappa_{i} }

\newcommand{\biVar }{h_{i} }
\newcommand{\barvarphiVar}{r_{max}}
\newcommand{\varphiVar}{r_{inf}}
\newcommand{\varpiVar}{r_{suc}}
\newcommand{\IqstarmaxVar}{I_{q,max}^{\dagger}}
\newcommand{\hatmuVar}{d_q}
\newcommand{\miVar}{\varphi_i}
\newcommand{\mVar}{\varphi}

\newcommand{\starVar}{\dagger}

\newtheorem{prop}{Proposition}[section] 
\newtheorem{dfn}{Definition}[section] 
\newtheorem{thm}{Theorem}[section] 
\newtheorem{lem}{Lemma}[section] 
\newtheorem{rmq}{Remark}[section] 
 
\newtheorem{cor}{Corollary}[section]

\title{Analysis of Control Measures for Vector-borne Diseases Using a  Multistage Vector Model with Multi-Host Sub-populations}
	\author{F. G. T. Kamba
$^{1}~$ , L.C. Eze$^{3}$,  J. C. Kamgang$^{2}$\thanks{Corresponding author Jean Claude Kamgang email: jckamgang@gmail.com} , C. P. Thron$^{4}$ 
\\$^1$ Department of Mathematics and Computer Sciences, Faculty of Science\\ University of N'Gaound\'er\'e, P. O. Box 454 N'Gaound\'er\'e (Cameroon)
\\$^2$ Department of Mathematics and Computer Sciences,\\ ENSAI -- University of N'Gaound\'er\'e, P. O. Box 455 N'Gaound\'er\'e (Cameroon)\\
$^3$ Department of Mathematics, African University of Science and Technology, Abuja (Nigeria)\\
$^4$ Department of Sciences and Mathematics, Texas A\& M University-Central Texas 79549 USA}

\begin{document}

\maketitle
    
\begin{abstract}
We propose and analyze  an epidemiological model for vector borne diseases that integrates a multi-stage vector population and several host sub-populations which may be characterized by a variety of compartmental model types: subpopulations all include Susceptible and Infected compartments, but may or may not include Exposed and/or Recovered compartments. The model was originally designed  to evaluate the effectiveness of various prophylactic measures in malaria-endemic areas, but can be applied as well to other vector-borne diseases. This model is expressed as a system of several differential equations, where the number of equations depends on the particular assumptions of the model.  We compute the basic reproduction number $\mathcal R_0$, 
and show  that if $\mathcal R_0\leqslant 1$, the disease free equilibrium (DFE)  is globally asymptotically stable (GAS) on the nonnegative orthant.  
If $\mathcal R_0>1$,  the system admits a unique endemic equilibrium (EE) that is GAS. We analyze the sensitivity of $R_0$ and the EE to different system parameters, and based on this analysis  we discuss the relative effectiveness  of different control measures.  
\end{abstract}

\noindent {\em Keywords:} Epidemiological model, Vector-borne, Basic reproduction number, Lyapunov function, Global asymptotic stability, Nonstandard finite difference scheme (NFDS), Simulation, Sensitivity, SEIRS

\noindent {\em 2000 MSC:}  34C60, 34D20, 34D23, 92D30
\section{Introduction}\label{sec.Intro}
 Vector-borne infectious diseases such as malaria are widely spread in tropical regions, including parts of America, Asia and much of Africa. For example, infected mosquitoes serve as vectors for a number of serious and potentially fatal diseases, including malaria, yellow fever, Zika, and dengue fever. The mosquito life cycle includes multiple stages of questing and resting:  infections are transmitted when a mosquito that has bitten an infected host during a previous questing stage later bites an uninfected host during a subsequent questing stage.  The chain of transmission for mosquito-borne diseases can be broken through the use of indoor or outdoor spraying, bed nets (which may be treated with insecticide) and prophylactic drugs, as well as other control strategies. Within the host population, individuals may be categorized into sub-populations based on the control measure(s) they use.

A previous paper of Kamgang et {\em al.}~\cite{jck-19} represents a system with host sub-populations and vectors under the assumption that the host sub-populations can all be described with a SIS model. This model is itself an enhancement of several previous models~\cite{Macdonald78, Chit_08, Ross1911, NgwaMCM00, Zongo09}. These assumptions may be justified in cases where vector density is high, and there is rapid detection and treatment of infectious host individuals. However, in many important situations these assumptions are not applicable. For example, many endemic areas for vector-borne diseases are in poor sub-Saharan countries with very low revenue and poor medical structures, so disease go untreated. Moreover, in sub Saharan regions, the climate permits mosquito breeding year-round, and infections may still occur during periods of low vector density. Furthermore, people may develop immunity against malaria infection~\cite{AlOdAl, BarKris}.  For these reasons,  it is more realistic to include Exposed(E) and Resistant/recovered(R) compartments of hosts, so that various sub-populations may obey SEIS, SIRS, or SEIRS models. Such a model has much wider applicability than the previous model in \cite{jck-19}, and can even be applied to rarer diseases with various vectors such as trypanosomiasis (borne by tsetse flies) and Leishmaniasis (borne by sandflies). These considerations provide motivation for the model developed in this paper. 

\noindent 
The paper is organized as follows. Section~\ref{sec.Bdnmdel} describes the new model and gives the corresponding system of  differential equations.  Section~\ref{dfestabana} establishes the well-posedness of the model by demonstrating invariance of the set of nonnegative states, as well as boundedness properties of the solution. The equilibria of the system, are calculated, and a threshold condition for the stability of the disease free equilibrium (DFE), which is based on the basic reproduction number $\mathcal R_0$ is calculated.  
Section~\ref{sec.analysis} analyzes the stability of equilibria. Section~\ref{subsec.dfestabanan}, contains a proof of the global asymptotic stability (GAS) of the disease free equilibrium (DFE) when $\mathcal R_0\leqslant 1$. Section~\ref{sec:eeqstana} has the proof of GAS of the endemic equilibrium (EE) when $\mathcal R_0>1$. Section~\ref{sec:secsam} presents a sensitivity analysis of $R_0$ and the EE with respect to various controllable system parameters. Section~\ref{sec.discuss} discusses the significance of our results. Finally, the Appendix contains detailed proofs and computations required by the analysis.

\section{Model description and mathematical specification}\label{sec.Bdnmdel}

\noindent The model assumes an area populated by $H(t)$ hosts and $M(t)$  vectors (disease vectors), where both populations may depend on time $t$.  Both host and vector populations are homogeneously mixed, so no spatial effects are present.  In the following subsections, we provide a detailed description of the population structure and dynamics of hosts and vectors.

\subsection{Host population structure and dynamics}\label{subsec.Assumption2}
The host population $H(t)$ is classified into $n$ subpopulations, indexed by $1,~2,~ \cdots,~ n$, depending on the way individuals in each subpopulation are disposed with respect to the infection. Individuals' disposition may depend on several factors, including protective measures employed; level of resistance to the disease; host species (if more than one species can serve as host); and so on.  
 At time $t$,we let $H_i(t)~(i=1,~\,\cdots,~\,n)$ denote the size of the $i^{th}$ sub-population, so that $H(t) = {\dsum_{i=1}^n} H_i(t)$. 

The dynamics of the $i^{th}$ host group ($i=1,\,\cdots,\,n$) is described by either a SIS, SEIS, SIRS or SEIRS-based compartment model  as shown in  Figure~\ref{fig:figMulticomAppli1} depending on the parameters $u_i$ and $v_i$ which may be  one or zero (note $\bar{u}_i = 1-u_i$ and $\bar{v}_i = 1-v_i$). The incidence of infection for  individuals in the $i^{th}$ group  is given by $a\varphi_i \frac{I_q}{H}$, where $a$ is the average number of interactions per vector per unit time (the entomological  inoculation rate); $I_q$ is the number of Infectious vectors; $\miVar$ is the infectivity of vectors relative to the individual of the $i^{th}$ group, which is the probability that a bite by an infectious vector on a susceptible individual of the $i^{th}$ group will transfer  infection to the individual. The transition rate from Infectious to Susceptible or Resistant state within the $i^{th}$  group is $\gamma_i$. The transition rate from Exposed  to Infectious state within the $i^{th}$  group is $\varepsilon_i$. The transition rate from Resistant to Susceptible state within the $i^{th}$  group is $\zeta_i$. The force of migration into the $i^{th}$  group is $\Lambda_i$. The incoming  $\widetilde \nu_i$ and outgoing $\nu_i$  rates in the $i^{th}$  group   include the effects of birth and death rates respectively, as well as the effects of hosts moving from one group to another. 

\subsection{Vector population structure and dynamics}\label{subsec.Assumption}

\noindent
The population of disease vectors is characterized by several classes, where each vector's class membership is determined by its own history of past and present activity. Before becoming infectious, vectors initially enter the Susceptible class: the rate of entry (that is, the recruitment rate) is $\Gamma$. Vectors alternate between two activities: {\it questing} (during which time the vectors interact with hosts, e.g. to bite for a blood meal) and {\it resting} (e.g. to lay down eggs, or to digest a blood meal). 
   
At any given instant $t$, questing vectors are equally likely to attempt to interact with any host individual, regardless of his/her disposition relative to the infection.
 Thus for any vector-host interaction, the time-dependent probability that the  host individual involved belongs to the $i^{th}$ group is 
$h_i(t)\equiv{H_i(t)}/{H(t)}$.  During an interaction involving an individual in the $i^{th}$ group, the  vector is killed with probability $\kappa_i$, while the probability the interaction is successful (i.e. the vector enters the resting state) is $\phi_i$ (so $\kappa_i + \phi_i \leqslant 1$).  Following a successful interaction, the vector moves to the resting state.
Letting $a$ denote the average number of interactions per vector per unit time (the {\em entomological  inoculation rate}), it follows that  at any given instant $t$, the incidence rate of successful interactions is $\varpiVar (t) \equiv \dsum_{i=1}^nah_i(t)\phi_i$, 
while the  additive death rate caused by the questing activity of vectors is $d(t)\equiv\dsum_{i=1}^nah_i(t)\kappa_i$. 

The infection rate for vectors may be computed as follows.
Let $I_i, \xi_i$ denote the number of infectious hosts in group $i$ and the corresponding probability of vector infection per interaction with one of these hosts. Similarly, let $R_i,\widetilde\xi_i$ denote  the number of resistant hosts in group $i$ (when $v_i = 1$) and the corresponding per-interaction vector infection probability.  Then the incidence rate for vectors becoming infected  is $r_{inf} (t) \equiv {\dsum_{i=1}^n} a\phi_i(\xi_i{I_i(t)}+v_i \widetilde\xi_i{R_i(t)})/{H(t)}$. 

Susceptible questing vectors that become infected enter the first Exposed resting class ($\mathrm{E}_r^{(\text 1)}$), while those which have not experienced successful infection stay uninfected and enter the Susceptible resting class ($\mathrm{S}_r$). Following initial infection, the vector must remain alive for a certain period before becoming Infectious (this period is called the {\em extrinsic incubation period} in the biological and medical literature ~\cite{010047862}).
During this period, the vector experiences a positive number of resting/questing cycles. In our model, we suppose that a vector becomes infectious after a fixed number $\ell$ of resting/questing cycles following initial infection.  These successive resting/questing cycles are modeled as a sequence of $2\ell$ Exposed states, and are denoted by $ \mathrm{E}^{(\text 1)}_q,\mathrm{E}^{(\text 2)}_r, \cdots, \mathrm{E}^{(\ell)}_q,\mathrm{E}^{(\ell+1)}_r$. If a vector survives through all of these states, it then enters the Infectious class, which is further divided into questing and resting sub-classes ($\mathrm{I}_q$ and $\mathrm{I}_r$, respectively). Once a vector enters the Infectious class, it remains there for the rest of its life, alternating between resting and questing states.

\begin{figure}[hpbt]
\begin{minipage}[c]{8cm}
\centerline{\hbox{\includegraphics[scale=0.34]{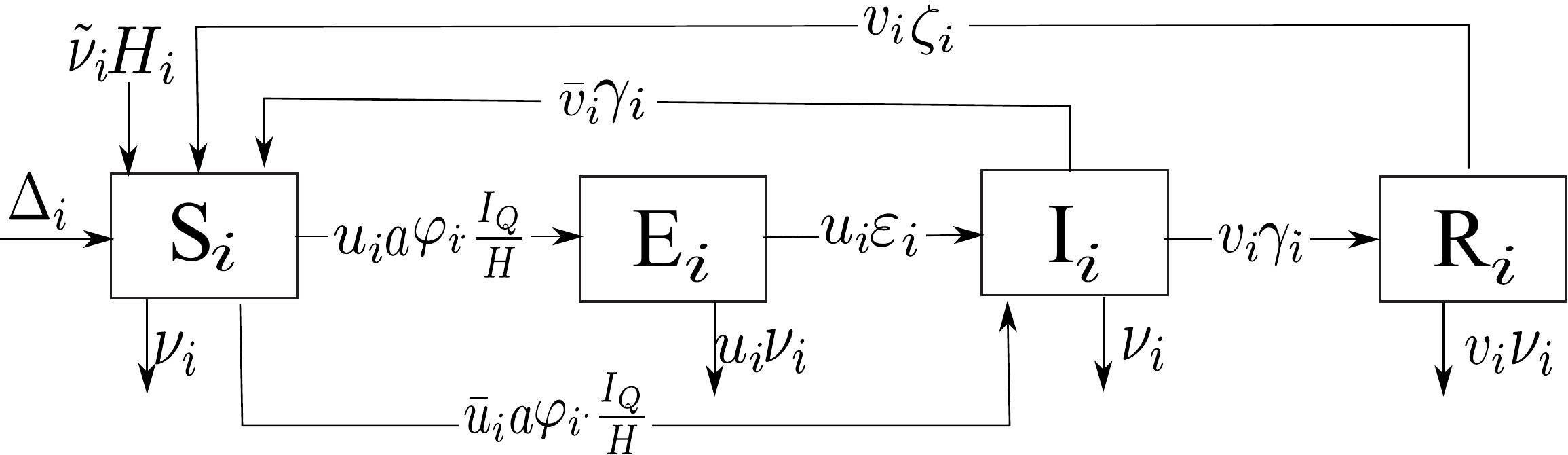}}} 
\caption{\footnotesize Dynamic in the $i^{th}$ human
subgroup\label{fig:figMulticomAppli1}}
\end{minipage}
\begin{minipage}{10cm}
\centerline{\hbox{\includegraphics[scale=0.43]{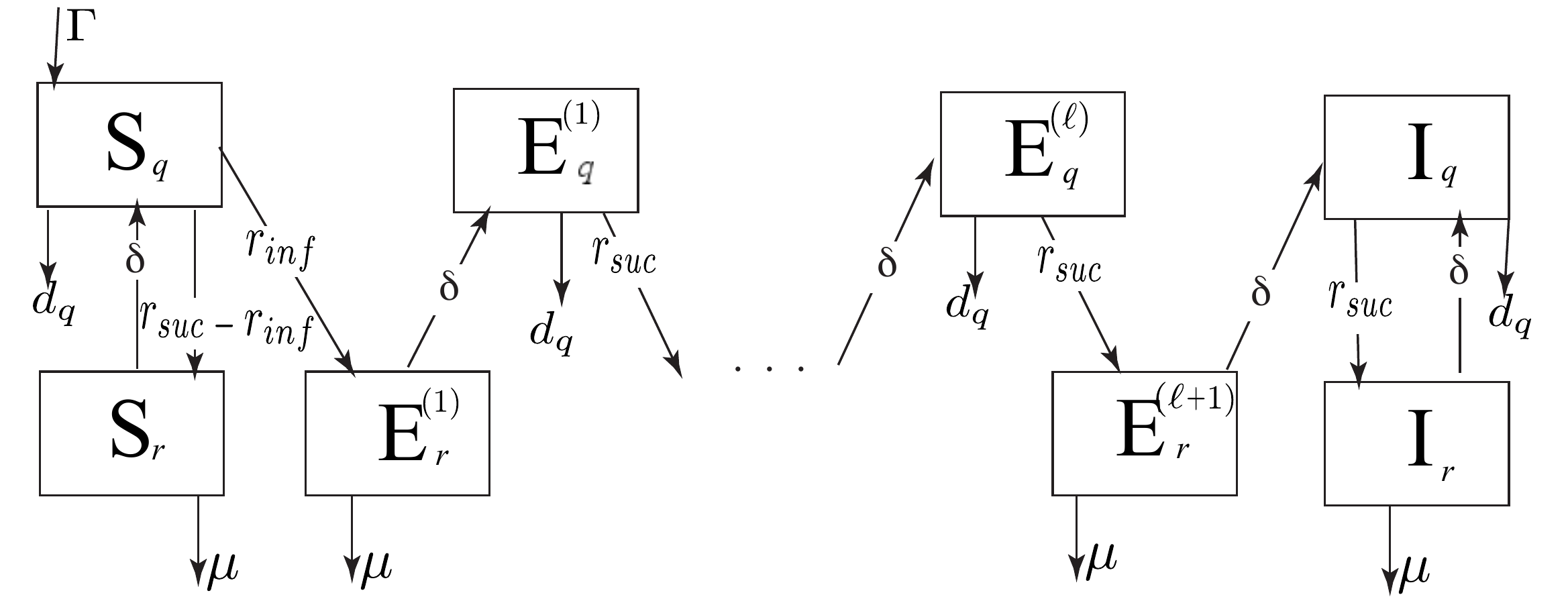}}} 
\caption{\footnotesize Dynamics of the vector  population\label{fig:figMulticomAppli2}}
\end{minipage}
\end{figure}

The overall dynamics of the vector population is depicted in the
multi-compartment diagram in  Figure~\ref{fig:figMulticomAppli2}: The fundamental model parameters are summarized in Table~\ref{tab.tabvd}, while derived parameters are  in Table~\ref{tab.tabvd2}. 

%

\begin{table}[htbp]

\captionsetup{skip=0pt}\caption{Fundamental model parameters}\label{tab.tabvd}

\begin{tabular}{p{1.2cm}p{15.25cm}}
\hline
Param. &  Description\hfill\,  \\
\hline
&\hfill Parameters that characterize the vector population\hfill\,  \\
\hline
$a$ & Vector-host interaction rate  
\\$\delta$ & Rate at which resting vectors move to the questing state
\\$\Gamma$ & Recruitment rate of vectors (vectors $/$ time ) 
\\$\mu$ & Natural death rate of vectors
\\\hline
&\hfill Parameters  that characterize vectors' interaction with hosts in the $i^{\textrm{th}}$ host group\hfill\,  \\
\hline
$\xi_i $ & Probability of infection for a vector that  successfully interacts with the $i^{th}$ Infectious group  (note $\overline{\xi_i} = 1 - \xi_i$)
\\$\widetilde\xi_i $ & Probability of infection for  a vector that  successfully interacts with the $i^{th}$ Resistant group  (note $\overline{\widetilde\xi_i} = 1 - \widetilde\xi_i$)
\\$\phi_i $ & Probability that a vector interaction with  the $i^{th}$ group is successful (i.e. the vector moves to the resting stage) 
\\$\kappa_i $ & Probability that a vector that interacts with the $i^{th}$ group is killed
\\\hline
&\hfill Parameters  that characterize the $i^{th}$ host group\hfill\,  \\
\hline
$\miVar $ & Probability that a vector interaction with a  host in group $i$ results in a host infection
\\$\Lambda_i$ & Migration  rate (hosts $/$ time)  
\\$\varepsilon_i $ & {\bftmp Proportional} transition rate from Exposed to Infectious  state  
\\$\gamma_i$ & {\bftmp Proportional} transition rate from Infectious to Susceptible or Resistant  state (i.e. cure rate)
\\$\zeta_i $ & {\bftmp Proportional} transition rate from Resistant to Susceptible   state 
\\$\nu_i$ & {\bftmp Proportional} outgoing rate for $i^{th}$ host group
\\$\widetilde\nu_i$ & {\bftmp Proportional} incoming rate for $i^{th}$ host group {\bftmp(note that $\widetilde\nu_i \ge \nu_i$)}
\\
\hline
\end{tabular}
\end{table}

\begin{table}[htbp]
\caption{Derived model parameters and time dependent functions}\label{tab.tabvd2}

\begin{tabular}{p{0.78cm}p{2.8cm}p{13.5cm}}
\hline
Param. &  Formula &  Description  \\
\hline
\\\\$h_i $ & $\frac{H_i}{H}$ &Proportion of hosts in group $i$ at a given time
\\$f_q$ & $\dfrac{r_{suc}}{d_q+r_{suc}}$ & Questing frequency of vectors (i.e. questing vector survival proportion) 
\\$f_r$ & $\dfrac{\delta}{\mu+\delta}$ & Resting frequency of vectors  (i.e. resting vector survival proportion)
\\$d_q$ & $\mu + \dsum_{i=1}^nah_i \kappa_i$ & Death rate of questing vectors
\\ $r_{inf}$ & ${\dsum_{i=1}^n} a \phi_i \frac{\xi_iI_i+v_i\widetilde\xi_iR_i}{H}$ &  Incidence rate of infection for questing susceptible vectors
\\ $r_{max} $ & ${\dsum_{i=1}^n} a\phi_ih_i(\xi_i+v_i\widetilde\xi_i)$ &  Maximum incidence rate of infection for questing susceptible vectors
\\ $r_{suc}$ & $\dsum_{i=1}^nah_i \phi_i $ & Incidence rate of successful interaction for questing vectors
\\
\hline
\end{tabular}
\end{table}

\subsection{Model equations}\label{subsec.Assumption3}
\noindent  The  system of ordinary differential equations that characterize the model are given as follows:
\begin{equation}
\left\{
\begin{aligned}
\dot S_i &~~=~~\Lambda_i+\widetilde \nu_i H_i - \left(\nu_i+a\varphi_i \frac{I_q}{H}\right)S_i + v_i\zeta_iR_i+\bar v_i\gamma_iI_i
& i=1,\;2,\;\cdots,\; n \\
\dot S_q &~~=~~\Gamma- (d_q+r_{suc}) S_q + \delta S_r & \,\\
\dot S_r &~~=~~ (r_{suc}-r_{inf}) S_q-(\mu+\delta)S_r\\
\dot E_i &~~=~~u_ia\varphi_i \frac{I_q}{H}S_i-u_i(\nu_i+\varepsilon_i) E_i &i=1,\;2,\;\cdots,\; n\\
\dot E^{(1)}_r &~~=~~r_{inf}  S_q-(\mu+\delta) E^{(1)}_r & \,\\
\dot E^{(j)}_q &~~=~~\delta E^{(j)}_r - (d_q+r_{suc})E^{(j)}_q & j = 1,\;2,\;\cdots,\; \ell \\
\dot  E^{(j+1)}_r &~~=~~r_{suc}  E^{(j)}_q-(\mu+\delta) E^{(j+1)}_r &j = 1,\;2,\;\cdots,\; \ell \\
\dot I_i &~~=~~\bar u_ia\varphi_i \frac{I_q}{H}S_i+u_i\varepsilon_iE_i - (\nu_i+\gamma_i)I_i & i = 1,\;2,\;\cdots,\; n \\
\dot R_i &~~=~~v_i\gamma_i I_i - v_i(\nu_i+\zeta_i)R_i&i=1,\;2,\;\cdots,\; n 
\\[6pt]
\dot I_q &~~=~~\delta E^{(l+1)}_r - (d_q+r_{suc}) I_q +\delta I_r& \, \\
\dot I_r &~~=~~r_{suc} I_q - (\mu+\delta)I_r & \,
\end{aligned}
\right.
\label{eq:eqbednet_}
\end{equation} 

\noindent The system \eqref{eq:eqbednet_} together with initial conditions completely specifies the evolution of the multi-compartment system shown in Figure~\ref{fig:figMulticomAppli1} and Figure~\ref{fig:figMulticomAppli2}. 

\section{Well-posedness, dissipativity and equilibria of the system}\label{dfestabana}

\noindent 
In this section we demonstrate well-posedness of the model by demonstrating invariance of the set of non-negative states, as well as boundedness properties of the solution. We also calculate the equilibria of the system, whose stability properties will be examined in the following section.

\subsection{Positive invariance of the non-negative cone in state space}\label{subsec.pinn}
 
The
system~\eqref{eq:eqbednet_} can be rewritten  in matrix form as 
\begin{equation} \label{eq:eqmodel}\dot{\mathbf x}=\mathbf A(\mathbf x)\mathbf x + \mathbf  b
 \Leftrightarrow \left\{\begin{array}{ccc}\dot{\mathbf x}_S & = & \mathbf A_S(\mathbf x)\,.\,\mathbf x_S  + \mathbf A_{S,\,I}(\mathbf x)\,.\,\mathbf x_I  +   \mathbf b_S\\
   \dot{\mathbf x}_I & = & \;\;\;\;\;\;\mathbf A_I(\mathbf x)\mathbf x_I
\end{array}\right. 
\Leftrightarrow \left\{\begin{array}{ccr}\dot{\mathbf x}_S & = & \mathbf A_S(\mathbf x)\,.\, \left(\mathbf x_S - \mathbf x^*_S\right)  +  \mathbf A_{S,\,I}(\mathbf x)\,.\,\mathbf x_I \\
   \dot{\mathbf x}_I & = & \;\;\;\;\;\;\mathbf A_I(\mathbf x)\,.\,\mathbf x_I
\end{array}\right. ,
\end{equation}
where 
\begin{equation}\label{eq:eqmodel1}
\mathbf A(\mathbf x)=\begin{pmatrix}
\mathbf A_S(\mathbf x) &\mathbf A_{S,\,I}(\mathbf x)\cr \mathbf 0&\mathbf A_I(\mathbf x)
\end{pmatrix};~~~~ \mathbf b = \left(\mathbf b_S;~\mathbf 0\right)~\text{with}~\mathbf b_{S}=\left(\Lambda_0;~\cdots;~\Lambda_n;~\Gamma;~0\right);~~~~
\mathbf x^*_S \equiv  A_S(\mathbf x^*)^{-1} \,.\, \mathbf b_S.
\end{equation}
$\mathbf x^*_S$ is a vector whose components are components of vector $\mathbf x_S$ in eq.~\eqref{eq:naive_and_non} at the disease free equilibrium; its computation is carried out in Proposition~\ref{prop:stdst0}.

 Equation~\eqref{eq:eqmodel} is defined for values of the state variable  $\mathbf x=(\mathbf x_S,\;\mathbf x_I)$ lying in the non-negative cone of $\R^u$ ($
u=4n+ 2\ell+5$), which we denote as $\R^u_+$.  Here $\mathbf x_S$  and  $\mathbf x_I$ represent respectively the naive and non-naive components of the system state: explicitly, 
\begin{equation}\label{eq:naive_and_non}
\mathbf x_S\equiv \left((S_i)_{1\leqslant i\leqslant
n},~~S_q,\;S_r\right) ; \qquad
\mathbf x_I\equiv \left(\;(E_{i})_{1\leqslant i\leqslant n},\;(E_r^{(j)},~~E_q^{(j)})_{1\leqslant j\leqslant \ell};~~E_r^{(\ell+1)};~~(I_i)_{1\leqslant i\leqslant n};~~(R_i)_{1\leqslant i\leqslant n};~~I_q;~~I_r\right).
\end{equation}
This notation is consistent with \cite{KamSal07}, and some results from this previous reference  are used in our analysis. 
	
\noindent The matrix $\mathbf A_S(\mathbf x) =\diag\left(\mathbf A_{S_h}(\mathbf x),\;\mathbf A_{S_v}(\mathbf x)\right)$ with 
\begin{equation}\label{eq:A_S_v}
\mathbf A_{S_h}(\mathbf x) = -\diag\left(\nu_i-\widetilde\nu_i+a\varphi_i \frac{I_q}{H}\right)_{1\leqslant i\leqslant n} \hbox{ and }\mathbf A_{S_v}(\mathbf x) =  \left(\begin{array}{cc} -(\hatmuVar +\varpiVar ) &   \delta \\ \varpiVar -\varphiVar  &  -(\mu+\delta) \end{array}\right),
\end{equation}
The $(n+2)\times (3n+2\ell+3)$ matrix $\mathbf A_{S,\,I}(\mathbf x)$ may be written in block form 
$$\mathbf A_{SI}(\mathbf x)=\begin{pmatrix}\mathbf  A_{S_hI_{E_h}} & \mathbf 0 &\mathbf  A_{S_hI_{I_h}}&\mathbf  A_{S_hI_{R_h}}&\mathbf 0\cr \mathbf 0&\mathbf 0&\mathbf 0&\mathbf 0&\mathbf 0\end{pmatrix} $$ with $\mathbf  A_{S_hI_{E_h}} = \mathrm{diag}(u_i\widetilde\nu_i)_{1\leqslant i\leqslant n},~ \mathbf  A_{S_hI_{I_h}} = \mathrm{diag}\left(\bar v_i\gamma_i+\widetilde\nu_i\right)_{1\leqslant i\leqslant n} \hbox{ and } \mathbf  A_{S_hI_{R_h}} = \mathrm{diag}(v_i(\zeta_i+\widetilde\nu_i))_{1\leqslant i\leqslant n}$  

\noindent The    matrix $\mathbf A_I(\mathbf x)$ may be written in block form  as 
\begin{equation}
\mathbf A_I(\mathbf x) =    \left(\begin{array}{cc} \mathbf A_{I_E}(\mathbf x)
&   \mathbf A_{I_{I,\,E}}(\mathbf x) \\ \mathbf A_{I_{E,\,I}}(\mathbf x) &  \mathbf
A_{I_I}(\mathbf x) \end{array}\right),
\label{eq:mati}
\end{equation}
where the four matrix blocks may be described as follows: 
\medskip

First, the  $(n+2\ell+1)\times(n+2\ell+1)$ matrix $\mb A_{I_E}(\mathbf x)$  expresses the interaction between exposed components of the system. It  may be written in block  form as $\mb A_{I_E}(\mathbf x) =\diag\left(\mb A_{I_{E_h}},\;\mb A_{I_{E_v}}(\mathbf x)\right)$ where $\mb A_{I_{E_h}} = \diag\left(u_i(\varepsilon_i+\nu_i)\right)_{1\leq i\leq n}$, and $\mb A_{I_{E_v}}(\mathbf x)$  is a 2-banded matrix whose diagonal and sub-diagonal elements are given by  the vectors $\mathbf d_0$ and $\mathbf d_{-1}$ respectively, defined by
\begin{equation}\label{eq.eqdiagelts}\mathbf d_0 = {\left(\right.}\underbrace{-(\mu+\delta),\; -(\hatmuVar +\varpiVar ),\;\cdots,-(\mu+\delta),\;
-(\hatmuVar +\varpiVar )}_{2\ell\;\; components},\;-(\mu+\delta) {\left.\right)};\qquad \mathbf d_{-1} = {\left(\right.} \underbrace{ \delta, \;\varpiVar , \;\cdots,
\;\delta,\;\varpiVar }_{2\ell\;\;components}{\left.\right.).}\end{equation} 

Next, the  $(n+2\ell+1)\times 2(n+1)$ matrix $\mathbf A_{I_{I,\,E}}(\mathbf x)$  expresses the dependence of the exposed components on the infectious components of the system. It  may be written in block  form as
\[
\mathbf A_{I_{I,\,E}}(\mathbf x) =    \left(\begin{array}{cc} \mathbf 0
&   \mathbf A_{I_{{I_v},\,{E_h}}}(\mathbf x) \\ \mathbf A_{I_{{I_h},\,{E_v}}}(\mathbf x) &  \mathbf
0 \end{array}\right),
\label{eq:matei}
\]
where the matrix $\mathbf A_{I_{{I_v},\,{E_h}}}(\mathbf x)$, that is $n\times2$ is  $\mathbf A_{I_{{I_v},\,{E_h}}}(\mathbf x) = \dfrac{a}{H}\begin{pmatrix}
u_1\varphi_1 S_1&0\cr\vdots&\vdots\cr u_n\varphi_n S_n&0
\end{pmatrix}$  gives the dependence of the exposed human's components $E_{i}\; (i=1,\cdots,\; n)$  on the infectious vector's components $I_q$; $\mathbf A_{I_{{I_h},\,{E_v}}}(\mathbf x)$ that is $2\ell+1\times n$ the matrix, given by  
\[
 \mathbf A_{I_{{I_h},\,{E_v}}}(\mathbf x)=a\frac{S_q}{H}\begin{pmatrix}\xi_1\phi_1&\cdots & \xi_n\phi_n & \widetilde\xi_1\phi_1&\cdots & \widetilde\xi_n\phi_n\\0 & \cdots&0 & 0 & \cdots&0\\\vdots&\cdots&\vdots & \vdots&\cdots&\vdots\\0 & \cdots&0 & 0 & \cdots&0\end{pmatrix}
\]
  gives the dependence of the exposed vector's components $E_r^{(j)}\; (j=1,\cdots,\; \ell+1)$, $E_q^{(j)},\;\; (j=1,\cdots,\; \ell)$ on the infectious human's components $I_i (i=1,\;\cdots,\; n)$, and resistant components  $R_i (i=1,\;\cdots,\; n)$ which here happens as carriers. 

Next, the $(2n+2)\times (n+2\ell+1)$ matrix  $ \mathbf A_{I_{E,\,I}}(\mathbf x)$  gives the dependence of infectious and resistant components on exposed components.  It  may be written in block  form as
\[
\mathbf A_{I_{E,\,I}}(\mathbf x) =    \begin{pmatrix} \mathbf  A_{I_{{E_h},\,{I_h}}}(\mathbf x) &   \mathbf 0 \\ \mathbf 0  &  \mathbf 0 \\ \mathbf 0 & \mathbf
 A_{I_{{E_v},\,{I_v}}}(\mathbf x) \end{pmatrix},
\label{eq:matie}
\]
where the $i$'th diagonal entry of the $n\times n$ matrix $\mathbf  A_{I_{{E_h},\,{I_h}}}(\mathbf x):= \mathrm{diag}(u_1\varepsilon_1,\;\cdots,\;u_n\varepsilon_n)$ is the transition rate from the Exposed class $E_i$ to the infectious class $I_i$. The $2\times 2\ell+1$ matrix $\mathbf
A_{I_{{E_v},\,{I_v}}}(\mathbf x)$ has all entries zero except the $(1,\; 2\ell+1)$ entry, which is equal to $\delta$ reflecting the transition rate of vectors from state $E^{(\ell+1)}_r$ to state $I_q$.
 
Finally, the $2(n+1)\times 2(n+1)$  matrix
$ \mathbf A_{I_I}(\mathbf x)$ may be written in block  form as $
\mathbf A_{I_I}(\mathbf x) =    \left(\begin{array}{cc} \mathbf A_{I_{I_h}}
&   \mathbf
A_{I_{{I_v},\,{I_h}}}(\mathbf x) \\ \mathbf 0 &  \mathbf
A_{I_{I_v}}(\mathbf x) \end{array}\right),
$
 
  where 
\begin{align*}
\mathbf A_{I_{I_v}}(\mathbf x) = \left(\begin{array}{cc}-\left(\hatmuVar +\varpiVar \right) &
\delta\\\varpiVar  & - \left( \mu+\delta\right)\end{array}\right); \qquad 
\mathbf A_{I_{{I_v},\,{I_h}}}(\mathbf x) = \dfrac{a}{H}\begin{pmatrix}
\bar u_1\varphi_1 S_1&0\cr\vdots&\vdots\cr\bar u_n\varphi_n S_n&0\cr0&0\cr\vdots&\vdots\cr0&0
\end{pmatrix};
\end{align*}
and
$\mathbf A_{I_{I_h}}$ is the $2n \times 2n$ two-banded matrix whose principal diagonal and $n^{th}$ sub-diagonal are given by the vectors $\mathbf a_0 $ and $\mathbf a_n$ respectively, where  
\begin{equation*}
\mathbf a_0 = -\left(
\nu_1+\gamma_1,\;\cdots,\;\nu_n+\gamma_n,\;v_1(\nu_1+\zeta_1),\;\cdots,\;v_n(\nu_n+\zeta_n)\right); \qquad\mathbf a_n = \left(v_1\gamma_1,\; \cdots,\; v_n\gamma_n\right).
\end{equation*}

For a given $\mathbf
x\in\mathbb
R^u_+$, the matrices  $\mathbf A_S(\mathbf x)$, $\mathbf A_I(\mathbf x)$ and $\mathbf A(\mathbf x)$ are Metzler matrices (see Appendix~\ref{appx:defs}), and the vector  $\mathbf b  \in \mathbb R_+^u$.

\noindent The following proposition establishes that  system~\eqref{eq:eqmodel} is epidemiologically well-posed. 
\begin{prop}\label{prop:invrnce} The non-negative cone $\mathbb R^u_+$  is positively invariant for
the system~\eqref{eq:eqmodel}.\end{prop}

\noindent \emph{Proof:}~~The proof is similar to the standard proof that systems determined by  Metzler matrices preserve invariance of the non-negative cone. It can be shown directly that if $\mathbf x$ is on the boundary of $\mathbb R_+^u$, then $\dot{\mathbf x}$ is pointing inside $\mathbb R_+^u$, hence the trajectories never leave $\mathbb R_+^u$.
\edem

\subsection{Disease-free equilibrium (DFE) of the system} 
The system~\eqref{eq:eqmodel} admits two steady states. The trivial steady state that is the DFE is established in  Proposition~\ref{prop:stdst0} below, while the nontrivial steady state will be established in Proposition~\ref{prop:stdst1} after some necessary preliminaries. 

Before characterizing the DFE, we first introduce some useful notation. The {\em questing frequency} $f_q$ and {\em resting frequency} $f_r$ are defined respectively as:
\[
f_q\equiv \frac{\varpiVar }{\hatmuVar+\varpiVar }; \qquad f_r\equiv \frac{\delta}{\mu+\delta}.
\]
$f_q$ may be interpreted as the frequency of questing vectors that pass on to the resting state; while $f_r$ is conversely the frequency of resting vectors that pass on to the questing state. In \cite{jckam201411} these parameters are constants of the model,  but in the current model $f_q$ depends on the system state. The value of $f_q$ at the DFE is denoted by $f^*_q$.   In the following we shall frequently make use of the following replacements:
\begin{equation}\label{eq:frfq}
\hatmuVar +\varpiVar =\frac{\varpiVar }{f_q}; \qquad \mu+\delta=
\frac{\delta}{f_r}.
\end{equation}
This new notation enables us to give a simple expression for the DFE and to shorten expressions in many other computations throughout this paper.
\begin{prop}\label{prop:stdst0}
The system~\eqref{eq:eqmodel} admits a trivial equilibrium (the disease-free equilibrium (DFE))  given by $\mathbf x^*=\left(\mathbf x_S^*;\;\mathbf x_I^*\right)\in\mathbb R^u_+$,  
 with $\mathbf x_I^*=\mathbf 0\in\mathbb R^{u-n-2}$;~ $\mathbf x_S^* =\left(\mathbf x_{S_h}^*;\;\mathbf x_{S_v}^*\right)$, where 
\begin{equation}\label{eq:eqexpdfe}\mathbf x_{S_h}^* = \left(\frac{\Lambda_1}{ \nu_1-\widetilde\nu_1};\,\frac{\Lambda_2}{ \nu_2-\widetilde\nu_2};\,\cdots,\,\frac{\Lambda_n}{ \nu_n-\widetilde\nu_n}\right)\;\;\hbox{ and }\;\; \mathbf x_{S_v}^* = (S_q^*;\,S_r^*) =\left( \frac{f^*_q\Gamma}{\varpiVar ^*(1-f^*_qf_r)},\;\frac{f_rf^*_q\Gamma}{\delta(1-f^*_qf_r)}\right).\end{equation} 
\end{prop}

\noindent{\em Proof }:~~
The DFE corresponds to  a state $\mathbf x^*$ in which all components representing non-naive classes are equal to zero: that is, $\mathbf x^* = \left(\mathbf x^*_S;\;\mathbf x^*_I\right)$ with $\mathbf x^*_I\equiv 0$. The steady-state equation for the system~\eqref{eq:eqmodel} with the constraint $\mathbf x_I\equiv0$ is  
\begin{equation}\label{eq:Asys}
\mathbf
A_S(\mathbf x_S\,;\mathbf 0)\,.\,\mathbf x_S  +  \mathbf b_S=\mathbf 0\Leftrightarrow \left\{\begin{array}{l}
\mathbf A_{S_h}(\mathbf x_S\,;\mathbf 0)\,.\,\mathbf x_{S_h}  +  \mathbf b_{S_h}=\mathbf 0\cr \mathbf A_{S_v}(\mathbf x_S\,;\mathbf 0)\,.\,\mathbf x_{S_v}  +  \mathbf b_{S_v}=\mathbf 0
\end{array}\right. .
\end{equation} 
This system may be solved in two stages, since the subsystem $\mathbf A_{S_h}(\mathbf x_S\,;\mathbf 0)\,.\,\mathbf x_{S_h}  +  \mathbf b_{S_h}=\mathbf 0$ is uncoupled. The solution of this subsystem is $\mathbf x^*_{S_h} = \left(\frac{\Lambda_1}{ \nu_1-\widetilde\nu_1};\,\frac{\Lambda_2}{ \nu_2-\widetilde\nu_2};\,\cdots,\,\frac{\Lambda_n}{\nu_n-\widetilde \nu_n}\right)$. Using this solution in the second subsystem of \eqref{eq:Asys}, we obtain the equation 
$$\mathbf A_{S_v}(\mathbf x_{S_h}\,,\mathbf x_{S_v}\,;\mathbf 0)\,.\,\mathbf x_{S_v}  +  \mathbf b_{S_v}=\mathbf 0
\implies
\mathbf x^*_{S_v}=-{\mathbf A_{S_v}(\mathbf x^*_{S_h}\,;\mathbf x_{S_v}\,;\mathbf 0)}^{-1}\,.\, \mathbf b_{S_v}.$$
Using expression~\eqref{eq:A_S_v} for $\mathbf A_{S_v}$ (with $\varphiVar =0$ at DFE), and recalling that $\mathbf b_{S_v} = (\Gamma ; 0)$, we find the solution
\begin{equation}\label{eq:xsv}
\mathbf x^*_{S_v} = (S_q^*;~S_r^*)= \left(\dfrac{\mu + \delta}{(\hatmuVar^* +\varpiVar^* )(\mu + \delta) - \varpiVar^*  \delta};\;\dfrac{\varpiVar^* }{(\hatmuVar^* +\varpiVar^* )(\mu + \delta) - \varpiVar^*  \delta}\right) = \left(\dfrac{f^*_q\Gamma}{\varpiVar ^*(1-f^*_qf_r)};\;\dfrac{f_rf^*_q\Gamma}{\delta(1-f^*_qf_r)}\right),
\end{equation}
where we have used the replacements \eqref{eq:frfq} to obtain the final equality in \eqref{eq:xsv}.

\noindent As a corollary we have

\begin{prop}\label{prop:stbsysred} The system \begin{equation}\label{eq.sysred}
\dot{\mathbf x} = \mathbf A_S(\mathbf x^*)\,.\,\left(\mathbf x-\mathbf x^*_S\right)
\end{equation}
 is GAS at $\mathbf x^*_S$ on $\mathbb R_+^{n+2}$.\end{prop}

\noindent{\em Proof }:~~ The system \eqref{eq.sysred} is a linear system with matrix $\mathbf A_S(\mathbf x^*)$ with eigenvalues $\tilde\nu_i-\nu_i$ $i=1,\cdots,n$ from the sub-matrix $\mathbf A_{S_h}(\mathbf x)$ and $-\dfrac{(\hatmuVar^*+\varpiVar^*+\mu+\delta)\pm\sqrt{(\hatmuVar^*+\varpiVar^*-\mu-\delta)^2+4\varpiVar^*\delta}}{2}$ from the sub-matrix $\mathbf A_{S_v}(\mathbf x)$ in \eqref{eq:A_S_v}. Assuming $\nu_i> \tilde\nu_i$ $i=1,\cdots,n$ the first $n$ eigenvalues are negatives;  and since $(\hatmuVar^*+\varpiVar^*-\mu-\delta)^2+4\varpiVar^*\delta < (\hatmuVar^*+\varpiVar^*+\mu+\delta)^2$, it comes out that the two last eigenvalues are negatives; hence the result\edem

\subsection{Boundedness and dissipativity of the trajectories}
The following proposition characterizes the long-term behavior of the system.
\begin{prop}
 \label{prop:dissip}  The simplex
\begin{equation}\Omega =\left\{\mathbf x\in\mathbb R^u_+\;\left|\;\left(S_q\leqslant S_q^* \right)\;\wedge \;\left(S_r\leqslant S_r^*\right)\;\wedge\;\left(H_i= H^*_i,\;\; 1\leqslant i\leqslant
n \right) \wedge\; \left( M_I\leqslant\frac{\barvarphiVar^* }{\mu}S_q^* \right) \; \right.
\right\}\label{eq.simplex}\end{equation} 
is a compact forward-invariant and absorbing set for the
system~\eqref{eq:eqbednet_},  where
\[
 M_I\equiv {\dsum_{j=1}^\ell} \left(E_q^{(j)} + E_r^{(j)}\right) + E_r^{(\ell+1)} + I_q + I_r~;~~~~\barvarphiVar^* \equiv a{\dsum_{i=1}^n}h_i^*\phi_i(\xi_i+v_i\widetilde\xi_i),
\]
and where $(S_q^*;\; S_r^*)$ are the DFE components for naive questing and resting vectors respectively (given in \eqref{eq:xsv}).
\end{prop}

Note that  $M_I$ is the overall population of non-naive vectors;  $\barvarphiVar^* $  is the maximum incidence rate of infection for questing susceptible vectors; and $h_i^* = H_i^*/H^*$. 
  
\noindent{\em Proof }:~~
Considering equations of the  system~\eqref{eq:eqbednet_} that describe the dynamic of the transmission in the vectors population, we write the system:  
\begin{equation}
\left\{
\begin{array}{lcl}
\dot S_q & = &\Gamma- (\hatmuVar +\varpiVar ) S_q+\delta S_r\\
\dot S_r & = & (\varpiVar -\varphiVar ) S_q-(\mu+\delta)S_r\\
\dot M_{I_v} & = & \varphiVar   S_q - \mu M_{I_v} - dM_{I_q}\\
\end{array}
\right. 
\label{eq:eqbednet_p}
\end{equation} 
constituted of the two first equations of the system~\eqref{eq:eqbednet_}, and the last made by adding equations that describe the evolution of all other components (components concerning non naive vectors) of the state of the model concerning vectors, to which we give the identification $M_{I_v}$.  $M_{I_q}$ stands  for the overall population of non naive vectors in the questing states. The evolution of other components of the state of the model are kept implicit since they are not needed in the proof.  Let $\left(S_q(t),\;S_r(t),\; M_{I_v}(t)\right)$ be the solution of the system~\eqref{eq:eqbednet_p} beginning at a given initial state $(S_q(0),\;S_r(0),M_{I_v}(0))\in\mathbb R^3_+$. Consider now the system
\begin{equation}
\left\{
\begin{array}{lcl}
\dot S_q & = &\Gamma- (\hatmuVar +\varpiVar ) S_q+\delta S_r\\
\dot S_r & = & \varpiVar  S_q-(\mu+\delta)S_r\\
\end{array}
\right. 
\label{eq:eqbednet_pu}
\end{equation} 

\noindent This is the canonical projection of the system~\eqref{eq:eqbednet_p} on the disease free sub-variety. As stated in Proposition~\ref{prop:stbsysred}, the unique equilibrium of system~\eqref{eq:eqbednet_pu} that is $\mathbf x_{S_v}^*$ is globally asymptotically stable in $\mathbb R^2_+$.  If $(\bar S_q(t),\;\bar S_r(t))$ is the solution of the system~\eqref{eq:eqbednet_pu}, with the initial state $(S_q(0);\;S_r(0))$,  
we have $0\leqslant S_r(t)\leqslant\bar S_r(t)$, since $(\varpiVar -\varphiVar )S_q-(\mu+\delta)S_r\leqslant \varpiVar  S_q-(\mu+\delta)S_r$; we have also $0\leqslant S_q(t)\leqslant\bar S_q(t)$ since at any $t$, $\Gamma-(\hatmuVar +\varpiVar )S_q+\delta S_r(t)\leqslant \Gamma-(\hatmuVar +\varpiVar )S_q+\delta \bar S_r(t)$. For any $\varepsilon>0$, there exists a $t_\varepsilon>0$ such that for all $t>t_\varepsilon$, we have  $0\leqslant S_r(t)\leqslant\bar S_r(t)<S_r^*+\varepsilon$, and $0\leqslant S_q(t)\leqslant\bar S_q(t)<S_q^*+\varepsilon$. For the third equation of eq.~\eqref{eq:eqbednet_p}, it follows that $0\leqslant M_{I}(t)\leqslant\bar M_{I}(t)<\frac{r^*_{max}}{\mu}S^*_q+\varepsilon$ where $\bar M_{I_v}(t)$ is the solution of the equation $\dot M_{I}=r^*_{max}  S_q^*-\mu M_{I}$ with initial condition $M_{I_v}^0$. This proves the attractiveness of the set $\Omega$. The invariance of $\Omega$ is straightforward. \edem 

As a result of Proposition~\ref{prop:dissip}, we may limit our study to the simplex specified in eq.~\eqref{eq.simplex}.

\subsection{Computation of the threshold condition}\label{sec.algo}
The following proposition gives a formula for the basic reproduction number $\mathcal R_0$, and  shows that the condition $\mathcal R_0<1$ is a necessary and sufficient condition  for local stability of the DFE. 

\begin{prop}\label{prop:basicrepn}
The basic reproduction number $\mathcal R_0$ of the system \eqref{eq:eqmodel} is 
\begin{equation}
\label{eq:R0} \mathcal
R_0 \equiv \frac{({f_q^*}f_r)^{\ell+1}}{(1-f_q^*f_r)^2}\dfrac{f_q^*}{{\varpiVar ^*}^2}\dfrac{\Gamma}{H^*}a^2\sum_{i=1}^nh_i^*\frac{\varphi_i\phi_i}{\gamma_i+\nu_i}\frac{\varepsilon_i+\bar u_i\nu_i }{\varepsilon_i+\nu_i}\left(\xi_{i}+\frac{v_i\widetilde \xi_{i}\gamma_i}{\zeta_i+\nu_i}\right),
\end{equation}  
where $f^*_q$ and $f_r$ are the questing and resting frequencies respectively of vectors at the DFE, $H^*\equiv {\dsum_{i=0}^n}\Lambda_i/({\nu_i-\widetilde \nu_i})\equiv {\dsum_{i=0}^n}H^*_i$ is the total host population at the DFE, and $\biVar^*\equiv {H^*_i}/{H^*}$ is the proportion of hosts in group $i$ at the DFE. Then $\mathcal R_0<1$ is a necessary and sufficient condition  for local stability of the DFE. 

\end{prop}

\noindent{\em Proof }:~~
We prove first the stability condition $\mathcal R_0 < 1$ for the system~\eqref{eq:eqmodel} at the DFE where $\mathcal R_0$ is given by~\eqref{eq:R0}. Then we show that $\mathcal R_0$ can be interpreted as the basic reproduction number. 

On the infection-free sub-variety of $\left(\mathbb R_+\right)^u$, system~\eqref{eq:eqmodel} is reduced to system~\eqref{eq.sysred} which  has a unique equilibrium $\mathbf x^*_S$ that is GAS according to Proposition~\ref{prop:stbsysred}. Thus to ensure local stability of~\eqref{eq:eqmodel} at the DFE, it is necessary and sufficient that  the sub-matrix $\mathbf A_I(\mathbf x^*)$ defined by~\eqref{eq:eqmodel1} is stable, since  $\mathbf A_I(\mathbf x^*)$ is the Jacobian matrix of the system~\eqref{eq:eqmodel} reduced to the infected sub-variety. 

\noindent The stability of $\mathbf A_I(\mathbf x^*)$ is established as follows. Since  $\mathbf A_I(\mathbf x^*)$ is a Metzler  matrix, according to Proposition~\ref{prop:blockdecomposition}  we have that $\mathbf A_I(\mathbf x^*)$ is Metzler stable if and only if  $\mathbf A_{I_E}(\mathbf x^*)$ and $\mathbf N(\mathbf x^*) \equiv \mathbf A_{I_I}(\mathbf x^*) - \mathbf A_{I_{I,\,E}}(\mathbf x^*){\mathbf A_{I_E}(\mathbf x^*)}^{-1}\mathbf A_{I_{E,\,I}}(\mathbf x^*)$ are Metzler stable. Since $\mathbf A_{I_E}(\mathbf x^*)$ is always Metzler stable, we may focus on $\mathbf N(\mathbf x^*)$, which is a $n+2\times n+ 2$ matrix that can be written as 
$$\mathbf N(\mathbf x^*)=\left(\begin{array}{cc} \mathbf N_{1\,1}(\mathbf x^*)&\mathbf N_{1\,2}(\mathbf x^*)\\\mathbf N_{2\,1}(\mathbf x^*)&\mathbf N_{2\,2}(\mathbf x^*) \end{array}\right),$$ 
with
$\mathbf N_{1\,1}(\mathbf x^*)=\mathbf A_{I_{I_h}}$, $\mathbf N_{2\,2}(\mathbf x^*)=\mathbf A_{I_{I_v}}$,
$\mathbf N_{1\,2}(\mathbf x^*)^{\mathrm T}=a\begin{pmatrix}
\frac{h_1^*\varphi_1(\varepsilon_1+\bar u_1\nu_1)}{\varepsilon_1+\nu_1} & \cdots & \frac{h_n^*\varphi_n(\varepsilon_n+\bar u_n\nu_n)}{\varepsilon_n+\nu_n}\cr 0&\cdots & 0
\end{pmatrix}$  and $$\mathbf N_{2\,1}(\mathbf x^*)=f_r^{\ell+1}{f_q^*}^\ell\dfrac{a}{H^*}S_q^*\begin{pmatrix}\phi_{1} \xi_{1} &  \cdots &  \phi_{n}  \xi_{n} &v_1\phi_{1}  \widetilde\xi_{1}   & \cdots & v_n\phi_{n} \widetilde\xi_{n}   \\
0 &  \cdots & 0 & 0 &  \cdots & 0 \end{pmatrix}.$$

\noindent Once again we may apply Proposition~\ref{prop:blockdecomposition}, this time to  $\mathbf N(\mathbf x^*)$, and conclude that 
$\mathbf N(\mathbf x^*)$ is Metzler stable if and only if $\mathbf N_{1\,1}(\mathbf x^*)$ and $\mathbf L(\mathbf x^*) \equiv  \mathbf N_{2\,2}(\mathbf x^*)-\mathbf N_{2\,1}(\mathbf x^*){\mathbf N_{1\,1}(\mathbf x^*)}^{-1}\mathbf N_{1\,2}(\mathbf x^*)$ are Metzler stable. Since $\mathbf N_{1\,1}(\mathbf x^*)$ is always Metzler stable, we only need to establish the Metzler stability of $\mathbf L(\mathbf x^*)$, which may be written more explicitly as
\begin{equation}\label{eq:Lmx}  
\mathbf L(\mathbf x^*) =\left(\begin{array}{cc}\varpi(\mathbf
x^*)-(\hatmuVar ^*+\varpiVar ^*)&\delta\\\varpiVar ^*&-(\mu+\delta)\end{array}\right),
\end{equation}
where
\begin{equation}\label{eq:coeffspec}
\varpi(\mathbf x^*)=f_r^{\ell+1}{f_q^*}^\ell \dfrac{S^*_q}{H^*}a^2\sum_{i=1}^n\frac{h^*_i\varphi_i\phi_i}{\gamma_i+\nu_i}\frac{\varepsilon_i+\bar u_i \nu_i}{\varepsilon_i+\nu_i}\left(\xi_{i}+\frac{v_i\widetilde \xi_{i}\gamma_i}{\zeta_i+\nu_i}\right)
\end{equation}
\noindent 
We make one final application of Proposition~\ref{prop:blockdecomposition}. Since $\mathbf L(\mathbf
x^*)_{2,\,2}$ is negative and hence Metzler stable,  the Metzler stability of   $\mathbf
A_I(\mathbf x^*)$ holds if and only if
$$\mathbf
L(\mathbf x^*)_{1\,1}-\mathbf L(\mathbf x^*)_{1\,2}\cdot{\mathbf L(\mathbf x^*)_{2\,2}}^{-1}\cdot\mathbf L(\mathbf x^*)_{2\,1}<0,$$  
which in view of (\ref{eq:Lmx}) and (\ref{eq:coeffspec}) may be rewritten as
\begin{equation}\label{eq:MetzlerCondition}
\dfrac{\mu+\delta}{(\mu+\delta)(\hatmuVar ^*+\varpiVar ^*)
	-\varpiVar ^*\delta}f_r^{\ell+1}{f_q^*}^\ell\dfrac{S^*_q}{H^*}a^2\sum_{i=1}^nh_i^*\frac{\varphi_i\phi_i}{\gamma_i+\nu_i}\frac{\varepsilon_i+\bar u_i\nu_i }{\varepsilon_i+\nu_i}\left(\xi_{i}+\frac{v_i\widetilde \xi_{i}\gamma_i}{\zeta_i+\nu_i}\right)<1.
\end{equation}
Using expressions from \eqref{eq:xsv}, we may simplify \eqref{eq:MetzlerCondition} to obtain the following necessary and sufficient condition for Metzler stability of the matrix $\mathbf A_I(\mathbf x^*)$ :

\begin{equation}\label{eq:thrshold}
\frac{({f_q^*}f_r)^{\ell+1}}{(1-f_q^*f_r)^2}\dfrac{f_q^*}{{\varpiVar ^*}^2}\dfrac{\Gamma}{H^*}a^2\sum_{i=0}^nh_i^*\frac{\varphi_i\phi_i}{\gamma_i+\nu_i}\frac{\varepsilon_i+\bar u_i\nu_i }{\varepsilon_i+\nu_i}\left(\xi_{i}+\frac{v_i\widetilde \xi_{i}\gamma_i}{\zeta_i+\nu_i}\right)<1.
\end{equation}
We now show directly that the coefficient in the left of the condition~\eqref{eq:thrshold} is the basic reproduction number $\mathcal R_0$, which can be represented as
\begin{equation}\label{eq:R0formula}
\mathcal R_0 = \sum_{i=1}^n \Big(\mathcal R_0^{vh_{I_i}} \mathcal R_0^{h_{I_i}v}+\mathcal R_0^{vh_{R_i}} \mathcal R_0^{h_{R_i}v}\Big),
\end{equation}
where $ \mathcal R_0^{vh_{R_i}}$ and $ \mathcal R_0^{vh_{I_i}}$ are the average number of hosts in group $i$ in the resistant and infectious state respectively, infected by a single infectious vector during the course of its remaining lifetime; $\mathcal R_0^{h_{R_i} v}$ and $\mathcal R_0^{h_{I_i} v}$  are the average number of vectors which become infectious due to interactions with a single resistant and infectious host respectively  in group $i$. We may characterize $\mathcal R_0^{vh_{I_i}}$ and  $\mathcal R_0^{vh_{R_i}}$ as
\begin{align*}
\mathcal R_0^{vh_i} =  a\miVar\frac{S_i^*}{H^*}\cdot\frac{1}{\hatmuVar ^*+ \varpiVar ^*}\sum_{j=0}^{\infty} (f_q^* f_r)^{j}  =   \frac{f_q^*} {1-f_q^* f_r} \cdot  \frac{a h_i^* \miVar}{\varpiVar ^*},  
\end{align*}
where $a\miVar\frac{S^*_i}{H^*}$ is the mean number of susceptible hosts in group $i$ per unit time that have caught the disease due to their contact with a single infectious questing vector; and  $\frac{1}{\hatmuVar ^*+ \varpiVar ^*}\dsum_{j=0}^{\infty} (f_q^* f_r)^{j}$ is the mean total questing duration for an infectious questing vector, since after each successful interaction  the infectious vector will enter the infectious resting state before returning to the infectious questing state, and the mean  duration of the $j^{th}$ infectious questing period is $\frac{(f_q^* f_r)^{j-1}}{\hatmuVar ^*+ \varpiVar ^*}\equiv\frac{f_q^*}{\varpiVar ^*}(f_q^* f_r)^{j-1}$.

\noindent We may also characterize $\mathcal R_0^{h_iv}$ as $\mathcal R_0^{h_iv}= \mathcal R_0^{h_{I_i}v}+\mathcal R_0^{h_{R_i}v}$ with
\begin{equation}\label{eq:R0hiiv}
\mathcal R_0^{h_{I_i}v} = \frac{a\phi_i\xi_i}{H^*}S_q^* \cdot  f_r(f_q^* f_r)^{\ell} \cdot \frac{\varepsilon_i}{\nu_i + \varepsilon_i}  \cdot \frac{1}{\nu_i + \gamma_i} = f_r(f_q^* f_r)^{\ell}\frac{a\varepsilon_i\phi_i\xi_iS_q^*}{H^*(\nu_i + \gamma_i)(\nu_i + \varepsilon_i)}.  
\end{equation}
\begin{equation}\label{eq:R0hirv}
\mathcal R_0^{h_{R_i}v} = \frac{a\phi_i\widetilde\xi_i}{H^*}S_q^* \cdot  f_r(f_q^* f_r)^{\ell} \cdot \frac{\varepsilon_i}{\nu_i + \varepsilon_i}  \cdot \frac{\gamma_i}{\nu_i + \gamma_i} \cdot \frac{1}{\nu_i + \zeta_i} = f_r(f_q^* f_r)^{\ell}\frac{a\varepsilon_i\gamma_i\phi_i\widetilde\xi_iS_q^*}{H^*(\nu_i + \gamma_i)(\nu_i + \varepsilon_i)(\nu_i + \zeta_i)}.  
\end{equation}
In these expressions, $\frac{a\xi_i\phi_iS_q^*}{H^*}$ and $\frac{a\widetilde\xi_i\phi_iS_q^*}{H^*}$ are rates at which susceptible questing vectors are infected through interactions with a single infectious and resistant individual of the $i^{th}$ host group respectively, and  $f_r(f_rf_q^*)^\ell$ is the frequency of survival through all $2\ell+1$ exposed states of the vector's dynamics. It follows that the product of $f_r(f_rf_q^*)^\ell$ with each of these two former factors gives the rate of production of infectious vectors due to the presence of a single individual in the $i^{th}$ host group that is a carrier of the parasite. The additional factors $\frac{\varepsilon_i}{\nu_i+\varepsilon_i}$ and $\frac{\gamma_i}{\nu_i+\gamma_i}$ in \eqref{eq:R0hiiv} and \eqref{eq:R0hirv} represent the average of host individuals of the $i^{th}$ host group that reach the infectious and resistant states respectively, while the factors $\frac{1}{\nu_i+\gamma_i}$ and $\frac{1}{\nu_i+\zeta_i}$ represent the mean duration of infectiousness of an infected host and a resistant host respectively.

Plugging these formulas for $\mathcal R_0^{h_iv}$ and $\mathcal R_0^{h_iv}$ into expression \eqref{eq:R0formula}
for $\mathcal R_0$ yields the left-hand side of \eqref{eq:thrshold}, thus completing the proof.\edem

\subsection{ Endemic equilibrium (EE) of the system} 
Our system has a unique endemic equilibrium that is  specified by the following proposition.
\begin{prop}\label{prop:stdst1}
System~\eqref{eq:eqmodel} admits a unique endemic equilibrium (EE) $\mathbf x^\dagger \in{\mathbb R_{>0}}^+$ with components given by

\begin{equation}\label{eq:eqexpMinfctstateb}
\begin{array}{l}
S_q^\dagger =\frac{f^*_q}{\varpiVar ^*(1-f^*_qf_r)}\Gamma - \frac{I_q^\dagger }{(f^*_qf_r)^\ell};\qquad S_r^\dagger =\frac{\delta f^*_q}{f_r(1-f^*_qf_r)}\Gamma - \frac{\varpiVar ^*}{\delta}\frac{(1-(f^*_qf_r)^{\ell+1}+f^*_qf_r)I_q^\dagger }{ f^*_q}; \\ 
E_q^{(j)\dagger }=\dfrac{1-f^*_qf_r }{(f^*_qf_r)^{\ell+1-j}}I_q^\dagger ; 
\qquad E_r^{(j)\dagger }=\frac{\varpiVar ^* }{\delta }\dfrac{1-f^*_qf_r}{ f^*_q(f^*_qf_r)^{\ell+1-j}} I_q^\dagger \qquad(1\leqslant j\leqslant \ell);
\\
I_r^\dagger =\frac{\varpiVar ^* }{\delta }f_rI_q^\dagger ; \qquad E_r^{(\ell+1)\dagger }=\frac{\varpiVar ^* }{\delta }\frac{1-f^*_qf_r}{f^*_q} I_q^\dagger ;\\
S_i^\dagger = \dfrac{(\varepsilon_i+\nu_i) (\gamma_i+\nu_i)(\zeta_i+\nu_i)H^*}{ (\varepsilon_i+\nu_i)(\gamma_i+\nu_i)(\zeta_i+\nu_i)H^* + \Big(u_i(\gamma_i+\nu_i)(\zeta_i+\nu_i) + (\bar u_i\nu_i+\varepsilon_i)(v_i\gamma_i + \zeta_i+\nu_i) \Big)a\varphi_i I^\dagger_q}H^*_i;\\I_i^\dagger  = \dfrac{a\varphi_i I^\dagger_q(\varepsilon_i + \bar u_i\nu_i)(\zeta_i+\nu_i)}{ (\varepsilon_i+\nu_i)(\gamma_i+\nu_i)(\zeta_i+\nu_i)H^* + \Big(u_i(\zeta_i+\nu_i)(\gamma_i+\nu_i) + (\bar u_i\nu_i+\varepsilon_i)(v_i\gamma_i + \zeta_i+\nu_i)\Big)a\varphi_i I^\dagger_q}H^*_i;\\ E^\dagger _i =  \dfrac{a\varphi_i I^\dagger_q(\gamma_i+\nu_i)(\zeta_i+\nu_i)}{ (\varepsilon_i+\nu_i)(\gamma_i+\nu_i)(\zeta_i+\nu_i)H^* + \Big(u_i(\gamma_i+\nu_i)(\zeta_i+\nu_i) + (\bar u_i\nu_i+\varepsilon_i)(v_i\gamma_i + \zeta_i+\nu_i)\Big)a\varphi_i I^\dagger_q}H^*_i,
\end{array}
\end{equation}

$(1\leqslant i\leqslant n)$, and where $I_q^\dagger \in~\big]0,\;\IqstarmaxVar \big[$ is the unique finite root of the equation \begin{equation}\small\label{eq:eqeqIQ}a^2\dsum_{i=1}^n\, \frac{h_i^*\phi_i\varphi_i (\varepsilon_i+\bar u_i\nu_i)\left(\widetilde\xi_iv_i\gamma_i+\xi_i(\zeta_i+\nu_i)\right)}{ (\varepsilon_i+\nu_i)(\gamma_i+\nu_i)(\zeta_i+\nu_i)H^* + \Big(u_i(\gamma_i+\nu_i)(\zeta_i+\nu_i) + (\bar u_i\nu_i+\varepsilon_i)(v_i\gamma_i + \zeta_i+\nu_i) \Big)a\varphi_i x}
=\dfrac{{\varpiVar ^*}^2(1-f^*_qf_r)^2 }{f^*_q(f^*_qf_r)^{\ell+1}\Gamma-\varpiVar ^* f^*_qf_r (1-f^*_qf_r)x }.
\end{equation}
and
\begin{equation}\label{eq:Iq-bar}
\IqstarmaxVar  \equiv \frac{\Gamma}{\varpiVar ^*}\frac{f^*_q(f^*_qf_r)^{\ell+1}}{1-f^*_qf_r}
\end{equation}
\end{prop}

\noindent{\em Proof }:~~
The purpose of Proposition~\ref{prop:stdst1} is to specify all  possible endemic equilibria (EE) of the system~\eqref{eq:eqbednet_}.
At an endemic state, at  least one of the infected or infectious components of the solution  $\mathbf x^\dagger $ must nonzero.  Since the continuing presence of disease requires questing infectious vectors that successfully transmits disease to a host in one of the host sub-population, we may assume that $I_q^\dagger \neq0$ where $I_q^\dagger $ is the component of $\mathbf x^\dagger $  corresponding to the Infectious questing vectors.

The dynamics of the overall population of vectors is given by the following set of differential equations: \begin{equation}\left\{\begin{array}{ccl}\dot M_q&=&\Gamma-(\hatmuVar +\varpiVar )M_q+\delta M_r\\\dot M_r &=&\varpiVar  M_q-(\mu+\delta)M_r,\end{array}\right.
\end{equation}
where $M_q$ and $M_r$ are the total numbers of questing and resting vectors, respectively.
It follows that the steady-state values  $M_q$,  $M_r$  must satisfy $M_q^*=\frac{(\mu+\delta)\Gamma}{(\mu+\delta)(\hatmuVar ^*+\varpiVar ^*)-\varpiVar ^*\delta}=\frac{f^*_q}{\varpiVar ^*(1-f^*_qf_r)}\Gamma = M_q^\dagger$  and   $M_r^* =\frac{\delta f^*_q}{f_r(1-f^*_qf_r)}\Gamma = M_r^\dagger$. Since $\hatmuVar$ and $\varpiVar $ depend only on the host dynamics and not on the presence or absence of infection, it follows that $\hatmuVar^*$ and $\varpiVar ^*$ for any possible EE must be the same as in the DFE.

Using system \eqref{eq:eqbednet_}, it is possible to solve for the various components of the infected vector population at the EE  in terms of $I_q^\dagger $:
\begin{equation}\label{eq:eqexpMinfctstate_}\begin{array}{ll}
I_r^\dagger =\frac{\varpiVar ^*}{(\mu+\delta)}I_q^\dagger =\frac{\varpiVar ^* }{\delta }f_rI_q^\dagger \hbox{, } & E_r^{(\ell+1)\dagger }=\frac{\varpiVar ^*}{\delta }\frac{1-f^*_qf_r}{f^*_q} I_q^\dagger ,\\ E_q^{(i)\dagger }=\frac{1-f^*_qf_r }{(f^*_qf_r)^{\ell+1-i}}I_q^\dagger \hbox{, } & E_r^{(i)\dagger }=\frac{\varpiVar ^*}{\delta }\frac{1-f^*_qf_r}{ f^*_q(f^*_qf_r)^{\ell+1-i}} I_q^\dagger \quad\hbox{, for } 1\leqslant i\leqslant \ell.\end{array}.
\end{equation}
From \eqref{eq:eqexpMinfctstate_} it follows that $\dsum_{i=1}^\ell E_q^{(i)\dagger }+I_q^\dagger =\frac{I_q^\dagger }{(f^*_qf_r)^\ell}$, and  since $M_q^\dagger=S^\dagger _q+\dsum_{i=1}^lE_q^{(i)\dagger }+I_q^\dagger $ we have
\begin{equation*}
S_q^\dagger =\frac{f^*_q}{\varpiVar ^*(1-f^*_qf_r)}\Gamma - \frac{I_q^\dagger }{(f^*_qf_r)^\ell}.\label{eq:eqexpMsucpqst}
\end{equation*}
Similarly, we find
\begin{equation}
S_r^\dagger =M_r^\dagger -\dsum_{i=1}^{\ell+1}E_r^{(i)\dagger }-I_r^\dagger =\frac{\delta f^*_q}{f_r(1-f^*_qf_r)}\Gamma - \frac{\varpiVar ^*}{\delta}\frac{(1-(f^*_qf_r)^{\ell+1}+f^*_qf_r)I_q^\dagger }{ f^*_q}\label{eq:eqexpMsucprst}.
\end{equation}
Equations \eqref{eq:eqexpMinfctstate_} and \eqref{eq:eqexpMsucprst} uniquely specify all vector components at EE in terms of $I_q^\dagger $.

Concerning host populations in the model at EE, for each $i$, the components $R^\dagger _i$, $I^\dagger _i$, $E^\dagger_i$ and $S^\dagger _i$ satisfy $a\,\varphi_i \frac{I_q^\dagger }{H^*}S^\dagger _i -(\nu_i+\varepsilon_i) E_i^\dagger =0$, $\bar u_ia\,\varphi_i \frac{I_q^\dagger }{H^*}S^\dagger _i+u_i\varepsilon_iE^\dagger_i=(\gamma_i+\nu_i)I^\dagger_i$, $\gamma_iI^\dagger_i=(\zeta_i+\nu_i)R^\dagger_i$, $\Lambda_i+\widetilde \nu_i H^*_i - \left(\nu_i+a\varphi_i \frac{I^\dagger_q}{H^*}\right)S^\dagger_i + v_i\zeta_iR^\dagger_i+\bar v_i\gamma_iI^\dagger_i=0$ and $R_i^\dagger +I_i^\dagger+E^\dagger_i +S^\dagger _i = H^*_i=\frac{\Lambda_i}{\nu_i-\widetilde \nu_i}$.

 This yields: \[R^\dagger _i =  \frac{a\varphi_i I^\dagger_q(\varepsilon_i+\bar u_i\nu_i)\gamma_i}{ (\varepsilon_i+\nu_i)(\gamma_i+\nu_i)(\zeta_i+\nu_i)H^* + \Big(u_i(\gamma_i+\nu_i)(\zeta_i+\nu_i) + (\bar u_i\nu_i+\varepsilon_i)(v_i\gamma_i + \zeta_i+\nu_i)\Big)a\varphi_i I^\dagger_q}H^*_i,\]
$$I_i^\dagger = \frac{a\varphi_i I^\dagger_q(\varepsilon_i + \bar u_i\nu_i)(\zeta_i+\nu_i)}{ (\varepsilon_i+\nu_i)(\gamma_i+\nu_i)(\zeta_i+\nu_i)H^* + \Big(u_i(\gamma_i+\nu_i)(\zeta_i+\nu_i) + (\bar u_i\nu_i+\varepsilon_i)(v_i\gamma_i + \zeta_i+\nu_i)\Big)a\varphi_i I^\dagger_q}H^*_i,
$$

$$E^\dagger _i =  \frac{a\varphi_i I^\dagger_q(\gamma_i+\nu_i)(\zeta_i+\nu_i)}{ (\varepsilon_i+\nu_i)(\gamma_i+\nu_i)(\zeta_i+\nu_i)H^* + \Big(u_i(\gamma_i+\nu_i)(\zeta_i+\nu_i) + (\bar u_i\nu_i+\varepsilon_i)(v_i\gamma_i + \zeta_i+\nu_i)\Big)a\varphi_i I^\dagger_q}H^*_i,$$ and $$S^\dagger _i  =  \frac{(\varepsilon_i+\nu_i)(\gamma_i+\nu_i)(\zeta_i+\nu_i)H^*}{ (\varepsilon_i+\nu_i)(\gamma_i+\nu_i)(\zeta_i+\nu_i)H^* + \Big(u_i(\gamma_i+\nu_i)(\zeta_i+\nu_i) + (\bar u_i\nu_i+\varepsilon_i)(v_i\gamma_i + \zeta_i+\nu_i) \Big)a\varphi_i I^\dagger_q}H^*_i.$$

\noindent All of the components at EE given above require the existence of  a feasible nonzero $I_q^\dagger $. To determine this component, we use  two  expressions of $\varphiVar $ that hold at the EE.
The first comes  from the equality $\varphiVar ^\dagger  S^\dagger _q-(\mu+\delta)E_r^{(1)\dagger }=0$ (third equation of~\eqref{eq:eqbednet_}), so that: \begin{equation}\label{eq:eqphi}r_{inf}^\dagger  = \frac{\delta}{f_r}\frac{E_r^{(1)\dagger }}{S^\dagger _q}=\frac{{r_{suc}^*}^2(1-f^*_qf_r)^2 I_q^\dagger }{f^*_q(f^*_qf_r)^{\ell+1}\Gamma-r_{suc}^* f^*_qf_r (1-f^*_qf_r)I_q^\dagger }.\end{equation}
The second comes from  the expression for $r_{inf}$ in Table~\ref{tab.tabvd2}:  $r_{inf}^\dagger = \frac{a}{H^*}\dsum_{i=1}^n\,\phi_i (\xi_i I_i^\dagger+v_i\widetilde\xi_i R_i^\dagger) $.   Rewriting this in terms of  $I^\dagger_q$ gives:
\begin{equation}\label{eq:eqphi1}r_{inf}^\dagger = a^2I^\dagger_q\dsum_{i=1}^n\, \frac{h_i^*\phi_i\varphi_i (\varepsilon_i+\bar u_i\nu_i)\left(\widetilde\xi_iv_i\gamma_i+\xi_i(\zeta_i+\nu_i)\right)}{ (\varepsilon_i+\nu_i)(\gamma_i+\nu_i)(\zeta_i+\nu_i)H^* + \Big(u_i(\gamma_i+\nu_i)(\zeta_i+\nu_i) + (\bar u_i\nu_i+\varepsilon_i)(v_i\gamma_i + \zeta_i+\nu_i) \Big)a\varphi_i I^\dagger_q}.\end{equation}

Setting~\eqref{eq:eqphi} equal to \eqref{eq:eqphi1} gives  the following equation with $I^\dagger _q$ as unknown: \begin{equation}\label{eq:eqeqIQ1}\small a^2\dsum_{i=1}^n\, \frac{h_i^*\phi_i\varphi_i (\varepsilon_i+\bar u_i\nu_i)\left(\widetilde\xi_iv_i\gamma_i+\xi_i(\zeta_i+\nu_i)\right)}{ (\varepsilon_i+\nu_i)(\gamma_i+\nu_i)(\zeta_i+\nu_i)H^* + \Big(u_i(\gamma_i+\nu_i)(\zeta_i+\nu_i) + (\bar u_i\nu_i+\varepsilon_i)(v_i\gamma_i + \zeta_i+\nu_i) \Big)a\varphi_i I^\dagger_q}
=\dfrac{{\varpiVar ^*}^2(1-f^*_qf_r)^2 }{f^*_q(f^*_qf_r)^{\ell+1}\Gamma-\varpiVar ^* f^*_qf_r (1-f^*_qf_r)I_q^\dagger },
\end{equation}
which is the determining equation for $I_q^*$ as specified in \eqref{eq:eqeqIQ}.

We now show that \eqref{eq:eqeqIQ1} has a unique biologically-feasible solution. For simplicity we express the solution in terms of the new variables 
\begin{equation}\label{eq:xdef}
x \equiv a\frac{I^\dagger_q}{H^*}; \qquad
\alpha_i \equiv \dfrac{h_i^*\phi_i (\varepsilon_i + \bar u_i\nu_i)\left(\widetilde\xi_iv_i\gamma_i +\xi_i(\zeta_i+\nu_i)\right)}{(\varepsilon_i+\nu_i)(\gamma_i+\nu_i)(\zeta_i+\nu_i)}; \qquad s \equiv \frac{\varpiVar ^*}{a}\frac{1-f^*_qf_r}{f^*_q(f^*_qf_r)^\ell}\frac{H^*}{\Gamma},
\end{equation}
where $x$ uniquely determines $I^\dagger _q$.  
From expression \eqref{eq:R0} for $\mathcal R_0$, it follows

\begin{equation}\label{eq:eqrelpo}s\dfrac{1-f^*_qf_r}{f^*_qf_r}\dfrac{\varpiVar ^*}{a}=\dfrac{{\varpiVar ^*}^2(1-f^*_qf_r)^2}{f^*_q(f^*_qf_r)^{\ell+1}}\dfrac{H^*}{a^2\,\Gamma}=\frac{1}{\mathcal R_0}\dsum_{i=0}^{n}\alpha_i \varphi_i.\end{equation}
The expression~\eqref{eq:eqeqIQ1} then becomes:
\begin{equation}\label{eq:eqeqIQnew}\dsum_{i=1}^{n}\alpha_i \miVar\left(\dfrac{1}{\mathcal R_0 (1-sx)}-\frac{(\varepsilon_i+\nu_i)(\gamma_i+\nu_i)(\zeta_i+\nu_i)}{(\varepsilon_i+\nu_i)(\gamma_i+\nu_i)(\zeta_i+\nu_i) + \left(u_i(\gamma_i+\nu_i)(\zeta_i+\nu_i) + (\bar u_i\nu_i+\varepsilon_i)(v_i\gamma_i + \zeta_i+\nu_i) \right)\varphi_i x} \right)=0\end{equation}

\noindent Using the same strategy as in~\cite{jckam201411}, we set:


\begin{eqnarray*}\label{eq:eqpolIQnew}
	T(x)& \equiv&{\dsum_{i=1}^{n}}\alpha_i \varphi_i\left(\dfrac{1}{\mathcal R_0 (1-sx)}-\frac{(\varepsilon_i+\nu_i)(\gamma_i+\nu_i)(\zeta_i+\nu_i)}{(\varepsilon_i+\nu_i)(\gamma_i+\nu_i)(\zeta_i+\nu_i) + \left(u_i(\gamma_i+\nu_i)(\zeta_i+\nu_i) + (\bar u_i\nu_i+\varepsilon_i)(v_i\gamma_i + \zeta_i+\nu_i) \right)\varphi_i x}\right).
\end{eqnarray*} 
Since $T(x)$ is a rational function, it is of class $\mathcal C^\infty$ on $\R_+\setminus\left\{\frac{1}{s}\right\}$. Furthermore, we  have  
$T(0)=\dfrac{1-\mathcal R_0}{\mathcal R_0}{\dsum_{i=1}^{n}}\alpha_i\miVar,$ 
which implies that $T(0)<0$ whenever $\mathcal R_0>1$. We have also that $\dlim_{x\rightarrow\frac{1}{s}^-}T(x)=+\infty$, $\dlim_{x\rightarrow\frac{1}{s}^+}T(x)=-\infty$, and $\dlim_{x\rightarrow+\infty}T(x)=0$.
The derivative of the function $T$ is 
$$\frac{d\,T}{dx}(x) = {\dsum_{i=1}^{n}}\alpha_i \miVar\left(\frac{s}{\mathcal R_0(1-sx)^2} + \frac{(\varepsilon_i+\nu_i)(\gamma_i+\nu_i) (\zeta_i + \nu_i)\left(u_i(\gamma_i+\nu_i)(\zeta_i+\nu_i) + (\bar u_i\nu_i+\varepsilon_i)(v_i\gamma_i + \zeta_i+\nu_i) \right)\varphi_i}{\left((\varepsilon_i+\nu_i)(\gamma_i+\nu_i)(\zeta_i+\nu_i) + \left(u_i(\gamma_i+\nu_i)(\zeta_i+\nu_i) + (\bar u_i\nu_i+\varepsilon_i)(v_i\gamma_i + \zeta_i+\nu_i) \right)\varphi_i x\right)^2} \right) ,$$ 
which is positive on $\R_+$ so that $T$ is an increasing function on $\mathbb R_+\setminus\left\{\frac{1}{s}\right\}$. The intermediate value theorem implies that there are two solutions for equation~\eqref{eq:eqeqIQnew}: a biologically feasible solution in the interval $\left]0,\;\;\frac{a}{\varpiVar ^*}\frac{f^*_q(f^*_qf_r)^\ell}{1-f^*_qf_r}\frac{\Gamma}{H^*}\right[$, and a second solution at  infinity, which is not biologically feasible. In view of the definition of $x$ in \eqref{eq:xdef}.

\noindent Moreover, from the formulas of components of the endemic  equilibrium of the system \eqref{eq:eqbednet_} (see eq.~\eqref{eq:eqexpMinfctstateb}),  it follows that 
\[S_q^\dagger  \varphiVar ^\dagger  = \Gamma \frac{f_q^*\varphiVar ^\dagger }{ \varpiVar ^*(1 - f_r f_q^*) +  f_q^*f_r \varphiVar^\dagger}. \]
Since $x / (a + bx)$ is a monotone increasing function of $x$ and $\varphiVar ^\dagger  < \barvarphiVar^* $, we have
\[S_q^\dagger  \varphiVar ^\dagger  < \Gamma \frac{f_q^*\barvarphiVar^*}{ \varpiVar ^*(1 - f_r f_q^*) +  f_q^*f_r \barvarphiVar^*}. \]
Using the equality $ \delta E_r^{(\ell+1)\dagger } = f_r (f_r f_q^*)^\ell S_q^\dagger  \varphiVar ^\dagger  = \frac{\varpiVar ^*}{f^*_q }(1-f^*_qf_r) I_q^\dagger $,   we have: 
\[
I_q^\dagger  < \Gamma \frac{f_q^*(f_r f_q^*)^{\ell+1}}{\varpiVar ^*(1 - f_q^*f_r)}  \left( \frac{ \barvarphiVar^*}{\varpiVar ^*(1 - f_r f_q^*) +  f_q^*f_r \barvarphiVar^*} \right),\]
that can also be written,
\[
I_q^\dagger  < \Gamma \frac{ f_q^*(f_r f_q^*)^{\ell+1}}{\varpiVar ^*(1 - f_q^*f_r)}  \left[ \frac{\barvarphiVar^*/\varpiVar ^*}{(1 -f_r f_q^*) +  f_r f_q^*\barvarphiVar^*/\varpiVar ^*  } \right].\]
Since $\barvarphiVar^*/\varpiVar ^* < 1$, we may once again use the monotonicity of $x/(a+bx)$ to obtain
\begin{equation}\label{eq:Iqbar_final}
I_q^\dagger  < \Gamma \frac{ f_q^*(f_r f_q^*)^{l+1}}{\varpiVar ^*(1 - f_q^*f_r)}\equiv \IqstarmaxVar .
\end{equation}
This ends the proof. \edem

\begin{rmq}\em \label{rmk.rplctm} According to \eqref{eq:eqeqIQ},  the dynamics of the vector population (reflected in the parameters $f_r, f_q$, and $\ell$) as well as the protection means used by the population (reflected in the parameter $\varpiVar $) 
strongly influence the location of the EE. This justifies our initial claim that vector dynamics and host protection means are important practical factors in determining the prevalence of infection.
\end{rmq}
%

\section{Stability of system equilibria}\label{sec.analysis}
In this section we analyze the stability of the system equilibria given in Propositions~\ref{prop:stdst0} and~~\ref{prop:stdst1}.

\subsection{Stability analysis of the disease free equilibrium (DFE)}\label{subsec.dfestabanan}
\noindent We have the following result about the global stability of the disease free equilibrium:
\begin{thm} \label{thm.defstab} When $\mathcal R_0 \leqslant 1$, then the DFE is GAS  in $\mathbb R_+^u$.
\end{thm}

\noindent{\em Proof }:~~
Our proof relies on Theorem 4.3 of \cite{KamSal07}, which establishes global asymptotic stability (GAS) for epidemiological systems that can be expressed in the matrix form~\eqref{eq:eqmodel}. This theorem is restated as Theorem~\ref{thm:kamsal} in the Appendix: for the proof, the reader may consult \cite{KamSal07}. To complete the proof, we need only to establish for the system~\eqref{eq:eqmodel} that the five conditions ({\bf h}1--{\bf h}5) required in Theorem~\ref{thm:kamsal} are satisfied when $\mathcal R_0\leqslant 1$:

\begin{enumerate}[({\bf h}1)]
\item 
This condition  is satisfied for the system~\eqref{eq:eqmodel} as a result of Proposition~\ref{prop:dissip}. 
\item
We note first that $n_S = n+2$, and the canonical projection of $\Omega$ on $\mathbb R_+^{n+2}$ is $\mathbb I = \{\mathbf x_{S_h}\}\times\left [0,\;\;S^*_q\right]\times \left [0,\;\;S^*_r\right]$. The system~\eqref{eq:eqmodel} reduced to this sub-variety is given in \eqref{eq.sysred}, and this system is GAS at its equilibrium ($\mathbf x^*_S$) as a result of  Proposition~\ref{prop:stbsysred}. 

\item
 We consider first the case $\ell=1$ and $n=2$. In this case, the matrix  $\mathbf A_I(\mathbf x)$ in the system \eqref{eq:eqmodel} is\\ $\small\left(\begin{array}{cccccccccccc}
 -\varepsilon_{1} - \nu_{1} & 0 & 0 & 0 & 0 & 0 & 0 & 0 & 0 & \frac{S_{1} a \varphi_{1}}{H} & 0 \\
 0 & -\varepsilon_{2} - \nu_{2} & 0 & 0 & 0 & 0 & 0 & 0 & 0 & \frac{S_{2} a \varphi_{2}}{H} & 0 \\
 0 & 0 & -\delta - \mu & 0 & 0 & \frac{S_{q} a \phi_{1} \xi_{1}}{H} & \frac{S_{q} a \phi_{2} \xi_{2}}{H} & \frac{S_{q} a \phi_{1}\widetilde\xi_1}{H} & \frac{S_{q} a\phi_{2}\widetilde  \xi_{2} }{H} & 0 & 0 \\
 0 & 0 & \delta & -d_{q} - r_{\mathit{suc}} & 0 & 0 & 0 & 0 & 0 & 0 & 0 \\
 0 & 0 & 0 & r_{\mathit{suc}} & -\delta - \mu & 0 & 0 & 0 & 0 & 0 & 0 \\
 \varepsilon_{1} & 0 & 0 & 0 & 0 & -\gamma_{1} - \nu_{1} & 0 & 0 & 0 & 0 & 0 \\
 0 & \varepsilon_{2} & 0 & 0 & 0 & 0 & -\gamma_{2} - \nu_{2} & 0 & 0 & 0 & 0 \\
 0 & 0 & 0 & 0 & 0 & \gamma_{1} & 0 & -\nu_{1} - \zeta_{1} & 0 & 0 & 0 \\
 0 & 0 & 0 & 0 & 0 & 0 & \gamma_{2} & 0 & -\nu_{2} - \zeta_{2} & 0 & 0 \\
 0 & 0 & 0 & 0 & \delta & 0 & 0 & 0 & 0 & -d_{q} - r_{\mathit{suc}} & \delta \\
 0 & 0 & 0 & 0 & 0 & 0 & 0 & 0 & 0 & r_{\mathit{suc}} & -\delta - \mu
 \end{array}\right).$ 

In this case, the two properties required for condition ($\mathbf h3$) follow immediately: the off-diagonal terms of the matrix   $\mathbf A_I(\mathbf x)$ are non-negative, and  the matrix is irreducible as can be seen from the associated directed graph $G(\mathbf A_I(\mathbf x))$ in Figure~\ref{graph}. 
For general $\ell$ and $n$, the proof of ({\bf h}3) is similar.

\begin{figure}[H]
\centering
\includegraphics[scale=0.50]{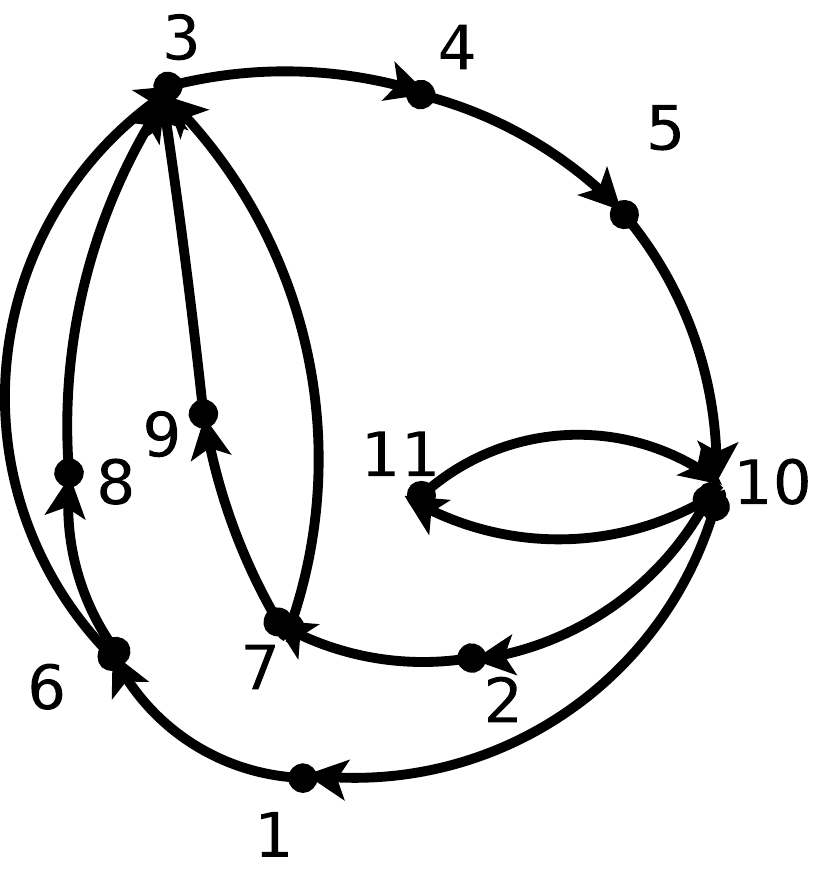}
\caption{Graph associated with the matrix $\mathbf A_I(\mathbf x)$}\label{graph}
\end{figure}

\item
Defining $\overline{\mathbf A}_I \equiv \mathbf A_I(\mathbf x^*)$, we have $\mathbf A_I(\mathbf x)\leqslant\overline{\mathbf A}_I$  $\forall \,\mathbf x\in\Omega$,   and $\mathbf x^*\in\left(\mathbb R_+^{n+2}\times\{\mathbf 0\}\right)\cap\Omega$. Thus the upper bound of $\mathfrak M$ is attained at the DFE which is a point on the boundary of $\Omega$, and condition ($\mathbf h 4$) is satisfied.

\item
We first observe that $\overline{\mathbf A}_I$ is the block matrix of the Jacobian matrix of the system~\eqref{eq:eqbednet_} corresponding to the Infected sub-manifold, taken at the DFE. As  noted in~\cite{KamSal07}, the condition that all eigenvalues of $\overline{\mathbf A}_I$ have negative real parts, which is equivalent to  the condition that $\overline{\mathbf A}_I$ is a stable Metzler matrix, is also equivalent to the condition $\mathcal R_0\leqslant1$. This fact is developed in the proof of Proposition~\ref{prop:basicrepn} (see Appendix) where we compute the value of $\mathcal R_0$ by establishing necessary and sufficient conditions for the stability of the Metzler matrix $\overline{\mathbf A}_I$.
\end{enumerate}
Since the five conditions of Theorem 4.3 in \cite{KamSal07} are satisfied, the theorem follows. \edem

\subsection{Stability analysis of the endemic equilibrium (EE)} \label{sec:eeqstana}
\noindent In this section we analyze the behavior of the system under the condition $\mathcal R_0>1$. From Proposition~\ref{prop:basicrepn} it follows  that the DFE is not stable in this case. As stated in Proposition~\ref{prop:stdst1}, the system~\eqref{eq:eqbednet_} also admits  a unique nontrivial biologically feasible  equilibrium  (or endemic equilibrium (EE)). So it remains to address the stability of the endemic equilibrium, which determines the behavior of the system when the disease persists. Our main result in this regard is the following theorem.

\begin{thm}\label{thm:thmstabee}
When $\mathcal R_0>1$,  the EE $\mathbf x^\dagger $ of the system~\eqref{eq:eqbednet_} defined in \eqref{eq:eqexpMinfctstateb} is GAS on  $\left(\mathbb R_{>0}\right)^u$.
\end{thm}

\begin{rmq}\label{rmq:rmqglstb}\em The above theorem implies that the GAS of the EE is in the nonnegative  cone  $\mathbb R^u_+$, since  the positive cone $\left(\mathbb R_{>0}\right)^u$ is absorbing for the system~\eqref{eq:eqbednet_}.\end{rmq}

\noindent {\em Proof }:~~Considering the system~\eqref{eq:eqbednet_} when $\mathcal R_0 > 1$, there is a unique endemic equilibrium $\mathbf x^\dagger $ with respective components given as in eq.~\eqref{eq:eqexpMinfctstateb}.
Let  the function $V_{ee}$ be defined on $\left(\mathbb R_{>0}\right)^u$ as follows:

\begin{equation}\label{eq:eqliapvee}
\begin{array}{rcl}V_{ee}(\mathbf x) & = & (S_q-S^\dagger _q\ln S_q)+(S_r-S^\dagger _r\ln S_r)+\, \sigma_r^{(1)}(E^{(1)}_r-E^{(1)\dagger }_r\ln E^{(1)}_r)\\&& +\,\dsum_{j=1}^\ell\left(\sigma_q^{(j)}(E^{(j)}_q-E^{(j)\dagger }_q\ln E^{(j)}_q)+\sigma_r^{(j+1)}(E^{(i+1)}_r-E^{(j+1)\dagger }_r\ln E^{(j+1)}_r)\right)+\,\tau_q(I_q-I_q^{\dagger }\ln \,I_q)\\&& +\tau_r(I_r-I_r^{\dagger }\ln \,I_r)+\,\dsum_{i=1}^{n}\upsilon_i\left((S_i-S_i^{\dagger }\ln \,S_i)+(E_i-E_i^{\dagger }\ln \,E_i)+(I_i-I_i^{\dagger }\ln \,I_i)\right),
\end{array}
\end{equation}
where  $\sigma^{(j)}_r = (f^*_qf_r)^{1-j}$, for $j=1,\;2,\;\cdots,\;\ell+1$, $\sigma^{(j)}_q = (f^*_qf_r)^{1-j}/f_r$, for $j=1,\;2,\;\cdots,\;\ell$, $\tau_q = (f^*_qf_r)^{-\ell}/f_r$,   $\tau_r =(f^*_qf_r)^{-\ell}$,  
$\upsilon_i=a\frac{S_q^\dagger }{H^*}\frac{f_i\bar k_i}{\nu_i - \widetilde \nu_i}$ for $i=1,\;2,\;\cdots,\;n$.
(The motivation for these coefficients, and the derivation of expression~\eqref{eq:eqliapeeder2} below for the derivative, are both provided in Appendix~\ref{sec.supplyap}.) $V_{ee}(\mathbf x)$ is a  $\mathcal C^{\infty}$ positive definite function defined on $\left(\mathbb R_{>0}\right)^u$, whose derivative along the trajectories of the system~\eqref{eq:eqbednet_} is given by:

\begin{equation}\label{eq:eqliapeeder2}
\begin{array}{rcl}
{\frac{dV_{ee}}{d\,t}}(\mathbf x(t)) & = & \hatmuVar  S^\dagger _q\left(2-\frac{S_q}{S^\dagger _q}-\frac{S^\dagger _q}{S_q}\right) + \delta S^\dagger _r  \left( \frac{S^\dagger _q}{S_q}+\frac{S_r}{S_r^\dagger }-\frac{S^\dagger _q}{S_q}\frac{S_r}{S_r^\dagger }-1\right)  +\,  \frac{\varpiVar  \sigma_r^{(1)}}{(f_qf_r)^{\ell}} I_q^\dagger\left(2 - \frac{I_r}{I_r^\dagger }\frac{I_q^\dagger }{I_q}- \frac{I_q}{I_q^\dagger }\frac{I^\dagger _r}{I_r}\right)\\
&&+\,{\dsum_{i=0}^n}\varpi_i\Bigg[ \hat\nu_i\Big[S_i^\dagger  \Big(4-\frac{S^\dagger _i}{S_i}-\frac{S^\dagger _q}{S_q}-\frac{S_i}{S^\dagger _i}\frac{S_q}{S^\dagger _q}\frac{S^\dagger _r}{S_r}-\frac{S_r}{S^\dagger _r}\Big) + u_i\Big[E^\dagger _i \Big(4 + \frac{I_q}{I^\dagger_q} - \frac{S^\dagger _q}{S_q} -\frac{S^\dagger _i}{S_i} -\frac{E_i}{E^\dagger _i} \frac{S_q}{S^\dagger _q}\frac{S^\dagger _r}{S_r}-\frac{S_r}{S^\dagger _r}-\frac{I_q}{I^\dagger_q} \frac{S_i}{S^\dagger_i} \frac{E^\dagger_i}{E_i}\Big)\\&& +\,\xi_i I^\dagger _i\Big(2\ell+6 -\frac{S^\dagger _i}{S_i}  -\frac{S_i}{S^\dagger _i}\frac{E^\dagger _i}{E_i}\frac{I_q}{I^\dagger _q} - \frac{I^\dagger_i}{I_i} \frac{E_i}{E^\dagger_i} -\frac{S^\dagger _q}{S_q} - \frac{I_i}{I^\dagger _i}\frac{S_q}{S^\dagger _q}\frac{E^{(1)\dagger }_r}{E^{(1)}_r}- {\dsum_{j=1}^{\ell}}\frac{E^{(j)}_q}{E^{(j)\dagger }_q}\frac{E^{(j+1)\dagger }_r}{E^{(j+1)}_r}-\, {\dsum_{j=1}^{\ell}}\frac{ E^{(j)}_r}{ E^{(j)\dagger }_r}\frac{E^{(j)\dagger }_q}{E^{(j)}_q} -\frac{E^{(\ell+1)}_r}{E^{(\ell+1)\dagger }_r}\frac{I^\dagger _q}{I_q} \Big) \\&& +\,\widetilde\xi_iv_iR^\dagger _i \Big(2\ell+7 - \frac{S^\dagger _i}{S_i} -\, \frac{I_q}{I^\dagger _q}\frac{S_i}{S^\dagger_i}\frac{E^\dagger_i}{E_i}-\frac{I^\dagger_i}{I_i} \frac{E_i}{E^\dagger_i} -\frac{I_i}{I^\dagger_i} \frac{R^\dagger_i}{R_i} -\frac{S^\dagger _q}{S_q} -\; \frac{R_i}{R^\dagger _i}\frac{S_q}{S^\dagger _q}\frac{E^{(1)\dagger }_r}{E^{(1)}_r}- {\dsum_{j=1}^{l}}\frac{E^{(j)}_q}{E^{(j)\dagger }_q}\frac{E^{(j+1)\dagger }_r}{E^{(j+1)}_r}-\, {\dsum_{j=1}^{\ell}}\frac{ E^{(j)}_r}{ E^{(j)\dagger }_r}\frac{E^{(j)\dagger }_q}{E^{(j)}_q} \\&& -\,\frac{E^{(l+1)}_r}{E^{(\ell+1)\dagger }_r}\frac{I^\dagger _q}{I_q} \Big)\Big] + \bar u_i\Big[\xi_i I^\dagger _i\Big(2\ell+5 -\frac{S^\dagger _i}{S_i}  -\frac{S_i}{S^\dagger _i}\frac{I^\dagger _i}{I_i}\frac{I_q}{I^\dagger _q}  -\frac{S^\dagger _q}{S_q} - \frac{I_i}{I^\dagger _i}\frac{S_q}{S^\dagger _q}\frac{E^{(1)\dagger }_r}{E^{(1)}_r}- {\dsum_{j=1}^{\ell}}\frac{E^{(j)}_q}{E^{(j)\dagger }_q}\frac{E^{(j+1)\dagger }_r}{E^{(j+1)}_r}-\, {\dsum_{j=1}^{\ell}}\frac{ E^{(j)}_r}{ E^{(j)\dagger }_r}\frac{E^{(j)\dagger }_q}{E^{(j)}_q} \\&& -\,\frac{E^{(\ell+1)}_r}{E^{(\ell+1)\dagger }_r}\frac{I^\dagger _q}{I_q} \Big) +\,\widetilde\xi_iv_iR^\dagger _i \Big(2\ell+6 - \frac{S^\dagger _i}{S_i} -\, \frac{I_q}{I^\dagger _q}\frac{S_i}{S^\dagger_i}\frac{I^\dagger_i}{I_i} -\frac{I_i}{I^\dagger_i} \frac{R^\dagger_i}{R_i} -\frac{S^\dagger _q}{S_q} -\; \frac{R_i}{R^\dagger _i}\frac{S_q}{S^\dagger _q}\frac{E^{(1)\dagger }_r}{E^{(1)}_r}- {\dsum_{j=1}^{l}}\frac{E^{(j)}_q}{E^{(j)\dagger }_q}\frac{E^{(j+1)\dagger }_r}{E^{(j+1)}_r} \\&&-\, {\dsum_{j=1}^{\ell}}\frac{ E^{(j)}_r}{ E^{(j)\dagger }_r}\frac{E^{(j)\dagger }_q}{E^{(j)}_q} -\,\frac{E^{(l+1)}_r}{E^{(\ell+1)\dagger }_r}\frac{I^\dagger _q}{I_q} \Big)\Big] +\,u_i\Big[\overline{\xi_i} I^\dagger _i\Big(5+\frac{I_q}{I^\dagger _q}- \frac{I^\dagger_i}{I_i}\frac{E_i}{E^\dagger_i}-\frac{S^\dagger _i}{S_i}-\frac{S_i}{S^\dagger _i}\frac{E^\dagger _i}{E_i}\frac{I_q}{I^\dagger _q}-\frac{S^\dagger _q}{S_q}-\frac{I_i}{I^\dagger _i}\frac{S_q}{S^\dagger _q}\frac{S^\dagger _r}{S_r}-\,\frac{S_r}{S^\dagger _r}\Big) \\&& +\, \overline{\widetilde\xi_i}v_iR^\dagger_i\Big(6 + \frac{I_q}{I^\dagger_q} - \frac{S^\dagger _i}{S_i} -\frac{I^\dagger_i}{I_i} \frac{E_i}{E^\dagger_i} -\frac{S^\dagger _q}{S_q} -\,\frac{S_r}{S^\dagger _r}-\,\frac{R_i}{R^\dagger _i}\frac{S_q}{S^\dagger _q}\frac{S^\dagger _r}{S_r}  -\frac{I_i}{I^\dagger_i} \frac{R^\dagger_i}{R_i} - \frac{I_q}{I^\dagger _q}\frac{S_i}{S^\dagger_i} \frac{E^\dagger_i}{E_i} \Big)\Big] + \bar u_i \Big[\overline{\xi_i}I^\dagger_i\Big(4 + \frac{I_q}{I_q^\dagger} - \frac{S^\dagger _i}{S_i}  -\frac{S^\dagger _q}{S_q} \\&& -\,\frac{I_i}{I^\dagger _i}\frac{S_q}{S^\dagger _q}\frac{S^\dagger _r}{S_r}-\,\frac{S_r}{S^\dagger _r} - \frac{I_q}{I_q^\dagger} \frac{S_i}{S_i^\dagger} \frac{I^\dagger_i}{I_i}\Big) + \overline{\widetilde\xi_i}v_iR^\dagger_i\Big(5 + \frac{I_q}{I_q^\dagger} - \frac{S^\dagger _i}{S_i}  -\frac{S^\dagger _q}{S_q}  -\,\frac{S_r}{S^\dagger_r} -\frac{R_i}{R^\dagger _i}\frac{S_q}{S^\dagger _q}\frac{S^\dagger _r}{S_r} - \frac{R^\dagger_i}{R_i} \frac{I_i}{I^\dagger_i} - \frac{I_q}{I_q^\dagger} \frac{S_i}{S_i^\dagger} \frac{I^\dagger_i}{I_i} \Big)\Big]\Big]\\&& +\; u_i\Big[\widetilde\nu_iE^\dagger_i\Big(1 + \frac{I_q}{I^\dagger_q}-\frac{I_q}{I^\dagger_q}\frac{S_i}{S^\dagger_i}\frac{E^\dagger_i}{E_i} - \frac{S^\dagger_i}{S_i} \frac{E_i}{E^\dagger_i}\Big) + (\widetilde\nu_i + \bar v_i\gamma_i)I^\dagger _i\left(2 + \frac{I_q}{I^\dagger _q} - \frac{I_i}{I^\dagger _i} \frac{S^\dagger _i}{S_i} - \frac{I^\dagger_i}{I_i}\frac{E_i}{E^\dagger_i} - \frac{I_q}{I^\dagger _q} \frac{S_i}{S^\dagger _i}\frac{E^\dagger _i}{E_i}\right) \\&& +\, v_i\hat\zeta_iR^\dagger_i\left(3 + \frac{I_q}{I^\dagger _q} - \frac{I_i}{I^\dagger _i} \frac{S^\dagger _i}{S_i} - \frac{I^\dagger_i}{I_i}\frac{E_i}{E^\dagger_i} - \frac{I_i}{I^\dagger_i} \frac{R^\dagger_i}{R_i} - \frac{I_q}{I^\dagger _q} \frac{S_i}{S^\dagger _i}\frac{E^\dagger _i}{E_i}\right)\Big]  +\; \bar u_i\Big[ (\widetilde\nu_i + \bar v_i\gamma_i)I^\dagger _i\left(1 + \frac{I_q}{I^\dagger _q} - \frac{I_i}{I^\dagger _i} \frac{S^\dagger _i}{S_i}  - \frac{I_q}{I^\dagger _q} \frac{S_i}{S^\dagger _i}\frac{I^\dagger _i}{I_i}\right) \\&& +\, v_i\hat\zeta_iR^\dagger_i\left(2 + \frac{I_q}{I^\dagger _q} - \frac{I_i}{I^\dagger _i} \frac{S^\dagger _i}{S_i}  - \frac{I_i}{I^\dagger_i} \frac{R^\dagger_i}{R_i} - \frac{I_q}{I^\dagger _q} \frac{S_i}{S^\dagger _i}\frac{I^\dagger _i}{I_i}\right)\Big] \Bigg]
\end{array}
\end{equation}

\noindent To show that $\frac{d\,V_{ee}}{dt}(\mathbf x(t))\leq0$ for $\mathbf x\in(\mathbb R_{>0})^u$, we split $(\mathbb R_{>0})^u$ into two overlapping subsets that are 
\begin{equation}\Omega_1=\left\{\mathbf x\in\left(\mathbb R_{>0}\right)^u\;|\;\frac{S^\dagger _q}{S_q}+\frac{S_r}{S_r^\dagger }-\frac{S^\dagger _q}{S_q}\frac{S_r}{S_r^\dagger }-1\leqslant 0\right\}\label{eq:subst1}\end{equation} and \begin{equation}\Omega_2 = \left\{\mathbf x\in\left(\mathbb R_{>0}\right)^u\;|\;\frac{S^\dagger _q}{S_q}+\frac{S_r}{S_r^\dagger }-\frac{S^\dagger _q}{S_q}\frac{S_r}{S_r^\dagger }-1\geqslant 0\right\}.\label{eq:subst2}\end{equation} We shall show separately that  $\frac{d\,V_{ee}}{dt}(\mathbf x(t))\leq0$ for $\mathbf x\in\Omega_k$, $k\in{1,~2}$.

\noindent On $\Omega_1$ defined in~\eqref{eq:subst1}, we rewrite~\eqref{eq:eqliapeeder2} as 

\begin{equation}\label{eq:eqliapeeder2rd1}
\frac{d\,V_{ee}}{dt}(\mathbf x(t)) :=    \delta S^\dagger _r  \left( \frac{S^\dagger _q}{S_q}+\frac{S_r}{S_r^\dagger }-\frac{S^\dagger _q}{S_q}\frac{S_r}{S_r^\dagger }-1\right) + f(\mathbf x)+{\dsum_{i=1}^n}\varpi_i G_{1,\,i}(\mathbf x) 
\end{equation}
where 
for each $i$, $\overline{\xi_i}$ and  $\overline{\widetilde\xi_i}$ are respectively the complementary probabilities of $\xi_i$ and $\widetilde\xi_i$ (i.e. $\overline{\xi_i}+\xi_i=1$ and $\overline{\widetilde\xi_i}+\widetilde\xi_i=1$), $ \hat \nu_i = \nu_i - \widetilde \nu_i$ and $\widehat{\zeta_i}\equiv\widetilde\nu_i+\zeta_i$;
\begin{equation*}
f(\mathbf x) = d_q S^\dagger _q\left(2-\frac{S_q}{S^\dagger _q}-\frac{S^\dagger _q}{S_q}\right)   +\, \frac{r_{suc}}{(f_qf_r)^{\ell}}I_q^\dagger  \left(2 - \frac{I_r}{I_r^\dagger }\frac{I_q^\dagger }{I_q}- \frac{I_q}{I_q^\dagger }\frac{I^\dagger _r}{I_r}\right);
\end{equation*}
and
\begin{equation}\label{eq:G1exp1}
G_{1,\,i}(\mathbf x) = g_{1,\,i}(\mathbf x)+h_{1,\,i}(\mathbf x)+p_{i}(\mathbf x),
\end{equation}
with
\begin{align}
g_{1,\,i}(\mathbf x)= &\hat\nu_i\Bigg[ S_i^\dagger  \Big(4-\frac{S^\dagger _i}{S_i}-\frac{S^\dagger _q}{S_q}-\frac{S_i}{S^\dagger _i}\frac{S_q}{S^\dagger _q}\frac{S^\dagger _r}{S_r}-\frac{S_r}{S^\dagger _r}\Big) + u_i\Big[\xi_i I^\dagger _i\Big(2\ell+6 - \frac{I^\dagger_i}{I_i}\frac{E_i}{E^\dagger_i}-\frac{S^\dagger _i}{S_i}-\frac{S_i}{S^\dagger _i}\frac{E^\dagger _i}{E_i}\frac{I_q}{I^\dagger _q} -\frac{S^\dagger _q}{S_q} - \frac{I_i}{I^\dagger _i}\frac{S_q}{S^\dagger _q}\frac{E^{(1)\dagger }_r}{E^{(1)}_r}\nonumber\\&- {\dsum_{j=1}^{\ell}}\frac{E^{(j)}_q}{E^{(j)\dagger }_q}\frac{E^{(j+1)\dagger }_r}{E^{(j+1)}_r}-\, {\dsum_{j=1}^{\ell}}\frac{ E^{(j)}_r}{ E^{(j)\dagger }_r}\frac{E^{(j)\dagger }_q}{E^{(j)}_q} -\frac{E^{(\ell+1)}_r}{E^{(\ell+1)\dagger }_r}\frac{I^\dagger _q}{I_q} \Big) + v_i\widetilde\xi_iR^\dagger_i\Big(2\ell+7 - \frac{S^\dagger _i}{S_i}-\frac{S_i}{S^\dagger _i}\frac{E^\dagger _i}{E_i}\frac{I_q}{I^\dagger _q}  - \frac{I^\dagger_i}{I_i}\frac{E_i}{E^\dagger_i}  - \frac{I_i}{I^\dagger_i}\frac{R^\dagger_i}{R_i} - \frac{S^\dagger _q}{S_q} \nonumber\\&- \frac{R_i}{R^\dagger _i}\frac{S_q}{S^\dagger _q}\frac{E^{(1)\dagger }_r}{E^{(1)}_r}- {\dsum_{j=1}^{\ell}}\frac{E^{(j)}_q}{E^{(j)\dagger }_q}\frac{E^{(j+1)\dagger }_r}{E^{(j+1)}_r}-\, {\dsum_{j=1}^{\ell}}\frac{ E^{(j)}_r}{ E^{(j)\dagger }_r}\frac{E^{(j)\dagger }_q}{E^{(j)}_q} -\frac{E^{(\ell+1)}_r}{E^{(\ell+1)\dagger }_r}\frac{I^\dagger _q}{I_q} \Big)\Big] + \bar u_i\Big[\xi_i I^\dagger _i\Big(2\ell+5 - \frac{S^\dagger _i}{S_i}-\frac{S_i}{S^\dagger _i}\frac{I^\dagger _i}{I_i}\frac{I_q}{I^\dagger _q} -\frac{S^\dagger _q}{S_q} \nonumber\\&- \frac{I_i}{I^\dagger _i}\frac{S_q}{S^\dagger _q}\frac{E^{(1)\dagger }_r}{E^{(1)}_r}- {\dsum_{j=1}^{\ell}}\frac{E^{(j)}_q}{E^{(j)\dagger }_q}\frac{E^{(j+1)\dagger }_r}{E^{(j+1)}_r}-\, {\dsum_{j=1}^{\ell}}\frac{ E^{(j)}_r}{ E^{(j)\dagger }_r}\frac{E^{(j)\dagger }_q}{E^{(j)}_q} -\frac{E^{(\ell+1)}_r}{E^{(\ell+1)\dagger }_r}\frac{I^\dagger _q}{I_q} \Big) + v_i\widetilde\xi_iR^\dagger_i\Big(2\ell+6 - \frac{S^\dagger _i}{S_i}-\frac{S_i}{S^\dagger _i}\frac{I^\dagger _i}{I_i}\frac{I_q}{I^\dagger _q}  - \frac{I_i}{I^\dagger_i}\frac{R^\dagger_i}{R_i} \nonumber\\& - \frac{S^\dagger _q}{S_q}- \frac{R_i}{R^\dagger _i}\frac{S_q}{S^\dagger _q}\frac{E^{(1)\dagger }_r}{E^{(1)}_r}- {\dsum_{j=1}^{\ell}}\frac{E^{(j)}_q}{E^{(j)\dagger }_q}\frac{E^{(j+1)\dagger }_r}{E^{(j+1)}_r}-\, {\dsum_{j=1}^{\ell}}\frac{ E^{(j)}_r}{ E^{(j)\dagger }_r}\frac{E^{(j)\dagger }_q}{E^{(j)}_q} -\frac{E^{(\ell+1)}_r}{E^{(\ell+1)\dagger }_r}\frac{I^\dagger _q}{I_q} \Big)\Big]
\Bigg],\\
h_{1,\,i}(\mathbf x)=& \hat\nu_i\Bigg[u_i\Big[E^\dagger _i \Big(4 + \frac{I_q}{I^\dagger_q} - \frac{S^\dagger _q}{S_q} -\frac{S^\dagger _i}{S_i} -\frac{E_i}{E^\dagger _i} \frac{S_q}{S^\dagger _q}\frac{S^\dagger _r}{S_r}-\frac{S_r}{S^\dagger _r}-\frac{I_q}{I^\dagger_q} \frac{S_i}{S^\dagger_i} \frac{E^\dagger_i}{E_i}\Big) + \overline{\xi_i} I^\dagger _i\Big(5+\frac{I_q}{I^\dagger _q}- \frac{I^\dagger_i}{I_i}\frac{E_i}{E^\dagger_i}-\frac{S^\dagger _i}{S_i}-\frac{S_i}{S^\dagger _i}\frac{E^\dagger _i}{E_i}\frac{I_q}{I^\dagger _q}-\frac{S^\dagger _q}{S_q}\nonumber\\&-\frac{I_i}{I^\dagger _i}\frac{S_q}{S^\dagger _q}\frac{S^\dagger _r}{S_r}-\,\frac{S_r}{S^\dagger _r}\Big) + v_i\overline{\widetilde\xi_i}R^\dagger_i\Big(6+\frac{I_q}{I^\dagger _q}- \frac{I^\dagger_i}{I_i}\frac{E_i}{E^\dagger_i}-\frac{I_i}{I^\dagger_i}\frac{R^\dagger_i}{R_i}-\frac{S^\dagger _i}{S_i}-\frac{S_i}{S^\dagger _i}\frac{E^\dagger _i}{E_i}\frac{I_q}{I^\dagger _q}-\frac{S^\dagger _q}{S_q}-\frac{R_i}{R^\dagger _i}\frac{S_q}{S^\dagger _q}\frac{S^\dagger _r}{S_r}-\,\frac{S_r}{S^\dagger _r}\Big)\Big] \nonumber\\&+ \bar u_i\Big[\overline{\xi_i} I^\dagger _i\Big(4+\frac{I_q}{I^\dagger _q}- \frac{S^\dagger _i}{S_i}-\frac{S_i}{S^\dagger _i}\frac{I^\dagger _i}{I_i}\frac{I_q}{I^\dagger _q}-\frac{S^\dagger _q}{S_q}-\frac{I_i}{I^\dagger _i}\frac{S_q}{S^\dagger _q}\frac{S^\dagger _r}{S_r}-\,\frac{S_r}{S^\dagger _r}\Big) + v_i\overline{\widetilde\xi_i}R^\dagger_i\Big(5+\frac{I_q}{I^\dagger _q}- \frac{I_i}{I^\dagger_i}\frac{R^\dagger_i}{R_i}-\frac{S^\dagger _i}{S_i}-\frac{S_i}{S^\dagger _i}\frac{I^\dagger _i}{I_i}\frac{I_q}{I^\dagger _q} \nonumber\\&-\frac{S^\dagger _q}{S_q}-\frac{R_i}{R^\dagger _i}\frac{S_q}{S^\dagger _q}\frac{S^\dagger _r}{S_r}-\,\frac{S_r}{S^\dagger _r}\Big)\Big]\Bigg],\\
p_{i}(\mathbf x)=& (\widetilde\nu_i +\bar v_i\gamma_i )I^\dagger _i\Bigg[u_i\Big(2 + \frac{I_q}{I^\dagger _q} - \frac{I_i}{I^\dagger _i} \frac{S^\dagger _i}{S_i} - \frac{I^\dagger_i}{I_i}\frac{E_i}{E^\dagger_i} - \frac{I_q}{I^\dagger _q} \frac{S_i}{S^\dagger _i}\frac{E^\dagger _i}{E_i}\Big) + \bar u_i\Big(1 + \frac{I_q}{I^\dagger _q} - \frac{I_i}{I^\dagger _i} \frac{S^\dagger _i}{S_i} -  \frac{I_q}{I^\dagger _q} \frac{S_i}{S^\dagger _i}\frac{I^\dagger _i}{I_i}\Big)\Bigg]\nonumber\\& +\,v_i\hat\zeta_iR^\dagger _i\Bigg[u_i\Big(3 + \frac{I_q}{I^\dagger _q} - \frac{R_i}{R^\dagger _i} \frac{S^\dagger _i}{S_i} - \frac{I^\dagger_i}{I_i}\frac{E_i}{E^\dagger_i}- \frac{I_i}{I^\dagger_i}\frac{R^\dagger_i}{R_i} - \frac{I_q}{I^\dagger _q} \frac{S_i}{S^\dagger _i}\frac{E^\dagger _i}{E_i}\Big) + \bar u_i\Big(2 + \frac{I_q}{I^\dagger _q} - \frac{I_i}{I^\dagger_i}\frac{R^\dagger_i}{R_i} -\frac{R_i}{R^\dagger _i} \frac{S^\dagger _i}{S_i} -  \frac{I_q}{I^\dagger _q} \frac{S_i}{S^\dagger _i}\frac{I^\dagger _i}{I_i}\Big)\Bigg]\nonumber\\& +\,u_i\widetilde\nu_iE^\dagger_i\left(1 + \frac{I_q}{I^\dagger_q}-\frac{I_q}{I^\dagger_q}\frac{S_i}{S^\dagger_i}\frac{E^\dagger_i}{E_i} - \frac{S^\dagger_i}{S_i} \frac{E_i}{E^\dagger_i}\right).\label{eq:frgmtliap1}
\end{align}

\noindent Alternatively for each $i$,  by adding and subtracting $u_i\nu_iE^\dagger _i \left( 1- \frac{S_i}{S^\dagger _i}\frac{E^\dagger_i}{E_i}\right) + \bar u_i\nu_iI^\dagger _i\left( 1- \frac{S_i}{S^\dagger _i}\frac{I^\dagger_i}{I_i}\right)$
the function $G_{1,\,i}(\mathbf x)$ in~\eqref{eq:G1exp1} may be rewritten as:
 
\begin{align}\label{eq:eqliapeeder20}
G_{1,\,i}(\mathbf x)  = &   \hat\nu_i\Bigg[S_i^\dagger  \Big(4-\frac{S^\dagger _i}{S_i}-\frac{S^\dagger _q}{S_q}-\frac{S_i}{S^\dagger _i}\frac{S_q}{S^\dagger _q}\frac{S^\dagger _r}{S_r}-\frac{S_r}{S^\dagger _r}\Big) + u_i\Big[E^\dagger _i \Big(5 - \frac{S^\dagger _q}{S_q} -\frac{S^\dagger _i}{S_i} -\frac{E_i}{E^\dagger _i} \frac{S_q}{S^\dagger _q}\frac{S^\dagger _r}{S_r}-\frac{S_r}{S^\dagger _r}- \frac{S_i}{S^\dagger_i} \frac{E^\dagger_i}{E_i}\Big) +\,I^\dagger _i\Big(\xi_i \Big(2\ell+7 - \frac{I^\dagger_i}{I_i} \frac{E_i}{E^\dagger_i}\nonumber\\&-\frac{S^\dagger _i}{S_i}-\;\frac{S_i}{S^\dagger _i}\frac{E^\dagger _i}{E_i} -\frac{S^\dagger _q}{S_q} - \frac{I_i}{I^\dagger _i}\frac{S_q}{S^\dagger _q}\frac{E^{(1)\dagger }_r}{E^{(1)}_r}- {\dsum_{j=1}^{\ell}}\frac{E^{(j)}_q}{E^{(j)\dagger }_q}\frac{E^{(j+1)\dagger }_r}{E^{(j+1)}_r}-\, {\dsum_{j=1}^{\ell}}\frac{ E^{(j)}_r}{ E^{(j)\dagger }_r}\frac{E^{(j)\dagger }_q}{E^{(j)}_q} -\frac{E^{(\ell+1)}_r}{E^{(\ell+1)\dagger }_r}\frac{I^\dagger _q}{I_q}  -\frac{I_q}{I^\dagger _q}\Big)  +\,\overline{\xi_i} \Big(6 - \frac{I^\dagger_i}{I_i} \frac{E_i}{E^\dagger_i} -\frac{S^\dagger _i}{S_i} \nonumber\\&-\frac{S_i}{S^\dagger _i}\frac{E^\dagger _i}{E_i} -\frac{S^\dagger _q}{S_q}-\frac{I_i}{I^\dagger _i}\frac{S_q}{S^\dagger _q}\frac{S^\dagger _r}{S_r}-\,\frac{S_r}{S^\dagger _r}\Big)\Big)
 +\; v_iR^\dagger_i\Big(\widetilde\xi_i\Big(2\ell+8 - \frac{S^\dagger _i}{S_i}-\frac{S_i}{S^\dagger _i}\frac{E^\dagger _i}{E_i}  - \frac{I^\dagger_i}{I_i}\frac{E_i}{E^\dagger_i}  - \frac{I_i}{I^\dagger_i}\frac{R^\dagger_i}{R_i} - \frac{S^\dagger _q}{S_q} - \frac{R_i}{R^\dagger _i}\frac{S_q}{S^\dagger _q}\frac{E^{(1)\dagger }_r}{E^{(1)}_r}-\frac{I_q}{I^\dagger _q} \nonumber\\&- {\dsum_{j=1}^{\ell}}\frac{E^{(j)}_q}{E^{(j)\dagger }_q}\frac{E^{(j+1)\dagger }_r}{E^{(j+1)}_r}-\, {\dsum_{j=1}^{\ell}}\frac{ E^{(j)}_r}{ E^{(j)\dagger }_r}\frac{E^{(j)\dagger }_q}{E^{(j)}_q}-\;\frac{E^{(\ell+1)}_r}{E^{(\ell+1)\dagger }_r}\frac{I^\dagger _q}{I_q} \Big)+ \overline{\widetilde\xi_i}\Big(7 - \frac{I^\dagger_i}{I_i} \frac{E_i}{E^\dagger_i} -\frac{I_i}{I^\dagger_i} \frac{R^\dagger_i}{R_i}-\frac{S^\dagger _i}{S_i}-\frac{S_i}{S^\dagger _i}\frac{E^\dagger _i}{E_i} -\frac{S^\dagger _q}{S_q}-\frac{R_i}{R^\dagger _i}\frac{S_q}{S^\dagger _q}\frac{S^\dagger _r}{S_r}-\,\frac{S_r}{S^\dagger _r}\Big)\Big) \Big] \nonumber\\& + \bar u_i\Big[I^\dagger _i\Big(\overline{\xi_i} \Big(5 - \frac{S^\dagger _q}{S_q} -\frac{S^\dagger _i}{S_i} -\frac{I_i}{I^\dagger _i} \frac{S_q}{S^\dagger _q}\frac{S^\dagger _r}{S_r}-\;\frac{S_r}{S^\dagger _r}- \,\frac{S_i}{S^\dagger_i} \frac{I^\dagger_i}{I_i}\Big) +\,\xi_i\Big(2\ell+6 - \frac{S^\dagger _i}{S_i}-\frac{S_i}{S^\dagger _i}\frac{I^\dagger_i}{I_i} -\frac{S^\dagger _q}{S_q} - \frac{I_i}{I^\dagger _i}\frac{S_q}{S^\dagger _q}\frac{E^{(1)\dagger }_r}{E^{(1)}_r}-\frac{I_q}{I^\dagger _q} \nonumber\\&- {\dsum_{j=1}^{\ell}}\frac{E^{(j)}_q}{E^{(j)\dagger }_q}\frac{E^{(j+1)\dagger }_r}{E^{(j+1)}_r}-\, {\dsum_{j=1}^{\ell}}\frac{ E^{(j)}_r}{ E^{(j)\dagger }_r}\frac{E^{(j)\dagger }_q}{E^{(j)}_q} -\frac{E^{(\ell+1)}_r}{E^{(\ell+1)\dagger }_r}\frac{I^\dagger _q}{I_q}  \Big) \Big)+\; v_iR^\dagger_i\Big(\widetilde\xi_i\Big(2\ell+7 - \frac{S^\dagger _i}{S_i}-\frac{S_i}{S^\dagger _i}\frac{I^\dagger_i}{I_i}  - \frac{I_i}{I^\dagger_i}\frac{R^\dagger_i}{R_i} - \frac{S^\dagger _q}{S_q} - \frac{R_i}{R^\dagger _i}\frac{S_q}{S^\dagger _q}\frac{E^{(1)\dagger }_r}{E^{(1)}_r}\nonumber\\& -\frac{I_q}{I^\dagger _q}- {\dsum_{j=1}^{\ell}}\frac{E^{(j)}_q}{E^{(j)\dagger }_q}\frac{E^{(j+1)\dagger }_r}{E^{(j+1)}_r}-\, {\dsum_{j=1}^{\ell}}\frac{ E^{(j)}_r}{ E^{(j)\dagger }_r}\frac{E^{(j)\dagger }_q}{E^{(j)}_q} -\;\frac{E^{(\ell+1)}_r}{E^{(\ell+1)\dagger }_r}\frac{I^\dagger _q}{I_q} \Big)+\; \overline{\widetilde\xi_i}\Big(7 - \frac{I^\dagger_i}{I_i} \frac{E_i}{E^\dagger_i} -\frac{I_i}{I^\dagger_i} \frac{R^\dagger_i}{R_i}-\frac{S^\dagger _i}{S_i}-\frac{S_i}{S^\dagger _i}\frac{E^\dagger _i}{E_i} -\frac{S^\dagger _q}{S_q}-\frac{R_i}{R^\dagger _i}\frac{S_q}{S^\dagger _q}\frac{S^\dagger _r}{S_r} \nonumber\\&-\,\frac{S_r}{S^\dagger _r}\Big)\Big)\Big]\Bigg]
 +\; (\widetilde\nu_i + v_i\gamma_i)I^\dagger _i\Big[u_i\Big(3  - \frac{I_i}{I^\dagger _i} \frac{S^\dagger _i}{S_i} - \frac{S_i}{S^\dagger_i} \frac{E^\dagger_i}{E_i} - \frac{I^\dagger_i}{I_i}\frac{E_i}{E^\dagger_i}  \Big)+\; \bar u_i\Big(2  - \frac{I_i}{I^\dagger _i} \frac{S^\dagger _i}{S_i} - \frac{S_i}{S^\dagger_i}\frac{I^\dagger_i}{I_i}   \Big)\Big] + v_i\hat\zeta_iR^\dagger _i\Big[u_i\Big(4 - \frac{R_i}{R^\dagger _i} \frac{S^\dagger _i}{S_i} \nonumber\\&- \frac{I^\dagger_i}{I_i}\frac{E_i}{E^\dagger_i}- \frac{I_i}{I^\dagger_i}\frac{R^\dagger_i}{R_i} - \frac{S_i}{S^\dagger _i}\frac{E^\dagger _i}{E_i}\Big) +\; \bar u_i\Big(3 - \frac{I_i}{I^\dagger_i}\frac{R^\dagger_i}{R_i} -\frac{R_i}{R^\dagger _i} \frac{S^\dagger _i}{S_i}  -  \frac{S_i}{S^\dagger _i}\frac{I^\dagger _i}{I_i}\Big)\Big]  + u_i\widetilde\nu_iE^\dagger_i\Big(2-\frac{S_i}{S^\dagger_i}\frac{E^\dagger_i}{E_i} - \frac{S^\dagger_i}{S_i} \frac{E_i}{E^\dagger_i}\Big) \nonumber\\&+\; u_i\bar\varepsilon_iE^\dagger_i\left(\frac{I_q}{I^\dagger _q} + \frac{S_i}{S^\dagger _i}\frac{E^\dagger _i}{E_i} - 1 - \frac{I_q}{I^\dagger _q} \frac{S_i}{S^\dagger _i}\frac{E^\dagger _i}{E_i}\right)+\; \bar u_i\bar\gamma_iI^\dagger_i\left(\frac{I_q}{I^\dagger _q} + \frac{S_i}{S^\dagger _i}\frac{I^\dagger _i}{I_i} - 1 - \frac{I_q}{I^\dagger _q} \frac{S_i}{S^\dagger _i}\frac{I^\dagger _i}{I_i}\right) \nonumber\\ :=&   \widetilde g_{1,\,i}(\mathbf x) + \widetilde p_{i}(\mathbf x),
\end{align}
`
with $\bar\varepsilon_i\equiv\nu_i+\varepsilon_i$ and  $\bar\gamma_i\equiv\nu_i+\gamma_i$,

\begin{align}
\widetilde{g}_{1,\,i}(\mathbf x)= &I^\dagger _i\Bigg[\hat\nu_i \Big[u_i\Big(\xi_i\Big(2\ell+7 - \frac{I^\dagger_i}{I_i} \frac{E_i}{E^\dagger_i}-\frac{S^\dagger _i}{S_i}-\frac{S_i}{S^\dagger _i}\frac{E^\dagger _i}{E_i} -\frac{S^\dagger _q}{S_q} - \frac{I_i}{I^\dagger _i}\frac{S_q}{S^\dagger _q}\frac{E^{(1)\dagger }_r}{E^{(1)}_r}- {\dsum_{j=1}^{\ell}}\frac{E^{(j)}_q}{E^{(j)\dagger }_q}\frac{E^{(j+1)\dagger }_r}{E^{(j+1)}_r}-\, {\dsum_{j=1}^{\ell}}\frac{ E^{(j)}_r}{ E^{(j)\dagger }_r}\frac{E^{(j)\dagger }_q}{E^{(j)}_q} -\frac{E^{(\ell+1)}_r}{E^{(\ell+1)\dagger }_r}\frac{I^\dagger _q}{I_q}  -\frac{I_q}{I^\dagger _q}\Big)\nonumber\\
& + \,\overline{\xi_i}\Big(6 - \frac{I^\dagger_i}{I_i} \frac{E_i}{E^\dagger_i} -\frac{S^\dagger _i}{S_i} -\frac{S_i}{S^\dagger _i}\frac{E^\dagger _i}{E_i} -\frac{S^\dagger _q}{S_q}-\frac{I_i}{I^\dagger _i}\frac{S_q}{S^\dagger _q}\frac{S^\dagger _r}{S_r}-\,\frac{S_r}{S^\dagger _r}\Big)\Big) + \bar u_i\Big(\xi_i\Big(2\ell+6 -\frac{S^\dagger _i}{S_i}-\frac{S_i}{S^\dagger _i} \frac{I^\dagger_i}{I_i} -\frac{S^\dagger _q}{S_q} - \frac{I_i}{I^\dagger _i}\frac{S_q}{S^\dagger _q}\frac{E^{(1)\dagger }_r}{E^{(1)}_r}-\frac{I_q}{I^\dagger _q}\nonumber\\
&- {\dsum_{j=1}^{\ell}}\frac{E^{(j)}_q}{E^{(j)\dagger }_q}\frac{E^{(j+1)\dagger }_r}{E^{(j+1)}_r}-\, {\dsum_{j=1}^{\ell}}\frac{ E^{(j)}_r}{ E^{(j)\dagger }_r}\frac{E^{(j)\dagger }_q}{E^{(j)}_q} -\frac{E^{(\ell+1)}_r}{E^{(\ell+1)\dagger }_r}\frac{I^\dagger _q}{I_q}  \Big) + \,\overline{\xi_i}\Big(5 - \frac{S^\dagger _i}{S_i} -\frac{S_i}{S^\dagger _i}\frac{I^\dagger_i}{I_i} -\frac{S^\dagger _q}{S_q}-\frac{I_i}{I^\dagger _i}\frac{S_q}{S^\dagger _q}\frac{S^\dagger _r}{S_r}-\,\frac{S_r}{S^\dagger _r}\Big)\Big)\Big]\nonumber\\&+ (\widetilde\nu_i + \gamma_i)\Big[u_i\Big(3  - \frac{I_i}{I^\dagger _i} \frac{S^\dagger _i}{S_i} - \frac{S_i}{S^\dagger_i}\frac{E^\dagger_i}{E_i} - \frac{I^\dagger_i}{I_i}\frac{E_i}{E^\dagger_i}  \Big) + \bar u_i\Big(2  - \frac{I_i}{I^\dagger _i} \frac{S^\dagger _i}{S_i} - \frac{S_i}{S^\dagger_i}\frac{I^\dagger_i}{I_i}   \Big)\Big] \Bigg] +\;v_iR^\dagger _i\Bigg[\hat\nu_i \Big[u_i\Big(\xi_i\Big(2\ell+8 - \frac{S^\dagger _i}{S_i}-\frac{S_i}{S^\dagger _i}\frac{E^\dagger _i}{E_i}\nonumber\\&  - \frac{I^\dagger_i}{I_i} \frac{E_i}{E^\dagger_i} -\frac{I_i}{I^\dagger_i} \frac{R^\dagger_i}{R_i} -\frac{S^\dagger _q}{S_q} - \frac{R_i}{R^\dagger _i}\frac{S_q}{S^\dagger _q}\frac{E^{(1)\dagger }_r}{E^{(1)}_r}-\frac{I_q}{I^\dagger _q} - {\dsum_{j=1}^{\ell}} \frac{E^{(j)}_q}{E^{(j)\dagger }_q}\frac{E^{(j+1)\dagger }_r}{E^{(j+1)}_r}-\, {\dsum_{j=1}^{\ell}}\frac{ E^{(j)}_r}{ E^{(j)\dagger }_r}\frac{E^{(j)\dagger }_q}{E^{(j)}_q}-\frac{E^{(\ell+1)}_r}{E^{(\ell+1)\dagger }_r}\frac{I^\dagger _q}{I_q}  \Big) + \,\overline{\xi_i}\Big(7  -\frac{S^\dagger _i}{S_i} -\frac{S_i}{S^\dagger _i}\frac{E^\dagger _i}{E_i}\nonumber\\
& - \frac{E_i}{E^\dagger_i}\frac{I^\dagger_i}{I_i} - \frac{I_i}{I^\dagger_i}\frac{R^\dagger_i}{R_i}  -\frac{S^\dagger _q}{S_q}-\frac{R_i}{R^\dagger _i}\frac{S_q}{S^\dagger _q}\frac{S^\dagger _r}{S_r}-\,\frac{S_r}{S^\dagger _r}\Big)\Big) + \bar u_i\Big(\xi_i\Big(2\ell+7 -\frac{S^\dagger _i}{S_i}-\frac{S_i}{S^\dagger _i} \frac{I^\dagger_i}{I_i}  -\frac{S^\dagger _q}{S_q} - \frac{I_i}{I^\dagger_i}\frac{R^\dagger_i}{R_i} - \frac{R_i}{R^\dagger _i}\frac{S_q}{S^\dagger _q}\frac{E^{(1)\dagger }_r}{E^{(1)}_r}-\frac{I_q}{I^\dagger _q}\nonumber\end{align}
\begin{align}  \nonumber\\
&- {\dsum_{j=1}^{\ell}}\frac{E^{(j)}_q}{E^{(j)\dagger }_q}\frac{E^{(j+1)\dagger }_r}{E^{(j+1)}_r}-\, {\dsum_{j=1}^{\ell}}\frac{ E^{(j)}_r}{ E^{(j)\dagger }_r}\frac{E^{(j)\dagger }_q}{E^{(j)}_q} -\frac{E^{(\ell+1)}_r}{E^{(\ell+1)\dagger }_r}\frac{I^\dagger _q}{I_q}  \Big) + \,\overline{\xi_i}\Big(6 - \frac{S^\dagger _i}{S_i} -\frac{S_i}{S^\dagger _i}\frac{I^\dagger_i}{I_i} -\frac{S^\dagger _q}{S_q} - \frac{I_i}{I^\dagger_i}\frac{R^\dagger_i}{R_i}-\frac{I_i}{I^\dagger _i}\frac{S_q}{S^\dagger _q}\frac{S^\dagger _r}{S_r}-\,\frac{S_r}{S^\dagger _r}\Big)\Big)\Big]\nonumber\\&  + \hat\zeta_i\Big[u_i\Big(4  - \frac{R_i}{R^\dagger _i} \frac{S^\dagger _i}{S_i} - \frac{S_i}{S^\dagger_i}\frac{E^\dagger_i}{E_i} - \frac{E_i}{E^\dagger_i}\frac{I^\dagger_i}{I_i} - \frac{I_i}{I^\dagger_i}\frac{R^\dagger_i}{R_i} \Big) + \bar u_i\Big(3  - \frac{R_i}{R^\dagger _i} \frac{S^\dagger _i}{S_i} - \frac{S_i}{S^\dagger_i}\frac{I^\dagger_i}{I_i}  - \frac{I_i}{I^\dagger_i}\frac{R^\dagger_i}{R_i} \Big)\Big] \Bigg] \nonumber\\& +\,\hat\nu_i\,S_i^\dagger  \left( 4-\frac{S^\dagger _i}{S_i} - \frac{S^\dagger _q}{S_q}-\frac{S_i}{S^\dagger _i}\frac{S_q}{S^\dagger _q}\frac{S^\dagger _r}{S_r}-\frac{S_r}{S^\dagger _r}\right) + \widetilde\nu_iE^\dagger_i\Big(2-\frac{S_i}{S^\dagger_i}\frac{E^\dagger_i}{E_i} - \frac{S^\dagger_i}{S_i} \frac{E_i}{E^\dagger_i}\Big),\\
\widetilde p_{i}(\mathbf x)=& u_i\bar\varepsilon_iE^\dagger _i\left(\frac{I_q}{I^\dagger _q} +  \frac{S_i}{S^\dagger_i} \frac{E^\dagger_i}{E_i}-1 - \frac{I_q}{I^\dagger _q} \frac{S_i}{S^\dagger _i}\frac{E^\dagger _i}{E_i}\right) + \bar u_i \bar\gamma_iI^\dagger _i\left(\frac{I_q}{I^\dagger _q} +  \frac{S_i}{S^\dagger_i} \frac{I^\dagger_i}{I_i}-1 - \frac{I_q}{I^\dagger _q} \frac{S_i}{S^\dagger _i}\frac{I^\dagger _i}{I_i}\right).,\label{eq:frgmtliap22}
\end{align}

\noindent It comes out that for $\mathbf x\in\Omega_1$, we may write $\frac{d\,V_{ee}}{dt}(\mathbf x(t))$ as \begin{align}
\frac{d\,V_{ee}}{dt}(\mathbf x(t)) =& \delta S_r^\dagger\Big(\frac{S^\dagger_q}{S_q} + \frac{S_r}{S^\dagger_r} - \frac{S_r}{S^\dagger_r}\frac{S^\dagger_q}{S_q}-1\Big) + f(\mathbf x)\nonumber\\& +\, \dsum_{\mathcal I_1}\varpi_i\Big(g_{1,\,i}(\mathbf x)+h_{1,\,i}(\mathbf x)+p_{i}(\mathbf x)\Big)+ \dsum_{\mathcal J_1}\varpi_i\Big(\widetilde g_{1,\,i}(\mathbf x) + \widetilde p_{i}(\mathbf x)\Big)\label{eq:rdcexp1vee}
\end{align} where $\{\mathcal I_i,~\mathcal J_1\}$ is any partition of $\{1,~2,~\ldots,~n\}$.

\noindent On the other hand, on $\Omega_2$ defined in~\eqref{eq:subst2}, using the identity $\delta S_r^\dagger  = (r_{suc} ^\dagger -r_{inf}^\dagger )S_q^\dagger -\mu S_r^\dagger $ we may write~\eqref{eq:eqliapeeder2} as

\begin{align}
{\dfrac{dV_{ee}}{d\,t}}(\mathbf
x(t))  = & d_q  S^\dagger _q\Big(2-\frac{S_q}{S^\dagger _q}-\frac{S^\dagger _q}{S_q}\Big) + \mu S^\dagger _r  \Big(1 + \frac{S^\dagger _q}{S_q}\frac{S_r}{S_r^\dagger } - \frac{S^\dagger _q}{S_q}-\frac{S_r}{S_r^\dagger }\Big)  +\, \frac{r_{suc}  }{(f_qf_r)^{l}}I_q^\dagger  \Big(2 - \frac{I_r}{I_r^\dagger }\frac{I_q^\dagger }{I_q}- \frac{I_q}{I_q^\dagger }\frac{I^\dagger _r}{I_r}\Big)\nonumber\\
&+\,{\dsum_{i=1}^n}\varpi_i\Bigg[ \hat\nu_i\Big[S_i^\dagger  \Big(3-\frac{S^\dagger _i}{S_i}-\frac{S^\dagger _q}{S_q}\frac{S_r}{S^\dagger _r} -\frac{S_i}{S^\dagger _i}\frac{S_q}{S^\dagger _q}\frac{S^\dagger _r}{S_r}\Big) + u_i\Big[E^\dagger _i \Big(3 + \frac{I_q}{I^\dagger_q}  -\frac{S^\dagger _i}{S_i}- \frac{S^\dagger _q}{S_q}\frac{S_r}{S^\dagger _r} -\frac{E_i}{E^\dagger _i} \frac{S_q}{S^\dagger _q}\frac{S^\dagger _r}{S_r}-\frac{I_q}{I^\dagger_q} \frac{S_i}{S^\dagger_i} \frac{E^\dagger_i}{E_i}\Big)\nonumber\\& +\,\xi_i I^\dagger _i\Big(2\ell+6 -\frac{S^\dagger _i}{S_i}  -\frac{S_i}{S^\dagger _i}\frac{E^\dagger _i}{E_i}\frac{I_q}{I^\dagger _q} - \frac{I^\dagger_i}{I_i} \frac{E_i}{E^\dagger_i} -\frac{S^\dagger _q}{S_q} - \frac{I_i}{I^\dagger _i}\frac{S_q}{S^\dagger _q}\frac{E^{(1)\dagger }_r}{E^{(1)}_r}- {\dsum_{j=1}^{\ell}}\frac{E^{(j)}_q}{E^{(j)\dagger }_q}\frac{E^{(j+1)\dagger }_r}{E^{(j+1)}_r}-\, {\dsum_{j=1}^{\ell}}\frac{ E^{(j)}_r}{ E^{(j)\dagger }_r}\frac{E^{(j)\dagger }_q}{E^{(j)}_q} -\frac{E^{(\ell+1)}_r}{E^{(\ell+1)\dagger }_r}\frac{I^\dagger _q}{I_q} \Big) \nonumber\\& +\,\widetilde\xi_iv_iR^\dagger _i \Big(2\ell+7 - \frac{S^\dagger _i}{S_i} -\, \frac{I_q}{I^\dagger _q}\frac{S_i}{S^\dagger_i}\frac{E^\dagger_i}{E_i}-\frac{I^\dagger_i}{I_i} \frac{E_i}{E^\dagger_i} -\frac{I_i}{I^\dagger_i} \frac{R^\dagger_i}{R_i} -\frac{S^\dagger _q}{S_q} -\; \frac{R_i}{R^\dagger _i}\frac{S_q}{S^\dagger _q}\frac{E^{(1)\dagger }_r}{E^{(1)}_r}- {\dsum_{j=1}^{l}}\frac{E^{(j)}_q}{E^{(j)\dagger }_q}\frac{E^{(j+1)\dagger }_r}{E^{(j+1)}_r}-\, {\dsum_{j=1}^{\ell}}\frac{ E^{(j)}_r}{ E^{(j)\dagger }_r}\frac{E^{(j)\dagger }_q}{E^{(j)}_q} \nonumber\\& -\,\frac{E^{(l+1)}_r}{E^{(\ell+1)\dagger }_r}\frac{I^\dagger _q}{I_q} \Big)\Big] + \bar u_i\Big[\xi_i I^\dagger _i\Big(2\ell+5 -\frac{S^\dagger _i}{S_i}  -\frac{S_i}{S^\dagger _i}\frac{I^\dagger _i}{I_i}\frac{I_q}{I^\dagger _q}  -\frac{S^\dagger _q}{S_q} - \frac{I_i}{I^\dagger _i}\frac{S_q}{S^\dagger _q}\frac{E^{(1)\dagger }_r}{E^{(1)}_r}- {\dsum_{j=1}^{\ell}}\frac{E^{(j)}_q}{E^{(j)\dagger }_q}\frac{E^{(j+1)\dagger }_r}{E^{(j+1)}_r}-\, {\dsum_{j=1}^{\ell}}\frac{ E^{(j)}_r}{ E^{(j)\dagger }_r}\frac{E^{(j)\dagger }_q}{E^{(j)}_q} \nonumber\\& -\,\frac{E^{(\ell+1)}_r}{E^{(\ell+1)\dagger }_r}\frac{I^\dagger _q}{I_q} \Big) +\,\widetilde\xi_iv_iR^\dagger _i \Big(2\ell+6 - \frac{S^\dagger _i}{S_i} -\, \frac{I_q}{I^\dagger _q}\frac{S_i}{S^\dagger_i}\frac{I^\dagger_i}{I_i} -\frac{I_i}{I^\dagger_i} \frac{R^\dagger_i}{R_i} -\frac{S^\dagger _q}{S_q} -\; \frac{R_i}{R^\dagger _i}\frac{S_q}{S^\dagger _q}\frac{E^{(1)\dagger }_r}{E^{(1)}_r}- {\dsum_{j=1}^{l}}\frac{E^{(j)}_q}{E^{(j)\dagger }_q}\frac{E^{(j+1)\dagger }_r}{E^{(j+1)}_r} \nonumber\\&-\, {\dsum_{j=1}^{\ell}}\frac{ E^{(j)}_r}{ E^{(j)\dagger }_r}\frac{E^{(j)\dagger }_q}{E^{(j)}_q} -\,\frac{E^{(l+1)}_r}{E^{(\ell+1)\dagger }_r}\frac{I^\dagger _q}{I_q} \Big)\Big] +\,u_i\Big[\overline{\xi_i} I^\dagger _i\Big(4+\frac{I_q}{I^\dagger _q}- \frac{I^\dagger_i}{I_i}\frac{E_i}{E^\dagger_i}-\frac{S^\dagger _i}{S_i}-\frac{S_i}{S^\dagger _i}\frac{E^\dagger _i}{E_i}\frac{I_q}{I^\dagger _q}-\frac{S^\dagger _q}{S_q}\frac{S_r}{S^\dagger _r}-\frac{I_i}{I^\dagger _i}\frac{S_q}{S^\dagger _q}\frac{S^\dagger _r}{S_r}\Big) \nonumber\\& +\, \overline{\widetilde\xi_i}v_iR^\dagger_i\Big(5 + \frac{I_q}{I^\dagger_q} - \frac{S^\dagger _i}{S_i} -\frac{I^\dagger_i}{I_i} \frac{E_i}{E^\dagger_i} -\frac{S^\dagger _q}{S_q} \frac{S_r}{S^\dagger _r}-\,\frac{R_i}{R^\dagger _i}\frac{S_q}{S^\dagger _q}\frac{S^\dagger _r}{S_r}  -\frac{I_i}{I^\dagger_i} \frac{R^\dagger_i}{R_i} - \frac{I_q}{I^\dagger _q}\frac{S_i}{S^\dagger_i} \frac{E^\dagger_i}{E_i} \Big)\Big] + \bar u_i \Big[\overline{\xi_i}I^\dagger_i\Big(3 + \frac{I_q}{I_q^\dagger} - \frac{S^\dagger _i}{S_i}  -\frac{S^\dagger _q}{S_q}\frac{S_r}{S^\dagger _r} \nonumber\\& -\,\frac{I_i}{I^\dagger _i}\frac{S_q}{S^\dagger _q}\frac{S^\dagger _r}{S_r}-\, \frac{I_q}{I_q^\dagger} \frac{S_i}{S_i^\dagger} \frac{I^\dagger_i}{I_i}\Big) + \overline{\widetilde\xi_i}v_iR^\dagger_i\Big(4 + \frac{I_q}{I_q^\dagger} - \frac{S^\dagger _i}{S_i}  -\frac{S^\dagger _q}{S_q}  \frac{S_r}{S^\dagger_r} -\frac{R_i}{R^\dagger _i}\frac{S_q}{S^\dagger _q}\frac{S^\dagger _r}{S_r} - \frac{R^\dagger_i}{R_i} \frac{I_i}{I^\dagger_i} - \frac{I_q}{I_q^\dagger} \frac{S_i}{S_i^\dagger} \frac{I^\dagger_i}{I_i} \Big)\Big]\Big] +\; u_i\Big[\widetilde\nu_iE^\dagger_i\Big(1 + \frac{I_q}{I^\dagger_q}\nonumber\\&-\frac{I_q}{I^\dagger_q}\frac{S_i}{S^\dagger_i}\frac{E^\dagger_i}{E_i} - \frac{S^\dagger_i}{S_i} \frac{E_i}{E^\dagger_i}\Big) + (\widetilde\nu_i + \bar v_i\gamma_i)I^\dagger _i\left(2 + \frac{I_q}{I^\dagger _q} - \frac{I_i}{I^\dagger _i} \frac{S^\dagger _i}{S_i} - \frac{I^\dagger_i}{I_i}\frac{E_i}{E^\dagger_i} - \frac{I_q}{I^\dagger _q} \frac{S_i}{S^\dagger _i}\frac{E^\dagger _i}{E_i}\right) +\, v_i\hat\zeta_iR^\dagger_i\Big(3 + \frac{I_q}{I^\dagger _q} - \frac{I_i}{I^\dagger _i} \frac{S^\dagger _i}{S_i}- \frac{I^\dagger_i}{I_i}\frac{E_i}{E^\dagger_i}\nonumber\end{align}\begin{align} \nonumber\\&  - \frac{I_i}{I^\dagger_i} \frac{R^\dagger_i}{R_i} - \frac{I_q}{I^\dagger _q} \frac{S_i}{S^\dagger _i}\frac{E^\dagger _i}{E_i}\Big)\Big]  +\; \bar u_i\Big[ (\widetilde\nu_i + \bar v_i\gamma_i)I^\dagger _i\left(1 + \frac{I_q}{I^\dagger _q} - \frac{I_i}{I^\dagger _i} \frac{S^\dagger _i}{S_i}  - \frac{I_q}{I^\dagger _q} \frac{S_i}{S^\dagger _i}\frac{I^\dagger _i}{I_i}\right)+\, v_i\hat\zeta_iR^\dagger_i\Big(2 + \frac{I_q}{I^\dagger _q} - \frac{I_i}{I^\dagger _i} \frac{S^\dagger _i}{S_i}    - \frac{I_i}{I^\dagger_i} \frac{R^\dagger_i}{R_i} \nonumber\\&- \frac{I_q}{I^\dagger _q} \frac{S_i}{S^\dagger _i}\frac{I^\dagger _i}{I_i}\Big)\Big] \Bigg]\nonumber\hfill\\
:=&   \mu S^\dagger _r  \Big(1 + \frac{S^\dagger _q}{S_q}\frac{S_r}{S_r^\dagger } - \frac{S^\dagger _q}{S_q}-\frac{S_r}{S_r^\dagger }\Big)  + f(\mathbf x)+{\dsum_{i=1}^n}\varpi_i G_{2,\,i},
\label{eq:eqliapeeder21}
\end{align}

with
\begin{equation}\label{eq:G2exp2}
G_{2,\,i} = g_{2,\,i}(\mathbf x)+h_{2,\,i}(\mathbf x)+p_{i}(\mathbf x),
\end{equation}
where $p_i$ is from~\eqref{eq:frgmtliap1} and 

\begin{align}
g_{2,\,i}(\mathbf x)= &\hat\nu_i\Bigg[ S_i^\dagger  \Big(3-\frac{S^\dagger _i}{S_i}-\frac{S^\dagger _q}{S_q}\frac{S_r}{S^\dagger _r}-\frac{S_i}{S^\dagger _i}\frac{S_q}{S^\dagger _q}\frac{S^\dagger _r}{S_r}\Big) + u_i\Big[\xi_i I^\dagger _i\Big(2\ell+6 - \frac{I^\dagger_i}{I_i}\frac{E_i}{E^\dagger_i}-\frac{S^\dagger _i}{S_i}-\frac{S_i}{S^\dagger _i}\frac{E^\dagger _i}{E_i}\frac{I_q}{I^\dagger _q} -\frac{S^\dagger _q}{S_q} - \frac{I_i}{I^\dagger _i}\frac{S_q}{S^\dagger _q}\frac{E^{(1)\dagger }_r}{E^{(1)}_r}\nonumber\\&- {\dsum_{j=1}^{\ell}}\frac{E^{(j)}_q}{E^{(j)\dagger }_q}\frac{E^{(j+1)\dagger }_r}{E^{(j+1)}_r}-\, {\dsum_{j=1}^{\ell}}\frac{ E^{(j)}_r}{ E^{(j)\dagger }_r}\frac{E^{(j)\dagger }_q}{E^{(j)}_q} -\frac{E^{(\ell+1)}_r}{E^{(\ell+1)\dagger }_r}\frac{I^\dagger _q}{I_q} \Big) + v_i\widetilde\xi_iR^\dagger_i\Big(2\ell+7 - \frac{S^\dagger _i}{S_i}-\frac{S_i}{S^\dagger _i}\frac{E^\dagger _i}{E_i}\frac{I_q}{I^\dagger _q}  - \frac{I^\dagger_i}{I_i}\frac{E_i}{E^\dagger_i}  - \frac{I_i}{I^\dagger_i}\frac{R^\dagger_i}{R_i} - \frac{S^\dagger _q}{S_q} \nonumber\\&- \frac{R_i}{R^\dagger _i}\frac{S_q}{S^\dagger _q}\frac{E^{(1)\dagger }_r}{E^{(1)}_r}- {\dsum_{j=1}^{\ell}}\frac{E^{(j)}_q}{E^{(j)\dagger }_q}\frac{E^{(j+1)\dagger }_r}{E^{(j+1)}_r}-\, {\dsum_{j=1}^{\ell}}\frac{ E^{(j)}_r}{ E^{(j)\dagger }_r}\frac{E^{(j)\dagger }_q}{E^{(j)}_q} -\frac{E^{(\ell+1)}_r}{E^{(\ell+1)\dagger }_r}\frac{I^\dagger _q}{I_q} \Big)\Big] + \bar u_i\Big[\xi_i I^\dagger _i\Big(2\ell+5 - \frac{S^\dagger _i}{S_i}-\frac{S_i}{S^\dagger _i}\frac{I^\dagger _i}{I_i}\frac{I_q}{I^\dagger _q} -\frac{S^\dagger _q}{S_q} \nonumber\\&- \frac{I_i}{I^\dagger _i}\frac{S_q}{S^\dagger _q}\frac{E^{(1)\dagger }_r}{E^{(1)}_r}- {\dsum_{j=1}^{\ell}}\frac{E^{(j)}_q}{E^{(j)\dagger }_q}\frac{E^{(j+1)\dagger }_r}{E^{(j+1)}_r}-\, {\dsum_{j=1}^{\ell}}\frac{ E^{(j)}_r}{ E^{(j)\dagger }_r}\frac{E^{(j)\dagger }_q}{E^{(j)}_q} -\frac{E^{(\ell+1)}_r}{E^{(\ell+1)\dagger }_r}\frac{I^\dagger _q}{I_q} \Big) + v_i\widetilde\xi_iR^\dagger_i\Big(2\ell+6 - \frac{S^\dagger _i}{S_i}-\frac{S_i}{S^\dagger _i}\frac{I^\dagger _i}{I_i}\frac{I_q}{I^\dagger _q}  - \frac{I_i}{I^\dagger_i}\frac{R^\dagger_i}{R_i} - \frac{S^\dagger _q}{S_q} \nonumber\\&- \frac{R_i}{R^\dagger _i}\frac{S_q}{S^\dagger _q}\frac{E^{(1)\dagger }_r}{E^{(1)}_r}- {\dsum_{j=1}^{\ell}}\frac{E^{(j)}_q}{E^{(j)\dagger }_q}\frac{E^{(j+1)\dagger }_r}{E^{(j+1)}_r}-\, {\dsum_{j=1}^{\ell}}\frac{ E^{(j)}_r}{ E^{(j)\dagger }_r}\frac{E^{(j)\dagger }_q}{E^{(j)}_q} -\frac{E^{(\ell+1)}_r}{E^{(\ell+1)\dagger }_r}\frac{I^\dagger _q}{I_q} \Big)\Big]
\Bigg],
\\h_{2,\,i}(\mathbf x)=& \hat\nu_i\Bigg[u_i\Big[E^\dagger _i \Big(3 + \frac{I_q}{I^\dagger_q} - \frac{S^\dagger _q}{S_q}\frac{S_r}{S^\dagger _r} -\frac{S^\dagger _i}{S_i} -\frac{E_i}{E^\dagger _i} \frac{S_q}{S^\dagger _q}\frac{S^\dagger _r}{S_r}-\frac{I_q}{I^\dagger_q} \frac{S_i}{S^\dagger_i} \frac{E^\dagger_i}{E_i}\Big) + \overline{\xi_i} I^\dagger _i\Big(4+\frac{I_q}{I^\dagger _q}- \frac{I^\dagger_i}{I_i}\frac{E_i}{E^\dagger_i}-\frac{S^\dagger _i}{S_i}-\frac{S_i}{S^\dagger _i}\frac{E^\dagger _i}{E_i}\frac{I_q}{I^\dagger _q}-\frac{S^\dagger _q}{S_q}\frac{S_r}{S^\dagger _r}-\frac{I_i}{I^\dagger _i}\frac{S_q}{S^\dagger _q}\frac{S^\dagger _r}{S_r}\Big)\nonumber\\& + v_i\overline{\widetilde\xi_i}R^\dagger_i\Big(5+\frac{I_q}{I^\dagger _q}- \frac{I^\dagger_i}{I_i}\frac{E_i}{E^\dagger_i}-\frac{I_i}{I^\dagger_i}\frac{R^\dagger_i}{R_i}-\frac{S^\dagger _i}{S_i}-\frac{S_i}{S^\dagger _i}\frac{E^\dagger _i}{E_i}\frac{I_q}{I^\dagger _q}-\frac{S^\dagger _q}{S_q}\frac{S_r}{S^\dagger _r}-\frac{R_i}{R^\dagger _i}\frac{S_q}{S^\dagger _q}\frac{S^\dagger _r}{S_r}\Big)\Big] + \bar u_i\Big[\overline{\xi_i} I^\dagger _i\Big(3+\frac{I_q}{I^\dagger _q}- \frac{S^\dagger _i}{S_i}-\frac{S_i}{S^\dagger _i}\frac{I^\dagger _i}{I_i}\frac{I_q}{I^\dagger _q}\nonumber\\&-\frac{S^\dagger _q}{S_q}\frac{S_r}{S^\dagger _r}-\frac{I_i}{I^\dagger _i}\frac{S_q}{S^\dagger _q}\frac{S^\dagger _r}{S_r}\Big) + v_i\overline{\widetilde\xi_i}R^\dagger_i\Big(4+\frac{I_q}{I^\dagger _q}- \frac{I_i}{I^\dagger_i}\frac{R^\dagger_i}{R_i}-\frac{S^\dagger _i}{S_i}-\frac{S_i}{S^\dagger _i}\frac{I^\dagger _i}{I_i}\frac{I_q}{I^\dagger _q}-\frac{S^\dagger _q}{S_q}\frac{S_r}{S^\dagger _r}-\frac{R_i}{R^\dagger _i}\frac{S_q}{S^\dagger _q}\frac{S^\dagger _r}{S_r}\Big)\Big]\Bigg],
\end{align}
\begin{align}
\end{align}

\noindent Alternatively, for each $i$, by adding and subtracting $u_i\nu_iE^\dagger _i \left( 1-\frac{S_i}{S^\dagger _i}\frac{E^\dagger _i}{E_i} \right) + \bar u_iI^\dagger _i \left( 1-\frac{S_i}{S^\dagger _i}\frac{I^\dagger _i}{I_i} \right)$ the function $G_{2,\,i}$ in~\eqref{eq:G2exp2}  may be rewritten as

\begin{align}
G_{2,\,i}(\mathbf x)  = &  \hat\nu_i\Bigg[S_i^\dagger  \Big(3-\frac{S^\dagger _i}{S_i}-\frac{S^\dagger _q}{S_q}\frac{S_r}{S^\dagger _r}-\frac{S_i}{S^\dagger _i}\frac{S_q}{S^\dagger _q}\frac{S^\dagger _r}{S_r}\Big) + u_i\Big[E^\dagger _i \Big(4  -\frac{S^\dagger _i}{S_i} - \frac{S_r}{S^\dagger _r} \frac{S^\dagger _q}{S_q} -\frac{E_i}{E^\dagger _i} \frac{S_q}{S^\dagger _q}\frac{S^\dagger _r}{S_r}- \frac{S_i}{S^\dagger_i} \frac{E^\dagger_i}{E_i}\Big) +\,I^\dagger _i\Big(\xi_i \Big(2\ell+7 - \frac{I^\dagger_i}{I_i} \frac{E_i}{E^\dagger_i}\nonumber\\&-\frac{S^\dagger _i}{S_i}-\;\frac{S_i}{S^\dagger _i}\frac{E^\dagger _i}{E_i} -\frac{S^\dagger _q}{S_q} - \frac{I_i}{I^\dagger _i}\frac{S_q}{S^\dagger _q}\frac{E^{(1)\dagger }_r}{E^{(1)}_r}- {\dsum_{j=1}^{\ell}}\frac{E^{(j)}_q}{E^{(j)\dagger }_q}\frac{E^{(j+1)\dagger }_r}{E^{(j+1)}_r}-\, {\dsum_{j=1}^{\ell}}\frac{ E^{(j)}_r}{ E^{(j)\dagger }_r}\frac{E^{(j)\dagger }_q}{E^{(j)}_q} -\frac{E^{(\ell+1)}_r}{E^{(\ell+1)\dagger }_r}\frac{I^\dagger _q}{I_q}  -\frac{I_q}{I^\dagger _q}\Big)  +\,\overline{\xi_i} \Big(5 - \frac{I^\dagger_i}{I_i} \frac{E_i}{E^\dagger_i} -\frac{S^\dagger _i}{S_i} \nonumber\\&-\frac{S_i}{S^\dagger _i}\frac{E^\dagger _i}{E_i} -\frac{S_r}{S^\dagger _r}\frac{S^\dagger _q}{S_q}-\frac{I_i}{I^\dagger _i}\frac{S_q}{S^\dagger _q}\frac{S^\dagger _r}{S_r}\Big)\Big)
+\; v_iR^\dagger_i\Big(\widetilde\xi_i\Big(2\ell+8 - \frac{S^\dagger _i}{S_i}-\frac{S_i}{S^\dagger _i}\frac{E^\dagger _i}{E_i}  - \frac{I^\dagger_i}{I_i}\frac{E_i}{E^\dagger_i}  - \frac{I_i}{I^\dagger_i}\frac{R^\dagger_i}{R_i} - \frac{S^\dagger _q}{S_q} - \frac{R_i}{R^\dagger _i}\frac{S_q}{S^\dagger _q}\frac{E^{(1)\dagger }_r}{E^{(1)}_r} -\frac{I_q}{I^\dagger _q}  \nonumber\end{align}\begin{align} &- {\dsum_{j=1}^{\ell}}\frac{E^{(j)}_q}{E^{(j)\dagger }_q}\frac{E^{(j+1)\dagger }_r}{E^{(j+1)}_r}-\, {\dsum_{j=1}^{\ell}}\frac{ E^{(j)}_r}{ E^{(j)\dagger }_r}\frac{E^{(j)\dagger }_q}{E^{(j)}_q}-\;\frac{E^{(\ell+1)}_r}{E^{(\ell+1)\dagger }_r}\frac{I^\dagger _q}{I_q}\Big)+ \overline{\widetilde\xi_i}\Big(6 - \frac{I^\dagger_i}{I_i} \frac{E_i}{E^\dagger_i} -\frac{I_i}{I^\dagger_i} \frac{R^\dagger_i}{R_i}-\frac{S^\dagger _i}{S_i}-\frac{S_i}{S^\dagger _i}\frac{E^\dagger _i}{E_i} -\frac{S_r}{S^\dagger _r}\frac{S^\dagger _q}{S_q}-\frac{R_i}{R^\dagger _i}\frac{S_q}{S^\dagger _q}\frac{S^\dagger _r}{S_r}\Big)\Big) \Big] \nonumber\\& + \bar u_i\Big[I^\dagger _i\Big(\overline{\xi_i} \Big(4  -\frac{S^\dagger _i}{S_i} - \,\frac{S_i}{S^\dagger_i} \frac{I^\dagger_i}{I_i} - \frac{S_r}{S^\dagger _r}\frac{S^\dagger _q}{S_q}-\frac{I_i}{I^\dagger _i} \frac{S_q}{S^\dagger _q}\frac{S^\dagger _r}{S_r}\Big) +\,\xi_i\Big(2\ell+6 - \frac{S^\dagger _i}{S_i}-\frac{S_i}{S^\dagger _i}\frac{I^\dagger_i}{I_i} -\frac{S^\dagger _q}{S_q} - \frac{I_i}{I^\dagger _i}\frac{S_q}{S^\dagger _q}\frac{E^{(1)\dagger }_r}{E^{(1)}_r}-\frac{I_q}{I^\dagger _q}\nonumber\\&- {\dsum_{j=1}^{\ell}}\frac{E^{(j)}_q}{E^{(j)\dagger }_q}\frac{E^{(j+1)\dagger }_r}{E^{(j+1)}_r}-\, {\dsum_{j=1}^{\ell}}\frac{ E^{(j)}_r}{ E^{(j)\dagger }_r}\frac{E^{(j)\dagger }_q}{E^{(j)}_q} -\frac{E^{(\ell+1)}_r}{E^{(\ell+1)\dagger }_r}\frac{I^\dagger _q}{I_q}  \Big) \Big) +\; v_iR^\dagger_i\Big(\widetilde\xi_i\Big(2\ell+7 - \frac{S^\dagger _i}{S_i}-\frac{S_i}{S^\dagger _i}\frac{I^\dagger_i}{I_i}  - \frac{I_i}{I^\dagger_i}\frac{R^\dagger_i}{R_i} - \frac{S^\dagger _q}{S_q} \nonumber\\&- \frac{R_i}{R^\dagger _i}\frac{S_q}{S^\dagger _q}\frac{E^{(1)\dagger }_r}{E^{(1)}_r}-\frac{I_q}{I^\dagger _q}- {\dsum_{j=1}^{\ell}}\frac{E^{(j)}_q}{E^{(j)\dagger }_q}\frac{E^{(j+1)\dagger }_r}{E^{(j+1)}_r}-\, {\dsum_{j=1}^{\ell}}\frac{ E^{(j)}_r}{ E^{(j)\dagger }_r}\frac{E^{(j)\dagger }_q}{E^{(j)}_q} -\;\frac{E^{(\ell+1)}_r}{E^{(\ell+1)\dagger }_r}\frac{I^\dagger _q}{I_q} \Big) +\; \overline{\widetilde\xi_i}\Big(6 - \frac{I^\dagger_i}{I_i} \frac{E_i}{E^\dagger_i} -\frac{I_i}{I^\dagger_i} \frac{R^\dagger_i}{R_i}-\frac{S^\dagger _i}{S_i}-\frac{S_i}{S^\dagger _i}\frac{E^\dagger _i}{E_i} \nonumber\\& -\frac{S_r}{S^\dagger _r}\frac{S^\dagger _q}{S_q}-\frac{R_i}{R^\dagger _i}\frac{S_q}{S^\dagger _q}\frac{S^\dagger _r}{S_r}\Big)\Big)\Big]\Bigg]
+\; (\widetilde\nu_i + v_i\gamma_i)I^\dagger _i\Big[u_i\Big(3  - \frac{I_i}{I^\dagger _i} \frac{S^\dagger _i}{S_i} - \frac{S_i}{S^\dagger_i} \frac{E^\dagger_i}{E_i} - \frac{I^\dagger_i}{I_i}\frac{E_i}{E^\dagger_i}  \Big)+\; \bar u_i\Big(2  - \frac{I_i}{I^\dagger _i} \frac{S^\dagger _i}{S_i} - \frac{S_i}{S^\dagger_i}\frac{I^\dagger_i}{I_i}   \Big)\Big]  \nonumber\\&+ v_i\hat\zeta_iR^\dagger _i\Big[u_i\Big(4 - \frac{R_i}{R^\dagger _i} \frac{S^\dagger _i}{S_i} - \frac{I^\dagger_i}{I_i}\frac{E_i}{E^\dagger_i}- \frac{I_i}{I^\dagger_i}\frac{R^\dagger_i}{R_i} - \frac{S_i}{S^\dagger _i}\frac{E^\dagger _i}{E_i}\Big) +\; \bar u_i\Big(3 - \frac{I_i}{I^\dagger_i}\frac{R^\dagger_i}{R_i} -\frac{R_i}{R^\dagger _i} \frac{S^\dagger _i}{S_i} -  \frac{S_i}{S^\dagger _i}\frac{I^\dagger _i}{I_i}\Big)\Big] +\; u_i\bar\varepsilon_iE^\dagger_i\Big(\frac{I_q}{I^\dagger _q} + \frac{S_i}{S^\dagger _i}\frac{E^\dagger _i}{E_i} - 1\nonumber\\& - \frac{I_q}{I^\dagger _q} \frac{S_i}{S^\dagger _i}\frac{E^\dagger _i}{E_i}\Big)+\; \bar u_i\bar\gamma_iI^\dagger_i\Big(\frac{I_q}{I^\dagger _q} + \frac{S_i}{S^\dagger _i}\frac{I^\dagger _i}{I_i} - 1 - \frac{I_q}{I^\dagger _q} \frac{S_i}{S^\dagger _i}\frac{I^\dagger _i}{I_i}\Big) + u_i\widetilde\nu_iE^\dagger_i\Big(2-\frac{S_i}{S^\dagger_i}\frac{E^\dagger_i}{E_i} - \frac{S^\dagger_i}{S_i} \frac{E_i}{E^\dagger_i}\Big)  \nonumber\\
:=&   \widetilde{g}_{2,\,i}(\mathbf x)+\widetilde p_{i}(\mathbf x),\label{eq:eqliapeeder23}
\end{align}
with $\widetilde p_{i}(\mathbf x)$ as in~\eqref{eq:frgmtliap22}, and

\begin{align}
\widetilde g_{2,\,i}(\mathbf x)= &I^\dagger _i\Bigg[\hat\nu_i \Big[u_i\Big(\xi_i\Big(2\ell+7 - \frac{I^\dagger_i}{I_i} \frac{E_i}{E^\dagger_i}-\frac{S^\dagger _i}{S_i}-\frac{S_i}{S^\dagger _i}\frac{E^\dagger _i}{E_i} -\frac{S^\dagger _q}{S_q} - \frac{I_i}{I^\dagger _i}\frac{S_q}{S^\dagger _q}\frac{E^{(1)\dagger }_r}{E^{(1)}_r}- {\dsum_{j=1}^{\ell}}\frac{E^{(j)}_q}{E^{(j)\dagger }_q}\frac{E^{(j+1)\dagger }_r}{E^{(j+1)}_r}-\, {\dsum_{j=1}^{\ell}}\frac{ E^{(j)}_r}{ E^{(j)\dagger }_r}\frac{E^{(j)\dagger }_q}{E^{(j)}_q} -\frac{E^{(\ell+1)}_r}{E^{(\ell+1)\dagger }_r}\frac{I^\dagger _q}{I_q}  -\frac{I_q}{I^\dagger _q}\Big)\nonumber\\
& + \,\overline{\xi_i}\Big(5 - \frac{I^\dagger_i}{I_i} \frac{E_i}{E^\dagger_i} -\frac{S^\dagger _i}{S_i} -\frac{S_i}{S^\dagger _i}\frac{E^\dagger _i}{E_i} -\frac{S_r}{S^\dagger _r}\frac{S^\dagger _q}{S_q}-\frac{I_i}{I^\dagger _i}\frac{S_q}{S^\dagger _q}\frac{S^\dagger _r}{S_r}\Big)\Big) + \bar u_i\Big(\xi_i\Big(2\ell+6 -\frac{S^\dagger _i}{S_i}-\frac{S_i}{S^\dagger _i} \frac{I^\dagger_i}{I_i} -\frac{S^\dagger _q}{S_q} - \frac{I_i}{I^\dagger _i}\frac{S_q}{S^\dagger _q}\frac{E^{(1)\dagger }_r}{E^{(1)}_r}-\frac{I_q}{I^\dagger _q}\nonumber\\
&- {\dsum_{j=1}^{\ell}}\frac{E^{(j)}_q}{E^{(j)\dagger }_q}\frac{E^{(j+1)\dagger }_r}{E^{(j+1)}_r}-\, {\dsum_{j=1}^{\ell}}\frac{ E^{(j)}_r}{ E^{(j)\dagger }_r}\frac{E^{(j)\dagger }_q}{E^{(j)}_q} -\frac{E^{(\ell+1)}_r}{E^{(\ell+1)\dagger }_r}\frac{I^\dagger _q}{I_q}  \Big) + \,\overline{\xi_i}\Big(4 - \frac{S^\dagger _i}{S_i} -\frac{S_i}{S^\dagger _i}\frac{I^\dagger_i}{I_i} -\frac{S_r}{S^\dagger _r}\frac{S^\dagger _q}{S_q}-\frac{I_i}{I^\dagger _i}\frac{S_q}{S^\dagger _q}\frac{S^\dagger _r}{S_r}\Big)\Big)\Big]\nonumber \\&+ (\widetilde\nu_i + \gamma_i)\Big[u_i\Big(3  - \frac{I_i}{I^\dagger _i} \frac{S^\dagger _i}{S_i} - \frac{S_i}{S^\dagger_i}\frac{E^\dagger_i}{E_i} - \frac{I^\dagger_i}{I_i}\frac{E_i}{E^\dagger_i}  \Big) + \bar u_i\Big(2  - \frac{I_i}{I^\dagger _i} \frac{S^\dagger _i}{S_i} - \frac{S_i}{S^\dagger_i}\frac{I^\dagger_i}{I_i}   \Big)\Big] \Bigg]\nonumber\\& +\;v_iR^\dagger _i\Bigg[\hat\nu_i \Big[u_i\Big(\xi_i\Big(2\ell+8 - \frac{S^\dagger _i}{S_i}-\frac{S_i}{S^\dagger _i}\frac{E^\dagger _i}{E_i}  - \frac{I^\dagger_i}{I_i} \frac{E_i}{E^\dagger_i} -\frac{I_i}{I^\dagger_i} \frac{R^\dagger_i}{R_i} -\frac{S^\dagger _q}{S_q} - \frac{R_i}{R^\dagger _i}\frac{S_q}{S^\dagger _q}\frac{E^{(1)\dagger }_r}{E^{(1)}_r}-\frac{I_q}{I^\dagger _q} - {\dsum_{j=1}^{\ell}} \frac{E^{(j)}_q}{E^{(j)\dagger }_q}\frac{E^{(j+1)\dagger }_r}{E^{(j+1)}_r}-\, {\dsum_{j=1}^{\ell}}\frac{ E^{(j)}_r}{ E^{(j)\dagger }_r}\frac{E^{(j)\dagger }_q}{E^{(j)}_q}\nonumber \\
&-\frac{E^{(\ell+1)}_r}{E^{(\ell+1)\dagger }_r}\frac{I^\dagger _q}{I_q}  \Big) + \,\overline{\xi_i}\Big(6  -\frac{S^\dagger _i}{S_i} -\frac{S_i}{S^\dagger _i}\frac{E^\dagger _i}{E_i} - \frac{E_i}{E^\dagger_i}\frac{I^\dagger_i}{I_i} - \frac{I_i}{I^\dagger_i}\frac{R^\dagger_i}{R_i}  -\frac{S_r}{S^\dagger _r}\frac{S^\dagger _q}{S_q}-\frac{R_i}{R^\dagger _i}\frac{S_q}{S^\dagger _q}\frac{S^\dagger _r}{S_r}\Big)\Big) + \bar u_i\Big(\xi_i\Big(2\ell+7 -\frac{S^\dagger _i}{S_i}-\frac{S_i}{S^\dagger _i} \frac{I^\dagger_i}{I_i}  -\frac{S^\dagger _q}{S_q}\nonumber\\
& - \frac{I_i}{I^\dagger_i}\frac{R^\dagger_i}{R_i} - \frac{R_i}{R^\dagger _i}\frac{S_q}{S^\dagger _q}\frac{E^{(1)\dagger }_r}{E^{(1)}_r}-\frac{I_q}{I^\dagger _q}- {\dsum_{j=1}^{\ell}}\frac{E^{(j)}_q}{E^{(j)\dagger }_q}\frac{E^{(j+1)\dagger }_r}{E^{(j+1)}_r}-\, {\dsum_{j=1}^{\ell}}\frac{ E^{(j)}_r}{ E^{(j)\dagger }_r}\frac{E^{(j)\dagger }_q}{E^{(j)}_q} -\frac{E^{(\ell+1)}_r}{E^{(\ell+1)\dagger }_r}\frac{I^\dagger _q}{I_q}  \Big) + \,\overline{\xi_i}\Big(5 - \frac{S^\dagger _i}{S_i} -\frac{S_i}{S^\dagger _i}\frac{I^\dagger_i}{I_i} -\frac{S_r}{S^\dagger _r}\frac{S^\dagger _q}{S_q} - \frac{I_i}{I^\dagger_i}\frac{R^\dagger_i}{R_i}\nonumber\\& -\frac{I_i}{I^\dagger _i}\frac{S_q}{S^\dagger _q}\frac{S^\dagger _r}{S_r}\Big)\Big)\Big] + \hat\zeta_i\Big[u_i\Big(4  - \frac{R_i}{R^\dagger _i} \frac{S^\dagger _i}{S_i} - \frac{S_i}{S^\dagger_i}\frac{E^\dagger_i}{E_i} - \frac{E_i}{E^\dagger_i}\frac{I^\dagger_i}{I_i} - \frac{I_i}{I^\dagger_i}\frac{R^\dagger_i}{R_i} \Big) + \bar u_i\Big(3  - \frac{R_i}{R^\dagger _i} \frac{S^\dagger _i}{S_i} - \frac{S_i}{S^\dagger_i}\frac{I^\dagger_i}{I_i}  - \frac{I_i}{I^\dagger_i}\frac{R^\dagger_i}{R_i} \Big)\Big] \Bigg] \nonumber\\& +\,\hat\nu_i\,S_i^\dagger  \Big( 3-\frac{S^\dagger _i}{S_i} - \frac{S_r}{S^\dagger _r}\frac{S^\dagger _q}{S_q}-\frac{S_i}{S^\dagger _i}\frac{S_q}{S^\dagger _q}\frac{S^\dagger _r}{S_r}\Big) + \widetilde\nu_iE^\dagger_i\Big(2-\frac{S_i}{S^\dagger_i}\frac{E^\dagger_i}{E_i} - \frac{S^\dagger_i}{S_i} \frac{E_i}{E^\dagger_i}\Big).
\end{align}

\noindent Thus, for $\mathbf x\in\Omega_2$, we may write $\frac{d\,V_{ee}}{dt}(\mathbf x(t))$ as: 

\begin{align}
\frac{d\,V_{ee}}{dt}(\mathbf x(t)) =& \mu S_r^\dagger\Big(1 + \frac{S_r}{S^\dagger_r}\frac{S^\dagger_q}{S_q} - \frac{S^\dagger_q}{S_q} - \frac{S_r}{S^\dagger_r}\Big) + f(\mathbf x)\nonumber\\& +\, \dsum_{\mathcal I_2}\varpi_i\Big(g_{2,\,i}(\mathbf x)+h_{2,\,i}(\mathbf x)+p_{i}(\mathbf x)\Big)+ \dsum_{\mathcal J_2}\varpi_i\Big(\widetilde g_{2,\,i}(\mathbf x) + \widetilde p_{i}(\mathbf x)\Big)\label{eq:rdcexp2vee}
\end{align} where $\{\mathcal I_2,~\mathcal J_2\}$ is any partition of $\{1,~2,~\ldots,~n\}$.


Using expressions~\eqref{eq:rdcexp1vee} and~\eqref{eq:rdcexp2vee}, we may now show that $\frac{d\,V_{ee}}{dt}(\mathbf x(t))\leqslant 0$ on $\Omega_1$ and $\Omega_2$ respectively. The initial terms in~\eqref{eq:rdcexp1vee} and~\eqref{eq:rdcexp2vee} are non-positive on  $\Omega_1$ and $\Omega_2$ respectively, by the definitions of $\Omega_1$ and $\Omega_2$. Corollary~\ref{cor.agmi} of the arithmetic--geometric means inequality (Lemma~\ref{lem.agmi}) implies $f_k(\mathbf x)\leqslant0$, $g_{k,\,i}(\mathbf x)\leq0$ and $\widetilde g_{k,\,i}(\mathbf x)\leqslant0$ for each $i$ and  $k\in\{1,~2\}$ when  $\mathbf x\in \left(\mathbb R_{>0}\right)^u$. For any $\mathbf x\in  \left(\mathbb R_{>0}\right)^u$ and for each $i$ such that $(I_i-I^\dagger _i)(I_q-I^\dagger _q)\geqslant0$  when $u_i=0$ or $(E_i-E^\dagger _i)(I_q-I^\dagger _q)\geqslant0$  when $u_i=1$, we may show that $\widetilde p_i\leqslant 0$ by applying Corollary~\ref{cor.agmi0} to the ratios $\frac{I^\dagger _q}{I_q}$ and $\frac{I_q}{I^\dagger _q}$ with respective associated weights $\frac{I_q}{I^\dagger _q}$ and $\frac{S_i}{S^\dagger _i}\frac{I^\dagger _i}{I_i}$  when $u_i=0$, or  to the ratios $\frac{I^\dagger _q}{I_q}$ and $\frac{I_q}{I^\dagger _q}$ with respective associated weights $\frac{I_q}{I^\dagger _q}$ and $\frac{S_i}{S^\dagger _i}\frac{E^\dagger _i}{E_i}$ when $u_i=1$. 
On the other hand, when  $(I_i-I^\dagger _i)(I_q-I^\dagger _q)\leqslant0$ (that corresponds to case of $i$ where $u_i=0$) or   $(E_i-E^\dagger _i)(I_q-I^\dagger _q)\leq0$ (that corresponds to case of $i$ where $u_i=1$) we may show $h_{1,\,i}(\mathbf x)\leqslant0$, $h_{2,\,i}(\mathbf x)\leqslant0$, $p_{i}(\mathbf x)\leqslant 0$, $s_{i}(\mathbf x)\leqslant 0$ and $t_{i}(\mathbf x)\leqslant 0$ by applying the same corollary to the following data (given in pairs as (number, weight)) in the following description of given pairs of ($u_i$ and $v_i$); \\For cases of $i$ with $u_i=0$ and $v_i=0$,    $\left(\frac{S_i^\dagger }{S_i},~1\right)$, $\left(\frac{I_i^\dagger }{I_i}\frac{S_i}{S^\dagger _i},~ \frac{I_q}{I^\dagger _q}\right)$, $\left(\frac{S^\dagger _q}{S_q},~ 1\right)$, $\left(\frac{I_i}{I^\dagger _i}\frac{S_q}{S^\dagger _q} \frac{S^\dagger _r}{S_r},~ 1 \right)$, $\left(\frac{S_r}{S^\dagger _r},~1\right)$ for $h_{1,\,i}$;  $\left(\frac{S_i^\dagger }{S_i},~1\right)$, $\left(\frac{I_i^\dagger }{I_i}\frac{S_i}{S^\dagger _i},~ \frac{I_q}{I^\dagger _q}\right)$, $\left(\frac{I_i}{I^\dagger _i}\frac{S_q}{S^\dagger _q} \frac{S^\dagger _r}{S_r},~1\right)$, $\left(\frac{S^\dagger _q}{S_q}\frac{S_r}{S^\dagger _r},~ 1\right)$ for $h_{2,\,i}$; and $\left(\frac{S_i}{S^\dagger _i} \frac{I^\dagger _i}{I_i},~\frac{I_q}{I^\dagger _q}\right)$, $\left(\frac{S^\dagger _i}{S_i}\frac{I_i}{I^\dagger _i},~1\right)$  for $p_i$, \\for cases of $i$ with $u_i=0$ and $v_i=1$,    $\left(\frac{S_i^\dagger }{S_i},~1\right)$, $\left(\frac{I_i^\dagger }{I_i}\frac{S_i}{S^\dagger _i},~ \frac{I_q}{I^\dagger _q}\right)$, $\left(\frac{S^\dagger _q}{S_q},~ 1\right)$, $\left(\frac{I_i}{I^\dagger _i}\frac{S_q}{S^\dagger _q} \frac{S^\dagger _r}{S_r},~ 1 \right)$, $\left(\frac{S_r}{S^\dagger _r},~1\right)$, $\left(\frac{R_i}{R^\dagger _i}\frac{S_q}{S^\dagger _q} \frac{S^\dagger _r}{S_r},~1\right)$, $\left(\frac{I_i}{I^\dagger_i}\frac{R^\dagger_i}{R _i},~1\right)$  for $h_{1,\,i}$;  $\left(\frac{S_i^\dagger }{S_i},~1\right)$, $\left(\frac{I_i^\dagger }{I_i}\frac{S_i}{S^\dagger _i},~ \frac{I_q}{I^\dagger _q}\right)$, $\left(\frac{I_i}{I^\dagger _i}\frac{S_q}{S^\dagger _q} \frac{S^\dagger _r}{S_r},~1\right)$, $\left(\frac{S^\dagger _q}{S_q}\frac{S_r}{S^\dagger _r},~ 1\right)$, $\left(\frac{R_i}{R^\dagger _i}\frac{S_q}{S^\dagger _q} \frac{S^\dagger _r}{S_r},~1\right)$, $\left(\frac{I_i}{I^\dagger_i}\frac{R^\dagger_i}{R _i},~1\right)$ for $h_{2,\,i}$; and $\left( \frac{I^\dagger _i}{I_i}\frac{S_i}{S^\dagger _i},~\frac{I_q}{I^\dagger _q}\right)$, $\left(\frac{I_i}{I^\dagger _i}\frac{R^\dagger _i}{R_i},~1\right)$, $\left(\frac{R_i}{R^\dagger _i}\frac{S^\dagger _i}{S_i},~1\right)$  for $p_i$, \\for cases of $i$ with $u_i=1$ and $v_i=0$,    $\left(\frac{S_i^\dagger }{S_i},~1\right)$, $\left(\frac{S_i}{S^\dagger _i}\frac{E_i^\dagger }{E_i},~ \frac{I_q}{I^\dagger _q}\right)$, $\left(\frac{E_i}{E^\dagger _i}\frac{I_i^\dagger }{I_i},~ 1\right)$, $\left(\frac{S^\dagger _q}{S_q},~ 1\right)$, $\left(\frac{I_i}{I^\dagger _i}\frac{S_q}{S^\dagger _q} \frac{S^\dagger _r}{S_r},~ 1 \right)$, $\left(\frac{S_r}{S^\dagger _r},~1\right)$  for $h_{1,\,i}$;  $\left(\frac{S_i^\dagger }{S_i},~1\right)$, $\left(\frac{S_i}{S^\dagger _i}\frac{E_i^\dagger }{E_i},~ \frac{I_q}{I^\dagger _q}\right)$, $\left(\frac{E_i}{E^\dagger_i}\frac{I^\dagger_i}{I_i},~1\right)$, $\left(\frac{I_i}{I^\dagger _i}\frac{S_q}{S^\dagger _q} \frac{S^\dagger _r}{S_r},~1\right)$, $\left(\frac{S^\dagger _q}{S_q}\frac{S_r}{S^\dagger _r},~ 1\right)$ for $h_{2,\,i}$; and $\left( \frac{S_i}{S^\dagger _i}\frac{E^\dagger _i}{E_i},~\frac{I_q}{I^\dagger _q}\right)$, $\left(\frac{E_i}{E^\dagger _i}\frac{I^\dagger _i}{I_i},~1\right)$, $\left(\frac{I_i}{I^\dagger _i}\frac{S^\dagger _i}{S_i},~1\right)$, $\left(\frac{E_i}{E^\dagger _i}\frac{S^\dagger _i}{S_i},~1\right)$  for $p_i$, \\for cases of $i$ with $u_i=1$ and $v_i=1$,    $\left(\frac{S_i^\dagger }{S_i},~1\right)$, $\left(\frac{S_i}{S^\dagger _i}\frac{E_i^\dagger }{E_i},~ \frac{I_q}{I^\dagger _q}\right)$, $\left(\frac{E_i}{E^\dagger _i}\frac{I_i^\dagger }{I_i},~ 1\right)$, $\left(\frac{S^\dagger _q}{S_q},~ 1\right)$, $\left(\frac{I_i}{I^\dagger _i}\frac{S_q}{S^\dagger _q} \frac{S^\dagger _r}{S_r},~ 1 \right)$, $\left(\frac{S_r}{S^\dagger _r},~1\right)$  for $h_{1,\,i}$;  $\left(\frac{S_i^\dagger }{S_i},~1\right)$, $\left(\frac{S_i}{S^\dagger _i}\frac{E_i^\dagger }{E_i},~ \frac{I_q}{I^\dagger _q}\right)$, $\left(\frac{E_i}{E^\dagger_i}\frac{I^\dagger_i}{I_i},~1\right)$, $\left(\frac{I_i}{I^\dagger _i}\frac{S_q}{S^\dagger _q} \frac{S^\dagger _r}{S_r},~1\right)$, $\left(\frac{S^\dagger _q}{S_q}\frac{S_r}{S^\dagger _r},~ 1\right)$ for $h_{2,\,i}$; and $\left( \frac{S_i}{S^\dagger _i}\frac{E^\dagger _i}{E_i},~\frac{I_q}{I^\dagger _q}\right)$, $\left(\frac{E_i}{E^\dagger _i}\frac{I^\dagger _i}{I_i},~1\right)$, $\left(\frac{I_i}{I^\dagger _i}\frac{S^\dagger _i}{S_i},~1\right)$, $\left(\frac{E_i}{E^\dagger _i}\frac{S^\dagger _i}{S_i},~1\right)$  for $p_i$. 

The arguments in the previous paragraph show that  $\frac{d\,V_{ee}}{dt}(\mathbf x(t))$ may be expressed as the sum of non positive terms for any $\mathbf x\in \Omega_1\cup\Omega_2\equiv(\mathbb R_{>0})^u$. We may conclude that ${\frac{dV_{ee}}{d\,t}}(\mathbf x(t)) \leqslant 0$ for all $\mathbf x(t) \in  (\mathbb R_{>0})^u$. 
 
%
%

\noindent In order to determine the subset of $(\mathbb R_{>0})^u$ where $\frac{dV_{ee}}{d\,t}(\mathbf x(t))= 0$, we make use of ~\eqref{eq:eqliapeeder2} to conclude that
$$\frac{dV_{ee}}{d\,t}(\mathbf x(t))= 0\;\Leftrightarrow(f_1(\mathbf x)=0)\wedge ({g_1}_i(\mathbf x),\;i=0,\;\cdots,\;n)\wedge ({h_1}_i(\mathbf x),\;i=0,\;\cdots,\;n)\wedge ({p_1}_i(\mathbf x),\;i=0,\;\cdots,\;n).$$ 
By Lemma~\ref{lem.agmi},  $f_1(\mathbf x)=0$ if and only if $S_q=S^\dagger _q\wedge  I_qI^\dagger _r=I^\dagger _qI_r$. 
Given that $f_1(\mathbf x)=0$, Lemma~\ref{lem.agmi} also implies that ${h_1}_i(\mathbf x)=0$ if and only if $(S_r=S^\dagger _r)\wedge  (S_i=S^\dagger _i) \wedge (I_i=I^\dagger _i) \wedge  (I_q=I^\dagger _q)$, and thus $I_r=I^\dagger _r$.
Finally, assuming $f_1(\mathbf x)=0$ and ${h_1}_i(\mathbf x)=0$, Lemma~\ref{lem.agmi} also gives  
$${g_1}_i(\mathbf x)=0\;\;\hbox{ if and only if }\;\;1=\frac{E^{(1)\dagger }_r}{E^{(1)}_r} = \frac{ E^{(1)}_r}{ E^{(1)\dagger }_r}\frac{E^{(1)\dagger }_q}{E^{(1)}_q}= \frac{E^{(1)}_q}{E^{(1)\dagger }_q}\frac{E^{(2)\dagger }_r}{E^{(2)}_r}=\cdots =  \frac{ E^{(l)}_r}{ E^{(l)\dagger }_r}\frac{E^{(l)\dagger }_q}{E^{(l)}_q}=  \frac{ E^{(l+1)\dagger }_r}{ E^{(l+1)}_r}\frac{E^{(l)}_q}{E^{(l)\dagger }_q} =\frac{E^{(l+1)}_r}{E^{(l+1)\dagger }_r},$$ 
which implies $E^{(j)}_r=E^{(j)\dagger }_r, \;j=1,\;\cdots,\;l+1$ and $E^{(j)}_q=E^{(j)\dagger }_q, \;j=1,\;\cdots,\;l$. Thus, $\frac{dV_{ee}}{d\,t}(\mathbf x(t))= 0$ if and only if $\mathbf x = \mathbf x^\dagger $. 

From the above discussion we may conclude that $V_{ee}$ is a strict Lyapunov function for the system~\eqref{eq:eqbednet_} on $(\mathbb R_{>0})^u$. By the LaSalle invariance principle, which implies the global asymptotic stability of the endemic equilibrium $\mathbf x^\dagger $ of the system~\eqref{eq:eqbednet_} on the set $(\mathbb R_{>0})^u$~\cite{Bhatia70, Las68, MR0481301, MR0594977}.\edem

\section{Sensitivity analysis of the model}\label{sec:secsam}

The analysis in the two previous sections provides two key epidemiological factors which characterize the behavior of the system, namely the reproduction number $\mathcal R_0$ and the EE  $\mathbf x^\dagger  = (\mathbf x_S^\dagger ;\;\mathbf x_I^\dagger )$. In systems where $\mathcal R_0<1$, the disease eventually dies out; otherwise, the component s of the EE  determine the latent prevalence of the disease. 
It follows that the ``holy grail'' in vector-borne disease control is to reduce $\mathcal R_0$ below 1: however if this is unfeasible, an alternative goal is to reduce EE levels as much as possible. A major purpose for this model is to evaluate the effectiveness of various  prophylactic measures in acheiving either of these two goals.

Various parameters in our model reflect the effects of possible control measures: examples include the killing and repelling capabilities $k_i$ and $r_i = 1-f_i~(i=0,1, \cdots, n)$.
The effectiveness of these measures  in eliminating and/or reducing the disease depends on  the sensitivities of $\mathcal R_0$ and the EE components relative to these input parameters.  This is the motivation for this section, which is devoted to the calculation of  sensitivities of $\mathcal R_0$ and the EE to various potential control parameters.  Since  all Infected components of the host state $\mathbf x^\dagger $  are increasing functions of the infected vector questing component $I_q^\dagger $, 
it follows that an effective strategy for reducing the EE is to reduce  $I^\dagger _q$.  Hence rather than computing the sensitivities of the separate EE components, instead we focus on the sensitivities of $I^\dagger _q$ with respect to various potential control parameters.

The sensitivities of interest  can be quantified using sensitivity indices. The \textit{normalized forward sensitivity index}  of a variable to a parameter is defined as the ratio of the relative change in the variable to the relative change in the parameter. When the variable is a differentiable function of the parameter, the sensitivity index may alternatively be defined using partial derivatives:
\begin{dfn} The normalized forward sensitivity index of a variable $u$ that depends differentiably on a parameter $p$ is defined as:
	\begin{equation}
	\label{eq:seind}
	\varUpsilon_p^u := \frac{\partial u}{\partial p}\times\frac{p}{u}.\end{equation}
\end{dfn}

\subsection{Calculation of sensitivities of the reproduction number }

The expression for $\mathcal{R}_0$ from \eqref{eq:thrshold} is repeated here for convenience:

\begin{equation}\label{eq:thrshold1}
\mathcal{R}_0 =
\frac{({f_q^*}f_r)^{\ell+1}}{(1-f_q^*f_r)^2}\dfrac{f_q^*}{{\varpiVar ^*}^2}\dfrac{\Gamma}{H^*}a^2\sum_{i=0}^nh_i^*\frac{\varphi_i\phi_i}{\gamma_i+\nu_i}\frac{\varepsilon_i+\bar u_i\nu_i }{\varepsilon_i+\nu_i}\left(\xi_{i}+\frac{v_i\widetilde \xi_{i}\gamma_i}{\zeta_i+\nu_i}\right).
\end{equation}

This result may be compared to the reproduction number given in \cite{jck-19} for a system without E and R compartments, which we denote as $\widetilde{\mathcal{R}}_0$:
\begin{equation}\label{eq:thrshold_old}
\widetilde{\mathcal{R}}_0 =
\frac{({f_q^*}f_r)^{\ell+1}}{(1-f_q^*f_r)^2}\dfrac{f_q^*}{{\varpiVar ^*}^2}\dfrac{\Gamma}{H^*}a^2\sum_{i=0}^nh_i^*\frac{\varphi_i\phi_i}{\gamma_i+\nu_i}\xi_{i}.
\end{equation}

We may define:
\begin{equation}
\mathcal{Q}_0 \equiv  \frac{({f_q^*}f_r)^{\ell+1}}{(1-f_q^*f_r)^2}\dfrac{f_q^*}{{\varpiVar ^*}^2}\dfrac{\Gamma}{H^*}a^2; ~~~
\mathcal{Q}_{1,i} \equiv h_i^*\frac{\varphi_i\phi_i}{\gamma_i+\nu_i};~~~
\mathcal{Q}_{2,i} \equiv \frac{\varepsilon_i+\bar u_i\nu_i }{\varepsilon_i+\nu_i};~~~
\mathcal{Q}_{3,i} \equiv \frac{v_i\widetilde \xi_{i}\gamma_i}{\zeta_i+\nu_i}.
\end{equation}

Using these definitions the two reproduction numbers $\widetilde{\mathcal{R}}_0$ and $\mathcal{R}_0$ may be re-expressed as:
\begin{equation}
\widetilde{\mathcal{R}}_0 = \mathcal{Q}_0 \sum_{i=0}^n \mathcal{Q}_{1,i} \xi_i ; ~~~
\mathcal{R}_0 = \mathcal{Q}_0 \sum_{i=0}^n \mathcal{Q}_{1,i} \mathcal{Q}_{2,i} \left(\xi_i + \mathcal{Q}_{3,i} \right)
\end{equation}

It follows that $\mathcal{R}_0$ reduces to $\widetilde{\mathcal{R}}_0$ when $\mathcal{Q}_{2,i}=1$ and $\mathcal{Q}_{3,i}=0$ for all $i$. 
Using the sensitivity calculation rules described in \cite{jck-19}, we may obtain the following expressions for sensitivities:
\begin{equation}
 \begin{aligned}
\Upsilon_p^{\widetilde{\mathcal{R}}_0} &=  \Upsilon_p^{\mathcal{Q}_0} + \frac{\sum_{i=0}^n \mathcal{Q}_{1,i} \xi_i \left(\Upsilon_p^{\mathcal{Q}_{1,i}} + \Upsilon_p^{\xi} \right) }{\sum_{i=0}^n \mathcal{Q}_{1,i} \xi_i} \\
\Upsilon_p^{\mathcal{R}_0} &=  \Upsilon_p^{\mathcal{Q}_0} + \frac{\sum_{i=0}^n \mathcal{Q}_{1,i}\mathcal{Q}_{2,i} \left(\xi_i + \mathcal{Q}_{3,i} \right)  \left(\Upsilon_p^{\mathcal{Q}_{1,i}} + \Upsilon_p^{\mathcal{Q}_{2,i}} +  \left(\xi_i \Upsilon_p^{\xi} + \mathcal{Q}_{3,i}\Upsilon_p^{\mathcal{Q}_{3,i}}\right)\left(\xi_i + \mathcal{Q}_{3,i}\right)^{-1} \right) }{\sum_{i=0}^n \mathcal{Q}_{1,i} \mathcal{Q}_{2,i} \left(\xi_i + \mathcal{Q}_{3,i} \right) } 
\end{aligned}
\end{equation} 
We may thus make use of the sensitivities for $\widetilde{\mathcal{R}}_0$ calculated in \cite{jck-19} to compute sensitivities of $\mathcal{R}_0$. Table~\ref{tab.tabvd3} gives a list of $\mathcal{R}_0$ sensitivities to important system parameters, including  $ \varepsilon_i $ and $\zeta_i $ which are transition rates from Exposed and Resistant classes respectively.
 The parameter $t$ which appears in the expression for is given by
\begin{equation}
t \equiv \ \frac{1-f_r}{1-f_q^*f_r} \quad (\text{Note  } 0 \leqslant t \leqslant 1).
\end{equation}
For the first four parameters listed  ($\Gamma, \varpiVar^*,\hatmuVar^*, $ and $\kVar_j$), the sensitivities for $\mathcal{R}_0$  are the same as for $\widetilde{\mathcal{R}}_0$. 

\begin{table}[htbp]
\caption{Sensitivities of $\mathcal{R}_0$ with respect to various parameters}\label{tab.tabvd3}
\begin{tabular}{p{8.5cm}p{1.3cm}p{5.8cm}}
\hline
Parameter & Symbol &  Formula   \\
\hline
Recruitment rate of vectors&$\varUpsilon_{\Gamma}^{\mathcal R_0}$ & $1$\\
Incidence rate of successful vector-host interactions& $\varUpsilon_{\varpiVar^*}^{\mathcal R_0}$ &  $\ell (1-f_q^*) - 2tf_q^*$\\
Death rate of questing vectors&
$\varUpsilon_{\hatmuVar^*}^{\mathcal R_0}$  &  $-2 - \varUpsilon_{\varpiVar^*}^{{\mathcal R}_0}$\\
 Kill rate of vectors per interaction&  $\varUpsilon_{\kVar_j }^{{\mathcal R}_0}$ &
$\biVar ^*\left(\frac{ a \kVar_j  }{\hatmuVar^*}\right)\varUpsilon_{\hatmuVar^*}^{{\mathcal R}_0}$\\
Host infection rate per interaction & $\varUpsilon_{\mVar_j}^{\mathcal R_0}$ &  $\frac{\mathcal{Q}_{1,j}\mathcal{Q}_{2,j} \left(\xi_j + \mathcal{Q}_{3,j} \right) }{\sum_{i=0}^n \mathcal{Q}_{1,i} \mathcal{Q}_{2,i} \left(\xi_i + \mathcal{Q}_{3,i} \right) }$ \\
Host cure rate& $\varUpsilon_{\gamma_i}^{\mathcal R_0}$ &  $\varUpsilon_{\mVar_j }^{{\mathcal R}_0}\left(\frac{\mathcal{Q}_{3,j} }{\xi_i + \mathcal{Q}_{3,i} } -\frac{\gamma_j}{\gamma_j+\nu_j }\right)$\\
Vector success rate per interaction& $\varUpsilon_{\phi_{j} }^{{\mathcal R}_0} $ &   $\left(\frac{ h_{j} ^*\phi_{j} }{\sum_i  \biVar ^* \phi_{i} }\varUpsilon_{\varpiVar^*}^{{\mathcal R}_0} + \varUpsilon_{\mVar_j }^{{\mathcal R}_0} \right)$ \\
Host proportional incoming rate& $\varUpsilon_{\nu_j}^{\mathcal R_0}$ & $\varUpsilon_{\mVar_j }^{{\mathcal R}_0}\left(  -\frac{u_j\nu_j}{\varepsilon_j + \nu_j} -\frac{\nu_j}{\gamma_j + \nu_j} -\frac{\nu_j\mathcal{Q}_{3,j}}{(\zeta_j + \nu_j)(\xi_j + \mathcal{Q}_{3,j})} \right)$\\
Transition rate Exposed $\rightarrow$ Infectious &$\varUpsilon_{\varepsilon_j}^{\mathcal R_0}$&   $\varUpsilon_{\mVar_j }^{{\mathcal R}_0}\left( \frac{u_j\nu_j}{\varepsilon_j + \nu_j}\right)$\\
Transition rate Resistant $\rightarrow$ Susceptible &$\varUpsilon_{\zeta_j}^{\mathcal R_0}$ & $- \varUpsilon_{\mVar_j }^{{\mathcal R}_0}\left(\frac{\mathcal{Q}_{3,j} }{\xi_i + \mathcal{Q}_{3,i} }\right)\left(\frac{\zeta_j}{\zeta_j + \nu_j}\right)$\\
\hline
\end{tabular}
\end{table}


\subsection{Sensitivities of endemic equilibrium questing vector component} 

Based on the results in \cite{jck-19}, we have the following expression for the sensitivity of $I_q^\starVar$ with respect to the vector recruitment rate $\Gamma$:
\begin{equation}\label{eq:sensitivityGamma}
 \varUpsilon_\Gamma^{I_q^\starVar} = \left( \frac{a I_q^\starVar}{Z} \sum_{i=0}^n \frac{\biVar ^* z_i^2}{\ciVar \fiVar  } + 
(1-t) \tilde{t}  \left(1 - \frac{a I_q^\starVar}{Z} \sum_{i=0}^n \frac{\biVar ^* z_i^2}{\ciVar \fiVar  } \right) 
\right)^{-1}, 
\end{equation}
where

\begin{equation}
z_i \equiv  \frac{A_i}{B_i + C_i};\label{eq:def_zi}\qquad 
 Z \equiv \sum_{i=1}^n \biVar ^*z_i ; \qquad 
\tilde{t} \equiv \frac{I_q^\starVar}{ \Gamma \widehat{\mathcal R}_0 \hatmuVar^*}=\frac{I_q^\starVar}{\IqstarmaxVar}\frac{f_q^*(1-f_q^* f_r)}{1-f_q^*},
\end{equation}
where $0 \leqslant \tilde{t} \leqslant 2$ and  
\begin{equation}
\begin{aligned}
A_i  &\equiv \phi_i\varphi_i (\varepsilon_i+\bar u_i\nu_i)\left(\widetilde\xi_iv_i\gamma_i+\xi_i(\zeta_i+\nu_i)\right);\\
B_i &\equiv   (\varepsilon_i+\nu_i)(\gamma_i+\nu_i)(\zeta_i+\nu_i)H^*;\\
C_i &\equiv  \Big(u_i(\gamma_i+\nu_i)(\zeta_i+\nu_i) + (\bar u_i\nu_i+\varepsilon_i)(v_i\gamma_i + \zeta_i+\nu_i) \Big)a\varphi_i I^\dagger_q.\label{eq:def_ABCi}\\ 
\end{aligned}
\end{equation}
(Note that \eqref{eq:sensitivityGamma} is identical to the corresponding expression in \cite{jck-19}, except  that the definition of $z_i$ in \eqref{eq:def_zi}  is different.)

In terms of  $ \varUpsilon_\Gamma^{I_q^\starVar}$, a general formula for the sensitivities with respect to $ p_i \in \{\varepsilon_i, \fiVar ,\gamma_i, \kiVar ,\miVar ,\nu_i,\xi_i,\widetilde\xi, \zeta_i\}$ may be found from \eqref{eq:eqeqIQ} (after extended calculations)  as:
\begin{equation}\label{eq:sensIq_wrt_pj}
\varUpsilon_{p_i}^{I_q^\starVar} =
\varUpsilon_{\Gamma}^{I_q^\starVar} \left(\varUpsilon_{p_i}^{\widehat{\mathcal R}_0}
+ t\varUpsilon_{p_i}^{t}\tilde{t}  + \varUpsilon_{p_i}^{\hatmuVar^*}(1-t)\tilde{t}   
+  \frac{\biVar ^*z_i}{Z} \varUpsilon_{p_i}^{z_i} 
\left(1 - (1-t)\tilde{t} \right)
\right),
\end{equation}
Other sensitivities may be computed by making use of the equation:
\begin{equation}
\varUpsilon_{p_i}^{z_i} = \varUpsilon_{p_i}^{A_i} - \frac{1}{B_i+C_i}\left(B_i\varUpsilon_{p_i}^{B_i} + C_i\varUpsilon_{p_i}^{C_i} \right).
\end{equation}
Table~\ref{tab.tabvd4} gives the complete list of sensitivities. The parameters $D$ and $E_i$ in the table are defined as:
\begin{equation}
\begin{aligned}
D &\equiv \ell (1-f_q^*) - 2tf_q^* + t\tilde{t}\left(\frac{f_rf_q^*(1-f_q^*)}{1-f_q^*f_r}\right)\\   
E_i &\equiv \biVar ^*\varUpsilon_{\Gamma}^{I_q^\starVar} \left(\frac{z_i}{Z}\right)(1 - (1-t)\tilde{t} ).
\end{aligned}
\end{equation}
The expressions for $\varUpsilon_{\kiVar }^{I_q^\starVar}$, $\varUpsilon_{\fiVar }^{I_q^\starVar}$ and $\varUpsilon_{\miVar }^{I_q^\starVar}$ in Table~\ref{tab.tabvd4} are identical to the corresponding expressions in \cite{jck-19} (see Eq.  (76-78)), except that $z_i$ and $Z$ have been redefined. Equations (77) and (78) in \cite{jck-19} have 
slight errors that have been corrected here.


\begin{table}[htbp]
\caption{Sensitivities of $I_q^\starVar$ with respect to various parameters}\label{tab.tabvd4}
\begin{tabular}{p{6.cm}p{0.8cm}p{9.5cm}}
\hline
Parameter & Sym. &  Formula   \\
\hline
Recruitment rate of vectors &
 $\varUpsilon_\Gamma^{I_q^\starVar}$ & $\left( \frac{a I_q^\starVar}{Z} \sum_{i=0}^n \frac{\biVar ^* z_i^2}{\ciVar \fiVar  } + 
(1-t) \tilde{t}  \left(1 - \frac{a I_q^\starVar}{Z} \sum_{i=0}^n \frac{\biVar ^* z_i^2}{\ciVar \fiVar  } \right) 
\right)^{-1}$\\ 
Kill rate of vectors per interaction & $\varUpsilon_{\kiVar }^{I_q^\starVar}$ 
&$\varUpsilon_{\Gamma}^{I_q^\starVar}\left( \frac{a \biVar ^* \kiVar  }{\hatmuVar^*} \right) \left( -D - 2  +  (1-t)\tilde{t}   \right)$\\

Host infection rate per interaction & $\varUpsilon_{\miVar }^{I_q^\starVar}$  & $E_i   \left(  1 - \frac{C_iz_i}{A_i} \right),$ \\

Host cure rate& 
$\varUpsilon_{\gamma_i}^{I_q^\starVar}$ & 
$\left(\frac{E_i \xi_iv_i \gamma_i}{v_i\gamma_i\widetilde\xi_i+\xi_i(\nu_i+\zeta_i)} - \frac{E_i}{B_i+C_i}\left(\frac{B_i\gamma_i}{\gamma_i+\nu_i} \right. \right.$\\
&~&\qquad $\left. \left. + a\varphi_i I^\dagger_q \gamma_i\left(v_i \varepsilon_i +  (u_i + \bar{u}_i v_i)\nu_i+ u_i\zeta_i \right) \right)\right)$\\

Vector success rate per interaction& $\varUpsilon_{\fiVar }^{I_q^\starVar}$ &
 $\varUpsilon_{\Gamma}^{I_q^\starVar} \left( \frac{a\biVar ^*\fiVar }{\varpiVar^*}\right)D + E_i$\\

Host proportional incoming rate& 
$\varUpsilon_{\nu_i}^{I_q^\starVar}$ & 
$\frac{E_i\bar u_i\nu_i}{\varepsilon_i+\bar u_i\nu_i} 
+ \frac{E_i\nu_i\xi_i}{v_i\gamma_i\widetilde\xi_i+\xi_i(\nu_i+\zeta_i)}$\\ & & ~~ $-\frac{E_i}{B_i+C_i} \left( \frac{B_i \nu_i}{\varepsilon_i+\nu_i} + \frac{B_i \nu_i}{\gamma_i+\nu_i} + \frac{B_i \nu_i}{\zeta_i+\nu_i}\right.$\\
&~&\qquad  
$\left.+ a\varphi_i I^\dagger_q\nu_i\left( (u_i+v_i)\gamma_i + (u_i+1)\varepsilon_i + (u_i+2)\nu_i + \zeta_i \right) \right)$\\

Transition rate from Exposed class&$\varUpsilon_{\varepsilon_i}^{I_q^\starVar}$ &  
$\frac{E_i \varepsilon_i}{\varepsilon_i+\bar u_i\nu_i} - \frac{E_i}{B_i+C_i}\left(\frac{B_i\varepsilon_i}{\varepsilon_i+\nu_i}
+  a\varphi_i I^\dagger_q\varepsilon_i(v_i\gamma_i +\nu_i+ \zeta_i)\right)$\\

Transition rate from Resistant class&
$\varUpsilon_{\zeta_i}^{I_q^\starVar}$ &$\varUpsilon_{\xi_i}^{I_q^\starVar} \frac{\zeta_i}{\nu_i+\zeta_i}
- \frac{ E_i}{B_i+C_i}\left( \frac{B_i  \zeta_i}{\nu_i+\zeta_i} + a \varphi_i    I^\dagger_q \zeta_i
\left( u_i\gamma_i + \nu_i   + \varepsilon_i \right)\right)$\\

Prob. of successful vector infection from interaction with Infectious host & 
$\varUpsilon_{\xi_i}^{I_q^\starVar}$ & 
$\frac{E_i\xi_i(\nu_i+\zeta_i)}{v_i\gamma_i\widetilde\xi_i+\xi_i(\nu_i+\zeta_i)}$\\

Prob. of successful vector infection from interaction with Resistant host & 
$\varUpsilon_{\widetilde\xi_i}^{I_q^\starVar}$ & $\varUpsilon_{\xi_i}^{I_q^\starVar}
\left(\frac{v_i\gamma_i\widetilde\xi_i}{\xi_i(\nu_i+\zeta_i)}\right)$\\

\hline

\end{tabular}
\end{table}

From these expressions for $\varUpsilon_{r_i}^{\bar I_q^\dagger }$ and $\varUpsilon_{k_i}^{\bar I_q^\dagger }$ we may make several significant observations:
\begin{itemize}
\item
Note that $\varUpsilon_{r_i}^{\bar I_q^\dagger }\leqslant0$ iff $\frac{f^*_q}{1-f^*_q}\frac{1-f_r}{1-f^*_qf_r}\leqslant l+1$, and $\varUpsilon_{k_i}^{\bar I_q^\dagger }\leqslant0$  for all $l\geqslant0$. 
In this case, it follows  that $\bar I^\dagger _q$ is a decreasing function of both $r_i$ and $k_i$, which is certainly to be expected.
\item
 For fixed $\hat \mu^\dagger $ and $\varpiVar ^\dagger $ both sensitivity indexes are proportional to  $b_i^*$, so that the sensitivity of $\bar I_q^\dagger $ with respect to $k_i$ and $r_i$  increases linearly as the equilibrium proportion of bed net users in group $i$ increases. This also agrees with our intuitive expectation that improving bed nets' prophylactic capabilities for a larger proportion of the population should have a greater impact in reducing the number of infected questing vectors.
\item
For fixed $\hat \mu^\dagger $ and $\varpiVar ^\dagger $, $\varUpsilon_{r_i}^{\bar I_q^\dagger }$  is proportional to  $\bar k_i$. When  $k_i=1$ we have $\varUpsilon_{r_i}^{\bar I_q^\dagger } = 0$, so that varying the repelling effect has no impact on $\bar I^\dagger _q$. On the other hand,  the magnitude of $\varUpsilon_{r_i}^{\bar I_q^\dagger }$ is maximized when $k_i=0$, so that the repelling effect has the greatest impact when the bed net has no killing effect. This is reasonable since the repelling effect only affects vectors that survive their bite attempts.
\item
For fixed $\hat \mu^\dagger $ and $\varpiVar ^\dagger $, $\varUpsilon_{k_i}^{\bar I_q^\dagger }$  depends affinely on $r_i$. The sensitivity is greatest when 
$r_i=0$ if and only if  $l+1  +  \frac{f_q^*f_r}{1-f_q^*f_r} \geqslant \frac{f_q^*}{1-f_q^*}$.
\item
The number of questing/resting cycles in the extrinsic incubation period of vectors has a large impact on the rate of decrease of $\bar I^*_q$  with increasing $r_i$ and $k_i$. For fixed $\hat \mu^\dagger $ and $\varpiVar ^\dagger $, $\varUpsilon_{k_i}^{\bar I_q^\dagger }$ and $\varUpsilon_{r_i}^{\bar I_q^\dagger }$   both are affine functions of $l$, with negative slope. It follows that increasing the extrinsic incubation period can enhance the effectiveness of increasing bed nets' killing and repelling capability. 
\end{itemize}
These results confirm our intuitive expectations for the  sensitivity of $I_q^*$ in relation to the parameters that characterize bed net effectiveness.
In summary, we have shown that for  host group $i$ of bed net users ($i=1,\cdots ,n$), the group's positive impact  in  weakening  transmission and prevalence of the disease  is enhanced by increasing  the equilibrium proportion $b^*_i$, the  killing capability $k_i$, and repelling capability $r_i$. All host groups that use bed nets have some influence on reducing the transmission and prevalence, even if the proportion is small and the killing capacity is zero. We thus conclude that widespread bed net usage should be encouraged,  regardless of  killing capability. In Section~\ref{sec:secsim}, we provide maps of $\mathcal R_0$ as function of inputs $f_i$ and $k_i$ for groups $i$ with high impact in various scenarios.

\section{Numerical simulations to be performed}\label{sec:secsim}

To illustrate  the behavior of the model, the system~\eqref{eq:eqbednet_}  will be simulated using  the parameter values and ranges specified in  Tables~\ref{tab.tab2} and \ref{tab.tab1} below. 
In all simulations, the hatching speed of vectors is sixty times the speed of human migration  (i.e.  $\Gamma=60\times \Lambda$), which reflects an episode of high endemicity of the disease. . The value of the migration speed of human in all group of hosts is the same in all simulations.

\noindent As explained in the preceding section, the $\phi_i$ ans $\kappa_i$ are input parameters of the model, and a main purpose of this paper is to analyze the impact of these parameters on the endemicity in the region of nterest. We thus write the parameters progression host rates, $\varepsilon_i$, $\zeta_i$, $\gamma_i$, probabilities $\xi_i$, $\widetilde\xi_i$ and $\miVar$ of the $i^{th}$ group of hosts as functions of the repelling parameter corresponding to protective properties used in the group $i$, namely $\rho_i\equiv1-\phi_i$. We assume that parameters  $\xi_i$, $\widetilde\xi_i$ and $\miVar$ are increasing functions of $\rho_i$, and $\gamma_i$, $\zeta_i$ decreasing function of $\varrho_i$: these assumptions are based on the principle that the greater the protective  property, the longer the time group members take to get the critical charge that can lead to sickness, and the shorter the duration of their infectiousness on the human side; on the vector side, the greater a group's protective property, the more is the contribution in keeping vectors in the naive state. Specifically, we assume 
$$\xi = \xi_{max}\mathrm e^{\phi+\alpha}+\beta~~\hbox{ with }~~\alpha=\ln\left(\frac{\xi_{max}-\xi_{min}}{\xi_{max}(\mathrm e-1)}\right)~~\hbox{ and }~~ \beta = \xi_{min}-\xi_{max}\mathrm e^\alpha$$ 
$$\widetilde\xi = \widetilde\xi_{max}\mathrm e^{\phi+\alpha}+\beta~~\hbox{ with }~~\alpha=\ln\left(\frac{\widetilde\xi_{max}-\widetilde\xi_{min}}{\widetilde\xi_{max}(\mathrm e-1)}\right)~~\hbox{ and }~~ \beta = \widetilde\xi_{min}-\widetilde\xi_{max}\mathrm e^\alpha$$
$$\varphi = \varphi_{max}\mathrm e^{\phi+\alpha}+\beta~~\hbox{ with }~~\alpha=\ln\left(\frac{\varphi_{max}-\varphi_{min}}{\varphi_{max}(\mathrm e-1)}\right)~~\hbox{ and }~~ \beta = \varphi_{min}-\varphi_{max}\mathrm e^\alpha$$ $$\gamma = \gamma_{min}\mathrm e^{\alpha-\phi}+\beta~~\hbox{ with }~~\alpha=\ln\left(\frac{\gamma_{min}-\gamma_{max}}{\gamma_{min}(1-\mathrm e^{-1})}\right)~~\hbox{ and }~~ \beta = \gamma_{min}(1-\mathrm e^\alpha)$$ 
$$\zeta = \zeta_{min}\mathrm e^{\alpha-\phi}+\beta~~\hbox{ with }~~\alpha=\ln\left(\frac{\zeta_{min}-\zeta_{max}}{\zeta_{min}(1-\mathrm e^{-1})}\right)~~\hbox{ and }~~ \beta = \zeta_{min}(1-\mathrm e^\alpha)$$ 
$$\varepsilon = \varepsilon_{min}\mathrm e^{\alpha-\phi}+\beta~~\hbox{ with }~~\alpha=\ln\left(\frac{\varepsilon_{min}-\varepsilon_{max}}{\varepsilon_{min}(1-\mathrm e^{-1})}\right)~~\hbox{ and }~~ \beta = \varepsilon_{min}(1-\mathrm e^\alpha)$$
$(\xi_{min},\;\xi_{max})$, $(\widetilde\xi_{min},\;\widetilde\xi_{max})$, $(\varphi_{min},\;\varphi_{max})$, $(\gamma_{min},\;\gamma_{max})$, $(\zeta_{min},\;\zeta_{max})$, $(\varepsilon_{min},\;\varepsilon_{max})$ are respectively upper and lower bounds for the values of $\xi$, $\widetilde\xi$, $\varphi$,  $\gamma$, $\zeta$  and $\varepsilon$ in Tables~\ref{tab.tab2} and \ref{tab.tab1}. The only constraint respected by these functions is that for each value of $f\in[0,\;\;1]$,    values of these functions are in between the bounds in the set of data in Tables~\ref{tab.tab2} and \ref{tab.tab1}. The parameter $\nu_i$ is the same for all groups, with $\nu_i=\frac{1}{49} year^{-1}$ and it is related to the current life expectancy of human in Cameroon. The $\widetilde{\nu}_i$ is varying relative to what proportion we want $b^*_i:=\frac{\Lambda}{\nu-\widetilde{\nu}_i}$ to be.

\noindent This paper is a step forward of the work presented in~\cite{jckam201411}. Here, we have enforced assumptions made there and introduced new parameters to better reflect the impact of bed net utilization on the dynamics of malaria transmission. Results of mathematical analysis in section~\ref{sec.analysis} are based on those carried in~\cite{jckam201411}, and thus trajectories of the model in this paper are nearly the same as those presented in the simulation section in~\cite{jckam201411}. In the current paper we focus our attention on how input parameters $\phi_i$ and $\kappa_i$ affect the weakening the endemicity of the malaria in the region of interest.  Previously we used simulated trajectories to show the global asymptotic stability of equilibriums. However,  in some situations it takes a very long time to reach equilibrium, so these studies do not show how input parameters impact on the level of the endemicity. Our goal in this paper is to generate maps of $\mathcal R_0$ as function of given pairs of input $(\phi_i,\;\kappa_i)$ and trajectories for significant pairs of $(\phi_i,\;\kappa_i)$ relative to the level of the endemicity of components  that express the infectiousness, i.e the infectious hosts components and infectious questing vectors component of the states of the model. 

\begin{table}[htbp]
	\captionsetup{skip=0pt}\caption{Parameter values for host population dynamics}\label{tab.tab2}
	\begin{tabular}{p{1.38cm}p{12cm}p{2.5cm}}
		\hline
		Param. & Description & Estimated \\
		&& Value/Ranges \\
		\hline
		$\Lambda$ & Force of migration in host population (humans/day) & $2.5$ \\
		$\miVar$ & infectivity coefficient of vector interactions with $i^{th}$ host group  
		& $0.072$ -- $0.64$ \\
		$\nu_i$ & Birth rate of hosts (year$^{-1}$) & $\frac{1}{49}$ \\
		$\widetilde \nu_i$ & Group $i$ host outgoing rate (day$^{-1}$) &  Variable \\
		
		$\gamma_i$ & Transition rate from $I \rightarrow S$ for  
		the $i^{th}$--host group (transitions/human/day) &$0.0014$ -- $0.017$\\
		
		\hline
	\end{tabular}
\end{table}

\begin{table}[phtb]
	\captionsetup{skip=0pt}\caption{Parameter values for vector population dynamics}\label{tab.tab1}
	\begin{tabular}{p{1.38cm}p{12cm}p{2.5cm}}
		\hline
		Param. & Description & Estimated \\ && Value/Ranges \\
		\hline
		$\Gamma$ & Recruitment force in the vector population~~(vectors/day) & $150$ \\
		$a$ & Host-vector interaction rate ~~(interactions/year/vector) & 150--200  \\
		$\mu$ & Natural death rate of vectors ~~(deaths/vector/day) & ${\frac{1}{30}}$ \\
		$\delta$ & Transition rate from any resting state to a questing state & Computed \\
		$\ell$ & Number of questing/resting cycles before infectivity & $6^{(\dag)},\;8,\;9$ \\
		$\xi_i$ & Probability that an interaction with the $i^{th}$ host group results in vector infection & $0.010$ -- $0.27$\\ 
		
		$\kappa_i$ & Probability of being killed during an interaction with the $i^{th}$ host group &Variable\\
		$\phi_i$ & Probability of  successful interaction with the $i^{th}$ host group & Variable\\
		\hline
	\end{tabular}
\vspace{-16pt}\noindent{\small Note. Source of the estimation:  $(\dag)$: \cite{010047862} 
}\end{table}

\noindent Here below, when we are in scenario with $n$ groups of bed net users, we assume inputs $\phi_i$ and $\kappa_i$ for the $n-1$ first groups constant; only inputs $\phi_n$ and $\kappa_n$ of the $n^{th}$ group are assume to vary. We present maps of values of $\mathcal R_0$ as function of  the two variables $(\phi_n,\;\kappa_n)$. Trajectories of components that express infectiousness in the state of the model are presented for three chosen initial states within a period of a hundred of days. Since graphical representation is possible only for graphs in $\mathbb R^k$ ($k=2,\,3$), as function of the two variables $(\phi_n,\;\kappa_n)$ trajectories are presented in animated maps. In visualization, only significant state components of the model that characterize the endemicity are considered; $I_q$ for the population of vectors, the $I_i$, $i = 1,2,\cdots,\,n$. We hope this makes more light on the impact of the utilization of the bed nets. The proportion of group users in the title of the sections refer to the proportion at the equilibrium, since the size of involved groups of bed net users are time dependent. 

\noindent  Scenarios are presented in a nested scheme depending at first level on $\ell$ (with the beginning to $\ell=6$ since it is the default value in the literature) the representation of the extrinsic incubation period of vectors; the second level is the proportion of users at the equilibria; the third level is values of the killing capability $\kappa_p$ of the protection that provides the utilization of the bed net in an increasing structure, the last level is the repelling capability $\varrho_p$ of protection provided by the bed nets (we recall that $\varrho_p = 1-\phi_p$, with $\phi_p$ the feed and survive probability of the vectors within the contact with protected host. In figures it is with the $\phi_p$ that the repelling capability is expressed since it is the factor used in the modeling). 


\section{Discussion}\label{sec.discuss}

This paper stands as a mathematical contribution in order to evaluate how effective the utilization of bed nets in the fight against malaria in endemic areas can be. We proposed a model of the dynamic of the transmission of the malaria involving a population of vectors and a human population  as hosts subdivided into several groups depending on the way they usually protect themselves against vector interactions. The model allows for an arbitrary number of different population groups (classified by protection level), but the model is sufficiently simple to capture essential system characteristics: namely how the protective factors ($f_i$ and $k_i$) and the transmission probabilities from vectors to hosts ($b_i$) and from host to vectors ($c_i$) influence the value of the basic reproduction number $\mathcal R_0$, as well as  the level of the endemicity in cases where $\mathcal R_0>1$. 

The level of the endemicity predicted by the model is given by  $I_q^\dagger $ in  \eqref{eq:eqexpMinfctstateb}. Although we do not have an explicit analytical  expression for $I_q^\dagger $, we do have an analytical expression for an upper bound $I_{q,max}^\dagger$ given by \eqref{eq:eqexpMinfctstateb}. This upper bound is  a decreasing function of the extrinsic incubation period  $l$, and depends also in a more delicate way on other parameters since they influence the values of the  frequencies $f_q$, $f_r$ and also $\varpiVar $. This dependence will be made explicit by the figures that will be generated by simulations. 
The value of $I_{q,max}^\dagger $ is smaller for smaller values of $\mathcal R_0$. Preliminary simulations show that how the reduction of endemicity in host sub-populations happen follows the reduction in endemicity in the vectors population,  and subsequently on the combination of parameters.  

If a large proportion of hosts uses insecticide treated bed nets with good protective capability (e.g. treated bed nets  with good repelling and killing properties), the level of the endemicity is markedly reduced,  even when some sub-populations use low level protection. In Cameroon, the media strongly encourage the use of long-lasting insecticide-treated bednets, known by the acronym  MILDA (i.e. Moustiquaire Impr\'egn\'ee \`a Longue Dur\'ee d'Action). Furthermore, MILDA are distributed freely and widely. The long-term killing effectiveness of MILDA has not been firmly established---however even if effectiveness is diminished, some reduction in enemicity is still realized by the protective capability.  The wide distribution of free MILDA means that it is feasible to encourage users to regularly replace them---more research is required to determine an optimal replacement schedule.


\section{Conclusion and perspective}
We have designed and analyzed the behavior of a detailed model of bed net usage in the fight against malaria. The proposed model takes into account multiple levels of protection with bed nets within human population; as well as multiple (questing, resting) steps between the first successful infected interaction and the infectious step of the mosquitoes.   We have obtained a general expression for the basic reproduction number, and we have  established that the DFE of the model is GAS providing that $\mathcal R_0\leqslant 1$. This is an improvement of the result  in \cite{MR2168743}, where the condition of the stability of the DFE was not based on the above inequality. We have also analyzed the behavior of the model when $\mathcal R_0>1$. In the latter case, we have established that there  is a unique endemic equilibrium, and we prove that this equilibrium is globally asymptotically stable for our system. 

These results are significant in light of the fact that malaria is one of the principal causes of death in African countries. We are hopeful that our results will prove useful and informative in the fight against malaria. Future work may include more accurate models of  death, birth and migration. 

\appendix{
\section{ Useful definitions and results}\label{appx:defs}
In order to make this paper self-contained, this appendix gives definitions and summarizes prior results from the literature that were used in the above discussion. Proofs of all results in this section may be found  in~\cite{McCluskey07} or \cite{KamSal07}, as indicated.

\subsection{Useful definitions}

\begin{dfn}[Metzler matrix, Metzler stable matrix~\cite{MR1298430, MR94c:34067, 0458.93001}] \label{dfn:dfnMzlr} A given $n\times n$ real matrix is said to be a Metzler matrix if all its off-diagonal terms are nonnegative.  The matrix is called Metzler stable if in addition all of the eigenvalues have negative real parts.   \end{dfn}

\noindent Note that a square matrix $\mathbf A$  is a Metzler matrix if and only if $-\mathbf A$ is a $Z$-matrix, and $\mathbf A$  is  Metzler stable  if and only if $-\mathbf A$ is a $M$-matrix. This paper makes use of the fact (which is not difficult to prove) that the positive cone is invariant for every dynamical system described by a system of ordinary differential equations whose Jacobian matrix is a Metzler matrix.

\begin{dfn}[Irreducible matrix] \label{dfn:dfnIrmtx} A given $n\times n$ matrix $\mathbf A$ is said to be a reducible matrix if there exists a  permutation matrix $\mathbf P$ such that $\mathbf P^{\mathbf T}\,\mathbf A\,\mathbf P$ has block matrix form:  $\mathbf P^{\mathbf T}\,\mathbf A\,\mathbf P=\left(\begin{array}{cc}\mathbf A_1 & \mathbf A_{1\,2}\\\mathbf 0 & \mathbf A_{2}
\end{array}\right)$, where $\mathbf A_1$ and $\mathbf A_2$ are square matrices. A matrix $\mathbf A$ that is not reducible is said to be irreducible.\end{dfn}

\noindent Irreducibility of $\mathbf A$ can be checked using the directed graph associated with $\mathbf A = (a_{k\,j} )$. This graph (denoted as $G(\mathbf A)$) has vertices ${1,\; 2,\; \cdots,\; n}$, and the directed arc $(k,\; j)$ from $k$ to $j$ is in $G(\mathbf A)$ iff $a_{k\,j}\neq 0$. $G(\mathbf A)$ is said to be \defn{strongly connected} if any two distinct vertices of $G(\mathbf A)$ are joined by an oriented path. The matrix $\mathbf A$ is irreducible if and only if $G(\mathbf A)$ is strongly connected~\cite{Guo_li_PAMS08}.

\subsection{Arithmetic-geometric means inequality, and consequences}In demonstrating that the Lyapunov derivative is nonpositive (see Section~\eqref{sec:eeqstana}), a key tool is the arithmetic-geometric means inequality, stated as follows: 

\begin{lem}[Weighted AM-GM]\label{lem.agmi}
	Let the positive numbers $z_1,\;\cdots,\;z_d$ and the positive weights $w_1,\;\cdots,\;w_d$ be given. Set $w = w_1+\;\cdots+\;w_d$. If $w>0$, then  
	$$\sqrt[w]{z_1^{w_1}\cdots z_d^{w_d}} \leqslant \frac{w_1z_1+\cdots+w_dz_d}{w}$$
	
	Furthermore, exact equality occurs iff $z_1=\cdots=z_d$
\end{lem}

\noindent The classical weighted AM--GM is more general than this lemma. 

\noindent An immediate consequence of weighted AM-GM is:

\begin{cor}[\cite{McCluskey07}]\label{cor.agmi} Let $z_1,\;\cdots,\;z_d$ be positive real numbers such that their product is 1.Then
	$$d - z_1 -\cdots - z_d\leqslant 0.$$  Furthermore, exact equality  occurs iff $z_1=\cdots=z_d$.
\end{cor}

\noindent This is obtained from Lemma~\ref{lem.agmi} by choosing $w_i=1$ for all  $i$.

\noindent Another useful consequence  is 

\begin{cor}\label{cor.agmi0} Let the positive numbers $z_1,\;\cdots,\;z_d$ and the positive weights $w_1,\;\cdots,\;w_d$ be given. Assume for a given $i$, $1\leq i \leq d$ there is a $v\in\mathbb R$ such that ${z_i}^v=z_1^{w_1}\cdots z_{i-1}^{w_{i-1}}z_{i+1}^{w_{i+1}}\cdots z_d^{w_d}$.  Set $w = w_1+\;\cdots+\;w_d$. If $w>0$ and $(z_i-1)(w+v)\leqslant 0$, then
	
	$$w - w_1z_1 -\cdots - w_dz_d\leqslant 0.$$  Furthermore, exact equality  occurs iff $z_1=\cdots=z_d$.
\end{cor}
\noindent {\em Proof }:~~ For numbers and associated weights as stated in Corollary, applying the Lemma~\ref{lem.agmi} gives

$$z_i^{\frac{w_i+v}{w}}\leqslant\frac{w_1z_1+\cdots+w_dz_d}{w};$$ since for given $x$ and $\beta$ positive real numbers and $\alpha$ a real number with $|\alpha|<\beta$ we have the relation $1\leqslant x^{\frac{\alpha}{\beta}}$ iff $x\leqslant 1\wedge\alpha\geqslant0$ or $x\geqslant 1\wedge\alpha\leqslant0$, which holds iff $(x-1)\alpha\leq0$. Thus assuming  $(z_i-1)(w_i+v)\leqslant 1$,  since this is equivalent to $1\leqslant z_i^{\frac{w_i+v}{w}}$, the  result follows straightforwardly. \edem

\subsection{Results on the global asymptotic stability (GAS) of the DFE of epidemiological model} Theorem~\ref{thm:kamsal} and Proposition~\ref{prop:blockdecomposition} given in this section  are key tools in demonstrating Theorem~\ref{thm.defstab} in Section
`\ref{subsec.dfestabanan} and Proposition~\ref{prop:basicrepn} in Section~\ref{sec.algo},respectively. 

Theorem~\ref{thm.defstab} is conventionally proven by constructing an adequate Lyapunov function for the model at the equilibrium concerned. Theorem~\ref{thm:kamsal} establishes the existence of such a Lyapunov function in cases similar to the situation in this paper.

\begin{thm}[\cite{KamSal07}]\label{thm:kamsal}Consider the system  
 \begin{equation} \label{eq:eqmodelc} \left\{\begin{array}{ccr}\dot{\mathbf x}_S & = & \mathbf
A_S(\mathbf x)\,.\, \left(\mathbf x_S - \mathbf x^*_S\right)  +  \mathbf A_{S,\,I}(\mathbf x)\,.\,\mathbf x_I \\
   \dot{\mathbf x}_I & = & \;\;\;\;\;\;\mathbf A_I(\mathbf x)\,.\,\mathbf x_I
\end{array}\right. \end{equation}
defined on a positively invariant set $\Omega\subset\mathbb R_+^{n_S\times n_I}$.  Given that:
\begin{enumerate}[{\bf h}1:]
\item The system is  dissipative on $\Omega$;
\item The equilibrium $\mathbf x_S^*$ of the subsystem $\dot{\mathbf x}_S = \mathbf A_S\left(\mathbf x_S,\;\mathbf 0\right)\,.\,(\mathbf x_S-\mathbf x^*_S)$ of system~\eqref{eq:eqmodelc} is globally asymptotically stable   on the canonical projection of $\Omega$ on $\mathbb R_+^{n_S}$;
\item The matrix $\mathbf A_I(\mathbf x)$ is a Metzler matrix  and irreducible for each $\mathbf x\in\Omega$;
\item There is an upper--bound matrix $\overline{\mathbf A}_I$ (in the sense of pointwise order) for the set of $n_I\times n_I$ square matrices  $\mathfrak{M} = \{\mathbf A_I(\mathbf x)\;/\;\mathbf x\in\Omega\}$ with the property that either $\overline{\mathbf A}_I\not \in\mathfrak{M}$ or if $\overline{\mathbf A}_I \in\mathfrak{M}$, then for any $\overline{\mathbf x}\in\Omega$ such that $\overline{\mathbf A}_I = \mathbf A_I(\overline{\mathbf x})$, we have $\overline{\mathbf x}\in\mathbb R_+^{n_S}\times\{\mathbf 0\}$;
\item $\alpha(\overline{\mathbf A}_I)\leq 0$.
\end{enumerate}

\noindent Then the DFE $\mathbf x^*$ is GAS for the system~\eqref{eq:eqmodelc} in $\overline{\Omega}$.
\end{thm}  

 Proposition~\ref{prop:blockdecomposition} shows that  the Metzler stability of matrix $\mathbf M$ is equivalent to the Metzler stability of two smaller matrices,which if properly chosen may be easier to compute with.

\begin{prop}[\cite{KamSal07}]
    Let  $\mathbf M$  be a Metzler matrix,  with block decomposition
    $\mathbf M=\left(\begin{array}{cc}
    \mathbf {A} & \mathbf {B}\\
    \mathbf {C} & \mathbf {D}
    \end{array}\right)$,
     where $ \mathbf {A}$ and $ \mathbf {D}$ are square matrices.
    \noindent Then      $\mathbf  M$ is Metzler stable if and only if 
    $\mathbf A$ and  $ \mathbf  D - \mathbf  C \mathbf A^{-1} \mathbf  B $ (or $\mathbf D$ and $\mathbf A-\mathbf B \mathbf D^{-1} \mathbf C$) are Metzler stable.
    \label{prop:blockdecomposition}
\end{prop}

 It comes out as it is usual while dealing with vector born diseases that $${\dsum_{i=0}^n}\mathcal R_0^{h_iv}\mathcal R_0^{vh_i}=\mathcal R_0={\dsum_{i=0}^n}\mathcal R_0^{vh_i}\mathcal R_0^{h_iv}$$ represents the average number of secondary cases of infectious vectors (respectively hosts) that are occasioned by one infectious vector (respectively host) introduced in a population of susceptible vectors (respectively hosts). i.e. $\mathcal R_0$ for the population of vectors and also for the population of hosts. When this number is computed with the technique of the next generation matrix of van den Driessche et {\it al.}~\cite{VddWat02}, it appears usually with a square root; it is so common to find, even if it is not computed with the next generation matrix technique a square root coming from nowhere appearing in the expression at the end on the number. There is a paper of J. Li et {\it al.}~\cite{527610} talking about possible failure of the next generation matrix technique. specially in cases of diseases with three actors or more, like vector borne diseases.  We have tried with the technique in~\cite{VddWat02} with reasonable choice of $l$ and $n$ and the  result was the square root of the $\mathcal R_0$ here above.

The above expression for $\mathcal R_0$, 
$$\mathcal R_0={\dsum_{i=0}^n}\mathcal R_0^{vh_i}\mathcal R_0^{h_iv},$$
occurs commonly when dealing with vector-borne diseases. 
 \edem


\section{The Lyapunov function in the  proof of the theorem ~\ref{thm:thmstabee}}\label{sec.supplyap}

\noindent The Lyapunov function technique is commonly used  in the study of the stability  of endemic equilibria of epidemiological systems in the literature~\cite{Guo_li_CAMQ_06, Guo_li_PAMS08, 0999.92036, KoroMMB04, koroMain04, 1022.34044,   MaLiuLi03, McClu06, 1008.92032, 1076.37012, 1056.92052, MR2518930, 2011.10.085}. For the current model, we define a function  $V_{ee}$ on the space state of the model, $\left(\mathbb R_{>0}\right)^u$:

\begin{equation}\label{eq:eqliapee}
\begin{array}{rcl}V_{ee}(\mathbf x) & \equiv & \left(S_q-S^\dagger _q\ln S_q\right) +\,\widetilde\sigma_r\left(S_r-S^\dagger _r\ln S_r\right)+\, \sigma_r^{(1)}\left(E^{(1)}_r-E^{(1)\dagger }_r\ln E^{(1)}_r\right)\\&& +\,\dsum_{j=1}^\ell\left(\sigma_q^{(j)}\left(E^{(j)}_q-E^{(i)\dagger }_q\ln E^{(j)}_q\right)+\sigma_r^{(j+1)}\left(E^{(j+1)}_r-E^{(j+1)\dagger }_r\ln E^{(j+1)}_r\right)\right)+\,\tau_q\left(I_q-I_q^{\dagger }\ln \,I_q\right)\\&& +\tau_r\left(I_r-I_r^{\dagger }\ln \,I_r\right)+\,\dsum_{i=0}^{n}\vartheta_i\left(S_i-S_i^{\dagger }\ln \,S_i\right)+\,\dsum_{\in\mc E}\theta_i\left(E_i-E_i^{\dagger }\ln \,E_i\right)+\,\dsum_{i=0}^{n}\varpi_i\left(I_i-I_i^{\dagger }\ln \,I_i\right)\\&&+\,\dsum_{i\in\mc R}\pi_i\left(R_i-R_i^{\dagger }\ln \,R_i\right),
\end{array}
\end{equation}
where the coefficients $\widetilde\sigma_r$; $\sigma_r^{(j)}$  for $j=1,\;2,\;\cdots,\; l+1$;  $\sigma_q^{(j)}$  for $j=1,\;2,\;\cdots,\; l$;  $\tau_r$; $\tau_q$;   $\upsilon_i$, $\vartheta_i$, for $i=0,\;1,\;\cdots,\; n$   are positive constants to be determined such that the derivative of $V_{ee}$ along the trajectories of the system~\eqref{eq:eqbednet_}  is negative. $\mc E = \{i,~0\leqslant i\leqslant n / u_i=1\}$, $\mc R = \{i,~0\leqslant i\leqslant n / v_i=1\}$, $\overline{\mc E} = \{i,~0\leqslant i\leqslant n / u_i=0\}$ and $\overline{\mc R} = \{i,~0\leqslant i\leq n / v_i=0\}$
This form of the Lyapunov function as well as some of the techniques used in our solution were inspired by  Guo $et~al.$~\cite{Guo_li_CAMQ_06, Guo_li_PAMS08}, who use a graph-theoretic approach to compute the derivative of the Lyapunov function.  In the system of Guo $et~al.$, each group has the same compartmental description, as well as the same mode of influence exchange with other groups (susceptible individuals of a given group are transferred to the next class of the group via contact with all infectious individuals in the system). In our model, not all groups have the same compartmental description: host groups have an SIS structure, while the vector group is more complex. Furthermore,  the mode of exchanging influence between groups is also different: there is no exchange of influence between hosts groups; the vector group influences hosts groups  through contact of susceptible hosts with infectious questing vectors; and all host groups influence questing vectors.


The function $V_{ee}$ defined in \eqref{eq:eqliapee} is $\mathcal C^\infty$ and is positive definite on $\left(\mathbb R_{>0}\right)^u$ as long as all coefficients are positive.  Its derivative  along the trajectories of the system~\eqref{eq:eqbednet_} is:
\begin{small}\begin{equation*}\begin{array}{rcl}\frac{dV_{ee}}{d\,t}(\mathbf
x(t)) & = & \left(1-\frac{S^\dagger _q}{S_q}\right)\left(\Gamma-(\hatmuVar +\varpiVar ) S_q+\delta S_r\right) +\, \widetilde\sigma_r\left(1-\frac{S^{\dagger }_r}{S_r}\right)\left(\left(\varpiVar -\varphiVar \right)S_q-\frac{\delta}{f_r} S_r \right)+\, \sigma_r^{(1)}\left(1-\frac{E^{(1)\dagger }_r}{E^{(1)}_r}\right)\left(\varphiVar  S_q-\frac{\delta}{f_r} E^{(1)}_r \right)\\&& +\,{\dsum_{j=1}^\ell}\sigma_q^{(j)}  \left(1-\frac{E^{(j)\dagger }_q}{E^{(j)}_q}\right) \left( \delta E^{(j)}_r - \frac{\varpiVar }{f_q}E^{(j)}_q \right) +\,{\dsum_{j=1}^\ell}\sigma_r^{(j+1)} \left(1-\frac{E^{(j+1)\dagger }_r}{E^{(j+1)}_r}\right) \left(   \varpiVar   E^{(j)}_q-\frac{\delta}{f_r}E^{(j+1)}_r  \right)\\&&+\,\tau_r\left(1-\frac{I^\dagger _r}{I_r}\right)\left(  \varpiVar  I_q - \frac{\delta}{f_r}I_r \right)  +\,\tau_q\left(1-\frac{I^\dagger _q}{I_q}\right)\left( \delta E^{(\ell+1)}_r - \frac{\varpiVar }{f_q} I_q +\delta I_r \right)+\,{\dsum_{i\in\mc E}}\theta_i\left(1-\frac{E^\dagger _i}{E_i}\right) \left( \frac{a}{H}\,\miVar S_i I_q -  (\nu_i+\varepsilon_i)E_i  \right)\\&&+\,{\dsum_{i=0}^n}\vartheta_i\left(1-\frac{S^\dagger _i}{S_i}\right)\left(\Lambda_i+\widetilde{\nu_i}H_i - \left(\nu_i+a\,\varphi_i \frac{I_q}{H}\right)S_i +v_i\zeta_i R_i +\bar v_i\gamma_iI_i\right) +\,{\dsum_{i=0}^n}\varpi_i\left(1-\frac{I^\dagger _i}{I_i}\right) \left( u_ia\,\varphi_i \frac{I_q}{H}S_i + \bar u_i \varepsilon_iE_i-  (\nu_i+\gamma_i)I_i  \right) \\&& +\,{\dsum_{i\in\mc R}}\pi_i\left(1-\frac{R^\dagger _i}{R_i}\right) \left( \gamma_iI_i -  (\nu_i+\zeta_i)R_i  \right).
\end{array}
\end{equation*}\end{small}

Since according to Proposition~\ref{prop:dissip} we may restrict ourselves to $\mathbf x \in \Omega$, we have  $H_i=S_i+u_iE_i+I_i+v_iR_i=H^*_i$ for each $i$. Furthermore, from Table~\ref{tab.tabvd2} we find that  $r_{suc} $, $d_q$ and $f_q$, and $h_i$ are all time-independent: hence  we may write $r_{suc} ^*=r_{suc} $, $d^*_q=d_q$, $f^*_q=f_q$ and $h^*_i=h_i$. After substituting $\Gamma=(d^*_q +r_{suc} ^*)S_q^\dagger -\delta S_r^\dagger $, $\Lambda_i = a\miVar\frac{I^\dagger _q}{H^*}S^\dagger _i + (\nu_i-\widetilde{\nu}_i)S^\dagger _i -u_i\widetilde\nu_iE^\dagger_i-\widetilde\nu_iI^\dagger_i -\bar v_i\gamma_iI^\dagger_i - v_i(\zeta_i+\widetilde \nu_i)R^\dagger _i$ for $i = 0,~1,~\ldots,~n$,   and rearranging we obtain

\begin{small}\begin{equation*}\begin{array}{rcl}\frac{dV_{ee}}{d\,t}(\mathbf
	x(t)) & = & d_q^* S_q^\dagger\left(2-\frac{S^\dagger _q}{S_q}-\frac{S _q}{S^\dagger_q}\right) + \left(1-\frac{S^\dagger _q}{S_q}\right)\left(\varpiVar ^*S_q^\dagger-\varpiVar S_q+\delta (S_r - S_r^\dagger ) \right) +\, \widetilde\sigma_r\left(1-\frac{S^{\dagger }_r}{S_r}\right)\left(\left(\varpiVar -\varphiVar \right)S_q-\frac{\delta}{f_r} S_r \right)+\, \sigma_r^{(1)}\left(1-\frac{E^{(1)\dagger }_r}{E^{(1)}_r}\right)\left(\varphiVar  S_q-\frac{\delta}{f_r} E^{(1)}_r \right)\\&& +\,{\dsum_{j=1}^\ell}\sigma_q^{(j)}  \left(1-\frac{E^{(j)\dagger }_q}{E^{(j)}_q}\right) \left( \delta E^{(j)}_r - \frac{\varpiVar }{f_q}E^{(j)}_q \right) +\,{\dsum_{j=1}^\ell}\sigma_r^{(j+1)} \left(1-\frac{E^{(j+1)\dagger }_r}{E^{(j+1)}_r}\right) \left(   \varpiVar   E^{(j)}_q-\frac{\delta}{f_r}E^{(j+1)}_r  \right)\\&&+\,\tau_r\left(1-\frac{I^\dagger _r}{I_r}\right)\left(  \varpiVar  I_q - \frac{\delta}{f_r}I_r \right)  +\,\tau_q\left(1-\frac{I^\dagger _q}{I_q}\right)\left( \delta E^{(\ell+1)}_r - \frac{r_{suc}}{f_q} I_q +\delta I_r \right) +\,{\dsum_{i\in\mc E}}\theta_i\left(1-\frac{E^\dagger _i}{E_i}\right) \left( \frac{a}{H}\,\varphi_i S_i I_q -  (\nu_i+\varepsilon_i)E_i  \right)\\&&+\,{\dsum_{i=0}^n}\vartheta_i\left(1-\frac{S^\dagger _i}{S_i}\right)\left(a\varphi_i\frac{I^\dagger _q}{H^*}S^\dagger _i + (\nu_i-\widetilde{\nu}_i)(S^\dagger_i - S_i)  -u_i\widetilde\nu_i(E^\dagger_i - E_i) -\widetilde\nu_i(I^\dagger_i - I_i) - \bar v_i\gamma_i(I^\dagger_i - I_i) - v_i(\zeta_i+\widetilde \nu_i)(R^\dagger_i-R_i) \right.\\&&\left. - a\,\miVar \frac{I_q}{H}S_i \right)+\,{\dsum_{i=0}^n} \varpi_i \left(1 - \frac{I^\dagger _i}{I_i}\right) \left(u_i \varepsilon_iE_i +\bar u_i a\,\varphi_i \frac{I_q}{H}S_i -  (\nu_i+\gamma_i)I_i \right)  +\,{\dsum_{i\in\mc R}}\pi_i\left(1-\frac{R^\dagger _i}{R_i}\right) \left( \gamma_iI_i -  (\nu_i+\zeta_i)R_i \right)
	\end{array}
	\end{equation*}\end{small}

\begin{equation*}
\begin{array}{rcl}\dfrac{dV_{ee}}{d\,t}(\mathbf x(t))   & = & \hatmuVar  S_q^\dagger \left(2-\frac{S_q^\dagger }{S_q}-\frac{S_q}{S_q^\dagger }\right) +\varpiVar ^*  S_q^\dagger -\varpiVar ^*  S_q^\dagger \frac{S^\dagger _q}{S_q}-\varpiVar  S_q+\varpiVar  S_q^\dagger  + \delta S^\dagger _r  \left( \frac{S^\dagger _q}{S_q}-\frac{S^\dagger _q}{S_q}\frac{S_r}{S_r^\dagger }\right) +\,\widetilde \sigma_r\left(\varpiVar -\varphiVar \right)S_q \left( 1 - \frac{S^\dagger _r}{S_r} \right)\\&& +\,\left(\frac{\widetilde\sigma_r}{f_r}-1\right)\delta \left(S_r^\dagger -S_r\right) +\, \sigma_r^{(1)}\varphiVar  S_q \left(1-\frac{E^{(1)\dagger }_r}{E^{(1)}_r}\right) +\, \sigma_r^{(1)}\frac{\delta}{f_r}\left(E^{(1)\dagger }_r- E^{(1)}_r\right)  +\,{\dsum_{i=1}^\ell} \sigma_q^{(i)} \delta E^{(i)}_r\left(1-\frac{E^{(i)\dagger }_q}{E^{(i)}_q}\right) \\&& +\,{\dsum_{j=1}^\ell} \sigma_q^{(j)}\frac{\varpiVar }{f_q}\left( E^{(j)\dagger }_q- E^{(j)}_q \right) +\,{\dsum_{j=1}^\ell}\sigma_r^{(j+1)}E^{(j)}_q\varpiVar   \left(1-\frac{E^{(j+1)\dagger }_r}{E^{(j+1)}_r}\right) +\,{\dsum_{j=1}^\ell}\sigma_r^{(j+1)}\frac{\delta}{f_r}\left(E^{(j+1)\dagger }_r-E^{(j+1)}_r\right)  \\&&   +\,\tau_q\left(E^{(\ell+1)}_r + I_r\right) \delta\left(1-\frac{I^\dagger _q}{I_q}\right) +\,\tau_q\frac{\varpiVar }{f_q}\left( I_q^\dagger  - I_q \right) +\,\tau_r \varpiVar  I_q \left(1- \frac{I^\dagger _r}{I_r} \right) +\,\tau_r \frac{\delta}{f_r}\left(I_r^\dagger - I_r\right) \\&&+\,{\dsum_{i=0}^{n}}\vartheta_i\left(\nu_i - \widetilde{\nu}_i\right)S^\dagger _i \left(2-\frac{S^\dagger _i}{S_i}-\frac{S_i}{S^\dagger_i} \right)  +\,{\dsum_{i=0}^{n}} \vartheta_i a\,\varphi_i \frac{I^\dagger_q}{H^*} S^\dagger _i\left(1-\frac{S^\dagger _i}{S_i} \right) +\,{\dsum_{i=0}^{n}} \vartheta_i  a\,\varphi_i \frac{I_q}{H} S^\dagger_i +\,{\dsum_{i=0}^n} (u_i\theta_i + \bar u_i\varpi_i - \vartheta_i) a\,\varphi_i \frac{I_q}{H} S_i \\&&  -\,{\dsum_{i=0}^n} \left(u_i\theta_i\frac{E^\dagger_i}{E_i} + \bar u_i\varpi_i\frac{I^\dagger _i}{I_i}\right)a \varphi_i\frac{I_q}{H}S_i   + \,{\dsum_{i\in\mc E}} (\theta_i\nu_i-\vartheta_i\widetilde\nu_i)(E^\dagger _i - E_i)    + \,{\dsum_{i\in\mc E}} \vartheta_i\widetilde\nu_i\frac{S^\dagger_i}{S_i} (E^\dagger _i - E_i)    + \,{\dsum_{i\in\mc E}} \varepsilon_i(\varpi_i-\theta_i) E_i  \\&& + \,{\dsum_{i\in\mc E}}\theta_i \varepsilon_i E^\dagger_i  - \,{\dsum_{i\in\mc E}} \varpi_i\varepsilon_i\frac{I^\dagger_i}{I_i} E_i + \,{\dsum_{i=0}^{n}} (\varpi_i\nu_i-\vartheta_i\widetilde\nu_i)(I^\dagger _i - I_i)   + \,{\dsum_{i=0}^{n}} (\varpi_i-\bar v_i\vartheta_i)\gamma_i I^\dagger_i + \,{\dsum_{i=0}^n}(v_i\pi_i + \bar v_i\vartheta_i -\varpi_i) \gamma_i I_i \\&&  +\,{\dsum_{i=0}^n} \vartheta_i\widetilde \nu_i\frac{S^\dagger _i}{S_i}(\widetilde\nu_i + \bar v_i\gamma_i)\left(I^\dagger_i - I_i\right) - \,{\dsum_{i\in\mc R}} \gamma_i\pi_i\frac{R^\dagger_i}{R_i} I_i + \,{\dsum_{i\in\mc R}} (\pi_i\nu_i-\vartheta_i\widetilde\nu_i) (R^\dagger _i - R_i)  + \,{\dsum_{i\in\mc R}} (\pi_i-\vartheta_i)\zeta_i (R^\dagger _i  - R_i) \\&& +\,{\dsum_{i\in\mc R}} \vartheta_i(\zeta_i + \widetilde \nu_i)\frac{S^\dagger _i}{S_i} \left(R^\dagger_i - R_i\right).  
\end{array}
\end{equation*}

We may write 
\begin{equation}\label{eq:Fexp}
\dfrac{dV_{ee}}{d\,t}(\mathbf x(t))    =  \hatmuVar  S_q^\dagger \left(2-\frac{S_q^\dagger }{S_q}-\frac{S_q}{S_q^\dagger }\right) + F(\mathbf x),
\end{equation}
where 
\begin{equation*}
\begin{array}{rcl}F(\mathbf x)   &=&   \varpiVar ^*  S_q^\dagger \left(1-\frac{S^\dagger _q}{S_q}\right)-\varpiVar  S_q+\varpiVar  S_q^\dagger  + \delta S^\dagger _r  \left( \frac{S^\dagger _q}{S_q}+\frac{S_r}{S_r^\dagger }-\frac{S^\dagger _q}{S_q}\frac{S_r}{S_r^\dagger }-1\right) +\widetilde \sigma_r\left(\varpiVar -\varphiVar \right)S_q \left( 1 - \frac{S^\dagger _r}{S_r} \right) +\,\widetilde\sigma_r\frac{\delta}{f_r} S^\dagger _r\left(1-\frac{S_r}{S_r^\dagger }\right) \\&&+ \, \sigma_r^{(1)}\varphiVar  S_q \left(1-\frac{E^{(1)\dagger }_r}{E^{(1)}_r}\right) +\,\delta{\dsum_{j=1}^\ell}\left(  E^{(j)\dagger }_r \left(\frac{\sigma_r^{(1)}}{f_r}- \sigma_q^{(j)}\frac{E^{(j)}_r}{E^{(j)\dagger }_r}\frac{E^{(j) \dagger }_q}{E^{(j)}_q}\right)  +\, \left(\sigma_q^{(j)} -\,\frac{\sigma_r^{(j)}}{f_r} \right)E^{(j)}_r \right) \\&&+\,\delta \left(\tau_q -\,\frac{\sigma_r^{(\ell+1)}}{f_r} \right)E^{(\ell+1)}_r  +\,\delta  E^{(\ell+1)\dagger }_r \left(\frac{\sigma_r^{(\ell+1)}}{f_r}- \tau_q\frac{E^{(\ell+1)}_r}{E^{(\ell+1)\dagger }_r}\frac{I^{ \dagger }_q}{I_q}\right)  +\, \delta\left(\tau_q -\frac{\tau_r}{f_r} \right)I_r +\, \delta I^\dagger _r\left(\frac{\tau_r}{f_r}-\tau_q  \frac{I_r}{I^\dagger _r}\frac{I^\dagger _q}{I_q}\right) \\&&+\,r_{suc} {\dsum_{j=1}^\ell}\left( E^{(j)\dagger }_q \left(\frac{\sigma_q^{(j)}}{f_q}-\sigma_r^{(j+1)}\frac{E^{(j)}_q}{E^{(j)\dagger }_q}\frac{E^{(j+1)\dagger }_r}{E^{(j+1)}_r} \right) +\,  \left(\sigma_r^{(j+1)}- \frac{\sigma_q^{(j)}}{f_q}\right)E^{(j)}_q\right) +\, r_{suc}  I^\dagger _q \left(\frac{\tau_q}{f_q}- \tau_r\frac{I_q}{I^\dagger _q}\frac{I^\dagger _r}{I_r} \right)  + \, r_{suc}  I_q\left( \tau_r- \frac{\tau_q}{f_q}\right) \\&&+\,{\dsum_{i=0}^{n}}\vartheta_i\left(\nu_i - \widetilde{\nu}_i\right)S^\dagger _i \left(2-\frac{S^\dagger _i}{S_i}-\frac{S_i}{S^\dagger_i} \right)  +\,{\dsum_{i=0}^{n}} \vartheta_i a\,\varphi_i \frac{I^\dagger_q}{H^*} S^\dagger _i\left(1-\frac{S^\dagger _i}{S_i} \right) +\,{\dsum_{i=0}^{n}} \vartheta_i  a\,\varphi_i \frac{I_q}{H} S^\dagger_i +\,{\dsum_{i=0}^n} (u_i\theta_i + \bar u_i\varpi_i - \vartheta_i) a\,\varphi_i \frac{I_q}{H} S_i \\&&  -\,{\dsum_{i=0}^n} \left(u_i\theta_i\frac{E^\dagger_i}{E_i} + \bar u_i\varpi_i\frac{I^\dagger _i}{I_i}\right)a \varphi_i\frac{I_q}{H}S_i   + \,{\dsum_{i\in\mc E}} (\theta_i\nu_i-\vartheta_i\widetilde\nu_i)(E^\dagger _i - E_i)    + \,{\dsum_{i\in\mc E}} \vartheta_i\widetilde\nu_i\frac{S^\dagger_i}{S_i} (E^\dagger _i - E_i)    + \,{\dsum_{i\in\mc E}} \varepsilon_i(\varpi_i-\theta_i) E_i  \\&& + \,{\dsum_{i\in\mc E}}\theta_i \varepsilon_i E^\dagger_i  - \,{\dsum_{i\in\mc E}} \varpi_i\varepsilon_i\frac{I^\dagger_i}{I_i} E_i + \,{\dsum_{i=0}^{n}} (\varpi_i\nu_i-\vartheta_i\widetilde\nu_i)(I^\dagger _i - I_i)   + \,{\dsum_{i=0}^{n}} (\varpi_i-\bar v_i\vartheta_i)\gamma_i I^\dagger_i + \,{\dsum_{i=0}^n}(v_i\pi_i + \bar v_i\vartheta_i -\varpi_i) \gamma_i I_i \\&&  +\,{\dsum_{i=0}^n} \vartheta_i\widetilde \nu_i\frac{S^\dagger _i}{S_i}(\widetilde\nu_i + \bar v_i\gamma_i)\left(I^\dagger_i - I_i\right) - \,{\dsum_{i\in\mc R}} \gamma_i\pi_i\frac{R^\dagger_i}{R_i} I_i + \,{\dsum_{i\in\mc R}} (\pi_i\nu_i-\vartheta_i\widetilde\nu_i) (R^\dagger _i - R_i)  + \,{\dsum_{i\in\mc R}} (\pi_i-\vartheta_i)\zeta_i (R^\dagger _i  - R_i) \\&& +\,{\dsum_{i\in\mc R}} \vartheta_i(\zeta_i + \widetilde \nu_i)\frac{S^\dagger _i}{S_i} \left(R^\dagger_i - R_i\right).
\end{array}
\end{equation*}
Note that  $\frac{dV_{ee}}{d\,t}(\mathbf x(t)) \leqslant  F(\mathbf x)$, since $2-\frac{S_q^\dagger }{S_q}-\frac{S_q}{S_q^\dagger }\leqslant 0$ for all $\mathbf x\in\left(\mathbb R_{>0}\right)^u$.

The expression for $F(\mathbf x)$ may be simplified by choosing
\begin{equation}
 \tau_q =\dfrac{\tau_r}{f_r}; ~\sigma^{(l+1)}_r = f_r\tau_q ;~  \sigma^{(j)}_q=f_q\sigma^{(j+1)}_r,  \sigma^{(j)}_r = f_r\sigma^{(j)}_q~ \textrm{~for~}j=l,~l-1,~\cdots 1,\\ ~\textrm{and~}\\  \theta_i=\vartheta_i = \varpi_i = \pi_i ~ \textrm{~for~} i = 0,~1,~\cdots,~n
\end{equation}
from which follows
 \begin{equation}\label{eq.coefvee} \begin{array}{l}\sigma^{(j)}_r = \sigma^{(1)}_r(f_qf_r)^{1-j},\;\hbox{ for} \;j=1,\;2,\;\cdots,\;l+1; \quad\sigma^{(j)}_q = \sigma^{(1)}_r(f_qf_r)^{1-j}f_r^{-1},\; \hbox{for}\; j=1,\;2,\;\cdots,\;l;\\   \tau_q = \sigma^{(1)}_rf_r^{-1}(f_qf_r)^{-l}; \qquad \tau_r =\sigma^{(1)}_r(f_qf_r)^{-l}.\end{array}
\end{equation} 
Using these substitutions and the fact that the size of the fraction of population in the $i^{th}$ hosts group $H_i=S_i+u_iE_i+I_i+v_iR_i$ is constant for $i=0,\;\cdots,\;n$, we get:

\begin{small}\begin{equation*}
\begin{array}{rcl}
F(\mathbf x)  &=&  \varpiVar ^\dagger  S_q^\dagger \left(1-\frac{S^\dagger _q}{S_q}\right)+\left((\sigma_r^{(1)}-\widetilde\sigma_r)\varphiVar +(\widetilde\sigma_r -1)\varpiVar \right) S_q+\varpiVar  S_q^\dagger  + \delta S^\dagger _r  \left( \frac{S^\dagger _q}{S_q}+\frac{S_r}{S_r^\dagger }-\frac{S^\dagger _q}{S_q}\frac{S_r}{S_r^\dagger } - 1\right)+\,\widetilde\sigma_r\frac{\delta}{f_r} S_r^\dagger \left(1-\frac{S_r}{S_r^\dagger }\right) \\&& -\,\widetilde \sigma_r\left(\varpiVar -\varphiVar \right)S_q \frac{S^\dagger _r}{S_r}    - \, \sigma_r^{(1)}\varphiVar  S_q \frac{E^{(1)\dagger }_r}{E^{(1)}_r} +\, \varpiVar  \sigma_r^{(1)}{\dsum_{j=1}^l}\frac{E^{(i)\dagger }_q}{(f_qf_r)^{j}} \left(1- \frac{E^{(j)}_q}{E^{(j)\dagger }_q} \frac{E^{(j+1)\dagger }_r}{E^{(j+1)}_r}\right)  +\,\delta \sigma_r^{(1)}{\dsum_{j=1}^l }\frac{ E^{(j)\dagger }_r}{f_r(f_qf_r)^{j-1}}\left(1-\, \frac{ E^{(j)}_r}{ E^{(j)\dagger }_r}\frac{E^{(j)\dagger }_q}{E^{(j)}_q}\right)\\&& +\delta  \sigma_r^{(1)} \frac{ E^{(l+1)\dagger }_r}{f_r(f_qf_r)^{\ell}} \left(1 - \frac{E^{(l+1)}_r}{ E^{(l+1)\dagger }_r} \frac{I^\dagger _q}{I_q}\right)  +\, \delta  \sigma_r^{(1)}\frac{I_r^\dagger }{f_r(f_qf_r)^{\ell}} \left(1 - \frac{I_r}{I_r^\dagger }\frac{I^\dagger _q}{I_q}\right) +\,\varpiVar  \sigma_r^{(1)} \frac{I_q^\dagger }{(f_qf_r)^{\ell}}   \left(\frac{1}{f_rf_q}- \frac{I_q}{I_q^\dagger }\frac{I^\dagger _r}{I_r} \right)  + \,\varpiVar   \sigma_r^{(1)} \frac{f_rf_q-1}{(f_qf_r)^{l+1}}I_q\\&&+\,{\dsum_{i=0}^{n}}\varpi_i\left(\nu_i - \widetilde{\nu}_i\right)S^\dagger _i \left(2-\frac{S^\dagger _i}{S_i}-\frac{S_i}{S^\dagger_i} \right)  +\,{\dsum_{i=0}^{n}} \varpi_i a\,\varphi_i \frac{I^\dagger_q}{H^*} S^\dagger _i\left(1-\frac{S^\dagger _i}{S_i} \right) +\,{\dsum_{i=0}^{n}} \varpi_i  a\,\varphi_i \frac{I_q}{H} S^\dagger_i  -\,{\dsum_{i=0}^n}\varpi_i \left(u_i\frac{E^\dagger_i}{E_i} + \bar u_i\frac{I^\dagger _i}{I_i}\right)a \varphi_i\frac{I_q}{H}S_i   \\&&  + \,{\dsum_{i\in\mc E}}\varpi_i (\nu_i-\widetilde\nu_i)(E^\dagger _i - E_i)    + \,{\dsum_{i\in\mc E}} \varpi_i\widetilde\nu_i\frac{S^\dagger_i}{S_i} (E^\dagger _i - E_i) + \,{\dsum_{i\in\mc E}}\varpi_i \varepsilon_i E^\dagger_i\left(1 - \frac{I^\dagger_i}{I_i}\frac{E_i}{E^\dagger_i} \right)  + \,{\dsum_{i=0}^{n}}\varpi_i (\nu_i-\widetilde\nu_i)(I^\dagger _i - I_i)  + \,{\dsum_{i=0}^{n}} \varpi_i(1-\bar v_i)\gamma_i I^\dagger_i  \\&&  +\,{\dsum_{i=0}^n} \varpi_i\widetilde \nu_i\frac{S^\dagger _i}{S_i}(\widetilde\nu_i + \bar v_i\gamma_i)\left(I^\dagger_i - I_i\right) - \,{\dsum_{i\in\mc R}}\varpi_i \gamma_i\frac{R^\dagger_i}{R_i} I_i + \,{\dsum_{i\in\mc R}}\varpi_i (\nu_i-\widetilde\nu_i) (R^\dagger _i - R_i)  +\,{\dsum_{i\in\mc R}}\varpi_i(\zeta_i + \widetilde \nu_i)\frac{S^\dagger _i}{S_i} \left(R^\dagger_i - R_i\right). 
\end{array}
\end{equation*}\end{small}

Using the fact that all time derivatives in system~\eqref{eq:eqbednet_} are zero at the EE, we find:  $\varpiVar ^\dagger  I_q^\dagger  = \frac{\delta}{f_r}I_r^\dagger $, $\delta S_r^\dagger =f_r\left(\varpiVar ^\dagger -\varphiVar ^\dagger \right)S_q^\dagger $, $\varpiVar  E_q^{(j)\dagger }=\frac{\delta}{f_r}E_r^{(j+1)\dagger }$  for $j = 1,\;2,\;\cdots,\; \ell$, $u_i\varepsilon_iE^\dagger_i + \bar u_i a\varphi_i\frac{I_q^\dagger }{H^*} S^\dagger _i = (\nu_i+\gamma_i)I_i^\dagger$ for $i=0,~\ldots,~n$ and $a\varphi_i\frac{I_q^\dagger }{H^*} S^\dagger _i = (\varepsilon_i + \nu_i)E^\dagger_i$   for $i\in\mc E$. This leads to

\begin{equation*}
\begin{array}{rcl}
F(\mathbf x)  &=&  \varpiVar ^\dagger  S_q^\dagger \left(1-\frac{S^\dagger _q}{S_q}\right) +\left((\sigma_r^{(1)}-\widetilde\sigma_r)\varphiVar +(\widetilde\sigma_r -1)\varpiVar \right) S_q +\varpiVar  S_q^\dagger  + \delta S^\dagger _r  \left( \frac{S^\dagger _q}{S_q}+\frac{S_r}{S_r^\dagger }-\frac{S^\dagger _q}{S_q}\frac{S_r}{S_r^\dagger }-1\right) +\,\widetilde\sigma_r\frac{\delta}{f_r} S_r^\dagger \left(1-\frac{S_r}{S_r^\dagger }\right)\\&&-\;\widetilde \sigma_r\left(\varpiVar -\varphiVar \right)S_q \frac{S^\dagger _r}{S_r}  - \, \sigma_r^{(1)}\varphiVar  S_q \frac{E^{(1)\dagger }_r}{E^{(1)}_r} +\,\sigma_r^{(1)} \frac{\delta}{f_r} E^{(1)\dagger }_r\left(1-\, \frac{ E^{(1)}_r}{ E^{(1)\dagger }_r}\frac{E^{(1)\dagger }_q}{E^{(1)}_q}\right)  +\,\varpiVar  \sigma_r^{(1)}  \frac{1-f_rf_q}{(f_qf_r)^{l+1}} \left(I^\dagger _q - I_q\right)    \\&&  +\, r_{suc}  \sigma_r^{(1)}{\dsum_{j=1}^{l-1}}\frac{E^{(i)\dagger }_q}{(f_qf_r)^{i}}\left(2- \frac{E^{(j)}_q}{E^{(j)\dagger }_q}\frac{E^{(j+1)\dagger }_r}{E^{(j+1)}_r}-\, \frac{ E^{(j+1)}_r}{ E^{(j+1)\dagger }_r}\frac{E^{(j+1)\dagger }_q}{E^{(j+1)}_q}\right) +\,  \varpiVar  \sigma_r^{(1)}\frac{E^{(\ell)\dagger }_q}{(f_qf_r)^{l}}\left(2- \frac{E^{(l)}_q}{E^{(\ell)\dagger }_q}\frac{E^{(\ell+1)\dagger }_r}{E^{(l+1)}_r}- \frac{E^{(\ell+1)}_r}{E^{(\ell+1)\dagger }_r}\frac{I^\dagger _q}{I_q}\right) \\&& +\, r_{suc}  \sigma_r^{(1)}\frac{I_q^\dagger }{(f_qf_r)^{l}} \left(2 - \frac{I_r}{I_r^\dagger }\frac{I_q^\dagger }{I_q}- \frac{I_q}{I_q^\dagger }\frac{I^\dagger _r}{I_r}\right) +\,{\dsum_{i=0}^{n}}\varpi_i\left(\nu_i - \widetilde{\nu}_i\right)S^\dagger _i \left(2-\frac{S^\dagger _i}{S_i} \right)  + \,{\dsum_{i\in\mc E}}\varpi_i (\nu_i-\widetilde\nu_i)E^\dagger _i\left(2 + \frac{I_q}{I^\dagger_q} -\frac{S^\dagger _i}{S_i}  - \frac{E_i}{E^\dagger_i} -  \frac{S _i}{S^\dagger_i}\frac{E^\dagger_i}{E_i} \frac{I_q}{I^\dagger_q}\right) \\&&  +\,{\dsum_{i\in\mc E}} \varpi_i \widetilde\nu_iE^\dagger_i\left(1 + \frac{I_q}{I^\dagger_q} -\frac{S^\dagger _i}{S_i}\frac{E_i}{E^\dagger_i} -  \frac{S _i}{S^\dagger_i}\frac{E^\dagger_i}{E_i}\frac{I_q}{I^\dagger_q} \right)   + \,{\dsum_{i=0}^{n}}\varpi_i (\nu_i-\widetilde\nu_i)I^\dagger_i\left(2 + u_i  + \frac{I_q}{I^\dagger_q} -\frac{S^\dagger _i}{S_i}  - u_i\frac{I^\dagger_i}{I_i}\frac{E_i}{E^\dagger_i}  - \frac{I_i}{I^\dagger_i} -  \frac{S _i}{S^\dagger_i}\left(u_i \frac{E^\dagger_i}{E_i} + \bar u_i\frac{I^\dagger _i}{I_i}\right)\frac{I_q}{I^\dagger_q}\right) \\&& +\,{\dsum_{i=0}^n} \varpi_i \widetilde\nu_i I^\dagger _i\left(1 + u_i + \frac{I_q}{I^\dagger_q} -\frac{S^\dagger _i}{S_i}\frac{I_i}{I^\dagger_i}  - u_i\frac{I^\dagger_i}{I_i}\frac{E_i}{E^\dagger_i} -  \frac{S _i}{S^\dagger_i}\left(u_i \frac{E^\dagger_i}{E_i} + \bar u_i\frac{I^\dagger _i}{I_i}\right)\frac{I_q}{I^\dagger_q} \right)  + \,{\dsum_{i=0}^{n}} \varpi_i\gamma_i I^\dagger_i\left(2 + u_i -\bar v_i + \frac{I_q}{I^\dagger_q} - v_i\frac{S^\dagger _i}{S_i} - \bar v_i\frac{S^\dagger _i}{S_i}\frac{I_i}{I^\dagger_i} \right.\\&& \left. -\, u_i\frac{I^\dagger_i}{I_i}\frac{E_i}{E^\dagger_i} -  \frac{S _i}{S^\dagger_i}\left(u_i \frac{E^\dagger_i}{E_i} + \bar u_i\frac{I^\dagger _i}{I_i}\right)\frac{I_q}{I^\dagger_q}  - \frac{R^\dagger_i}{R_i} \frac{I_i}{I^\dagger_i}\right)  + \,{\dsum_{i\in\mc R}}\varpi_i (\nu_i-\widetilde\nu_i) (R^\dagger _i - R_i)  +\,{\dsum_{i\in\mc R}}\varpi_i(\zeta_i + \widetilde \nu_i)\frac{S^\dagger _i}{S_i} \left(R^\dagger_i - R_i\right) -\,{\dsum_{i=0}^{n}}\varpi_i\left(\nu_i - \widetilde{\nu}_i\right)S_i.
\end{array}
\end{equation*}
Using now the fact that all time derivatives in system~\eqref{eq:eqbednet_} are zero at the EE, we find:  $a\varphi_i\frac{I_q^\dagger }{H^*} S^\dagger _i = (\nu_i+\varepsilon_i)E^\dagger_i$ and $\gamma_i I^\dagger_i = (\nu_i+\zeta_i)E_i^\dagger$  for $i\in\mc R$ makes the previous evolute in

\begin{equation*}
\begin{array}{rcl}
F(\mathbf x)  &=&  \varpiVar ^\dagger  S_q^\dagger \left(1-\frac{S^\dagger _q}{S_q}\right) +\left((\sigma_r^{(1)}-\widetilde\sigma_r)\varphiVar +(\widetilde\sigma_r -1)\varpiVar \right) S_q +\varpiVar  S_q^\dagger  + \delta S^\dagger _r  \left( \frac{S^\dagger _q}{S_q}+\frac{S_r}{S_r^\dagger }-\frac{S^\dagger _q}{S_q}\frac{S_r}{S_r^\dagger }-1\right) +\,\widetilde\sigma_r\frac{\delta}{f_r} S_r^\dagger \left(1-\frac{S_r}{S_r^\dagger }\right)\\&&-\;\widetilde \sigma_r\left(\varpiVar -\varphiVar \right)S_q \frac{S^\dagger _r}{S_r}  - \, \sigma_r^{(1)}\varphiVar  S_q \frac{E^{(1)\dagger }_r}{E^{(1)}_r} +\,\sigma_r^{(1)} \frac{\delta}{f_r} E^{(1)\dagger }_r\left(1-\, \frac{ E^{(1)}_r}{ E^{(1)\dagger }_r}\frac{E^{(1)\dagger }_q}{E^{(1)}_q}\right)  +\,\varpiVar  \sigma_r^{(1)}  \frac{1-f_rf_q}{(f_qf_r)^{l+1}} \left(I^\dagger _q - I_q\right)    \\&&  +\, r_{suc}  \sigma_r^{(1)}{\dsum_{j=1}^{l-1}}\frac{E^{(i)\dagger }_q}{(f_qf_r)^{i}}\left(2- \frac{E^{(j)}_q}{E^{(j)\dagger }_q}\frac{E^{(j+1)\dagger }_r}{E^{(j+1)}_r}-\, \frac{ E^{(j+1)}_r}{ E^{(j+1)\dagger }_r}\frac{E^{(j+1)\dagger }_q}{E^{(j+1)}_q}\right) +\,  \varpiVar  \sigma_r^{(1)}\frac{E^{(l)\dagger }_q}{(f_qf_r)^{l}}\left(2- \frac{E^{(l)}_q}{E^{(l)\dagger }_q}\frac{E^{(l+1)\dagger }_r}{E^{(l+1)}_r}- \frac{E^{(l+1)}_r}{E^{(l+1)\dagger }_r}\frac{I^\dagger _q}{I_q}\right) \\&& +\, r_{suc}  \sigma_r^{(1)}\frac{I_q^\dagger }{(f_qf_r)^{l}} \left(2 - \frac{I_r}{I_r^\dagger }\frac{I_q^\dagger }{I_q}- \frac{I_q}{I_q^\dagger }\frac{I^\dagger _r}{I_r}\right) +\,{\dsum_{i=0}^{n}}\varpi_i\left(\nu_i - \widetilde{\nu}_i\right)S^\dagger _i \left(2-\frac{S^\dagger _i}{S_i} \right) + \,{\dsum_{i\in\mc E}}\varpi_i (\nu_i-\widetilde\nu_i)E^\dagger _i\left(2 + \frac{I_q}{I^\dagger_q} -\frac{S^\dagger _i}{S_i}   -  \frac{S _i}{S^\dagger_i}\frac{E^\dagger_i}{E_i} \frac{I_q}{I^\dagger_q}\right) \\&&  +\,{\dsum_{i\in\mc E}} \varpi_i \widetilde\nu_iE^\dagger_i\left(1 + \frac{I_q}{I^\dagger_q} -\frac{S^\dagger _i}{S_i}\frac{E_i}{E^\dagger_i} -  \frac{S _i}{S^\dagger_i}\frac{E^\dagger_i}{E_i}\frac{I_q}{I^\dagger_q} \right)   + \,{\dsum_{i=0}^{n}}\varpi_i (\nu_i-\widetilde\nu_i)I^\dagger_i\left(2 + u_i  + \frac{I_q}{I^\dagger_q} -\frac{S^\dagger _i}{S_i}  - u_i\frac{I^\dagger_i}{I_i} \frac{E_i}{E^\dagger_i}  -  \frac{S _i}{S^\dagger_i}\left(u_i \frac{E^\dagger_i}{E_i} + \bar u_i\frac{I^\dagger _i}{I_i}\right)\frac{I_q}{I^\dagger_q}\right) \\&& +\,{\dsum_{i=0}^n} \varpi_i \widetilde\nu_i I^\dagger _i\left(1 + u_i + \frac{I_q}{I^\dagger_q} -\frac{S^\dagger _i}{S_i}\frac{I_i}{I^\dagger_i}  - u_i\frac{I^\dagger_i}{I_i}\frac{E_i}{E^\dagger_i} -  \frac{S _i}{S^\dagger_i}\left(u_i \frac{E^\dagger_i}{E_i} + \bar u_i\frac{I^\dagger _i}{I_i}\right)\frac{I_q}{I^\dagger_q} \right)  + \,{\dsum_{i\in\overline{\mc R}}} \varpi_i\gamma_i I^\dagger_i\left(1 + u_i + \frac{I_q}{I^\dagger_q} - \frac{S^\dagger _i}{S_i}\frac{I_i}{I^\dagger_i} -\, u_i\frac{I^\dagger_i}{I_i}\frac{E_i}{E^\dagger_i} \right.\\&& \left. -  \frac{S _i}{S^\dagger_i}\left(u_i \frac{E^\dagger_i}{E_i} + \bar u_i\frac{I^\dagger _i}{I_i}\right)\frac{I_q}{I^\dagger_q} \right)  + \,{\dsum_{i\in\mc R}}\varpi_i (\nu_i-\widetilde\nu_i) R^\dagger _i\left(3 + u_i + \frac{I_q}{I^\dagger_q} - \frac{S^\dagger _i}{S_i}  -\, u_i\frac{I^\dagger_i}{I_i}\frac{E_i}{E^\dagger_i} -  \frac{S _i}{S^\dagger_i}\left(u_i \frac{E^\dagger_i}{E_i} + \bar u_i\frac{I^\dagger _i}{I_i}\right)\frac{I_q}{I^\dagger_q}  - \frac{R^\dagger_i}{R_i} \frac{I_i}{I^\dagger_i}\right)\\&&  + \,{\dsum_{i\in\mc R}} \varpi_i(\zeta_i+\widetilde\nu_i)R^\dagger_i\left(2 + u_i + \frac{I_q}{I^\dagger_q}   -\, u_i\frac{I^\dagger_i}{I_i}\frac{E_i}{E^\dagger_i} -  \frac{S _i}{S^\dagger_i}\left(u_i \frac{E^\dagger_i}{E_i} + \bar u_i\frac{I^\dagger _i}{I_i}\right)\frac{I_q}{I^\dagger_q}  -  \frac{S^\dagger_i}{S_i}\frac{R_i}{R^\dagger_i} - \frac{R^\dagger_i}{R_i} \frac{I_i}{I^\dagger_i}\right)  -\,{\dsum_{i=0}^{n}}\varpi_i\left(\nu_i - \widetilde{\nu}_i\right)H_i.
\end{array}
\end{equation*}
 Using  relations between components of $\mathbf x^\dagger $  given in~\eqref{eq:eqexpMinfctstateb} we may derive $$\varphiVar ^\dagger  S^\dagger _q = \frac{\delta E^{(1)\dagger }_r}{f_r} = \frac{\delta E^{(2)\dagger }_r}{f_r(f_qf_r)} = \cdots = \frac{\delta E^{(\ell+1)\dagger }_r}{f_r(f_qf_r)^l} = \frac{\varpiVar (1-f_qf_r) I^{\dagger }_q}{(f_qf_r)^{\ell+1}} \hbox{ and } \frac{\delta E^{(j+1)\dagger }_r}{f_r(f_qf_r)^j} = \frac{\varpiVar  E^{(j)\dagger }_q}{(f_qf_r)^j},\;\;\; j=1,\;\cdots,\;\ell,$$ 
and after few algebraic rearrangements, the above becomes:

\begin{equation*}
\begin{array}{rcl}
F(\mathbf x)  &=&  \left((\sigma_r^{(1)}-\widetilde\sigma_r)\varphiVar +(\widetilde\sigma_r -1)\varpiVar \right) S_q +\varpiVar ^\dagger  S_q^\dagger \left(1-\frac{S^\dagger _q}{S_q}\right) +\varpiVar  S_q^\dagger + \delta S^\dagger _r  \left( \frac{S^\dagger _q}{S_q} + \frac{S_r}{S_r^\dagger }-\frac{S^\dagger _q}{S_q}\frac{S_r}{S_r^\dagger }-1\right)+\,\widetilde\sigma_r\frac{\delta}{f_r} S^\dagger _r  \left(1-\frac{S_r}{S_r^\dagger }\right) \\&&-\;\widetilde \sigma_r\left(\varpiVar -\varphiVar \right)S_q \frac{S^\dagger _r}{S_r}  - \, \sigma_r^{(1)}\varphiVar  S_q \frac{E^{(1)\dagger }_r}{E^{(1)}_r}+\, \sigma_r^{(1)}\varphiVar ^\dagger  S^\dagger _q\left(2l+2- {\dsum_{j=1}^{l}}\frac{E^{(j)}_q}{E^{(j)\dagger }_q}\frac{E^{(j+1)\dagger }_r}{E^{(j+1)}_r}-\, {\dsum_{j=1}^{l}}\frac{ E^{(j)}_r}{ E^{(j)\dagger }_r}\frac{E^{(j)\dagger }_q}{E^{(j)}_q}- \frac{E^{(l+1)}_r}{E^{(l+1)\dagger }_r}\frac{I^\dagger _q}{I_q} - \frac{I_q}{I^\dagger _q}\right)      \\&&+\, \frac{\varpiVar  \sigma_r^{(1)}}{(f_qf_r)^{l}}I_q^\dagger  \left(2 - \frac{I_r}{I_r^\dagger }\frac{I_q^\dagger }{I_q}- \frac{I_q}{I_q^\dagger }\frac{I^\dagger _r}{I_r}\right) +\,{\dsum_{i=0}^{n}}\varpi_i\left(\nu_i - \widetilde{\nu}_i\right)S^\dagger _i \left(2-\frac{S^\dagger _i}{S_i} \right)   + \,{\dsum_{i\in\mc E}}\varpi_i (\nu_i-\widetilde\nu_i)E^\dagger _i\left(2 + \frac{I_q}{I^\dagger_q} -\frac{S^\dagger _i}{S_i}   -  \frac{S _i}{S^\dagger_i}\frac{E^\dagger_i}{E_i} \frac{I_q}{I^\dagger_q}\right) \\&&  +\,{\dsum_{i\in\mc E}} \varpi_i \widetilde\nu_iE^\dagger_i\left(1 + \frac{I_q}{I^\dagger_q} -\frac{S^\dagger _i}{S_i}\frac{E_i}{E^\dagger_i} -  \frac{S _i}{S^\dagger_i}\frac{E^\dagger_i}{E_i}\frac{I_q}{I^\dagger_q} \right)   + \,{\dsum_{i=0}^{n}}\varpi_i (\nu_i-\widetilde\nu_i)I^\dagger_i\left(2 + u_i  + \frac{I_q}{I^\dagger_q} -\frac{S^\dagger _i}{S_i}  - u_i\frac{I^\dagger_i}{I_i} \frac{E_i}{E^\dagger_i}  -  \frac{S _i}{S^\dagger_i}\left(u_i \frac{E^\dagger_i}{E_i} + \bar u_i\frac{I^\dagger _i}{I_i}\right)\frac{I_q}{I^\dagger_q}\right) \\&& +\,{\dsum_{i=0}^n} \varpi_i \widetilde\nu_i I^\dagger _i\left(1 + u_i + \frac{I_q}{I^\dagger_q} -\frac{S^\dagger _i}{S_i}\frac{I_i}{I^\dagger_i}  - u_i\frac{I^\dagger_i}{I_i}\frac{E_i}{E^\dagger_i} -  \frac{S _i}{S^\dagger_i}\left(u_i \frac{E^\dagger_i}{E_i} + \bar u_i\frac{I^\dagger _i}{I_i}\right)\frac{I_q}{I^\dagger_q} \right)  + \,{\dsum_{i\in\overline{\mc R}}} \varpi_i\gamma_i I^\dagger_i\left(1 + u_i + \frac{I_q}{I^\dagger_q} - \frac{S^\dagger _i}{S_i}\frac{I_i}{I^\dagger_i} -\, u_i\frac{I^\dagger_i}{I_i}\frac{E_i}{E^\dagger_i} \right.\\&& \left. -  \frac{S _i}{S^\dagger_i}\left(u_i \frac{E^\dagger_i}{E_i} + \bar u_i\frac{I^\dagger _i}{I_i}\right)\frac{I_q}{I^\dagger_q} \right)  + \,{\dsum_{i\in\mc R}}\varpi_i (\nu_i-\widetilde\nu_i) R^\dagger _i\left(3 + u_i + \frac{I_q}{I^\dagger_q} - \frac{S^\dagger _i}{S_i}  -\, u_i\frac{I^\dagger_i}{I_i}\frac{E_i}{E^\dagger_i} -  \frac{S _i}{S^\dagger_i}\left(u_i \frac{E^\dagger_i}{E_i} + \bar u_i\frac{I^\dagger _i}{I_i}\right)\frac{I_q}{I^\dagger_q}  - \frac{R^\dagger_i}{R_i} \frac{I_i}{I^\dagger_i}\right)\\&&  + \,{\dsum_{i\in\mc R}} \varpi_i(\zeta_i+\widetilde\nu_i)R^\dagger_i\left(2 + u_i + \frac{I_q}{I^\dagger_q}   -\, u_i\frac{I^\dagger_i}{I_i}\frac{E_i}{E^\dagger_i} -  \frac{S _i}{S^\dagger_i}\left(u_i \frac{E^\dagger_i}{E_i} + \bar u_i\frac{I^\dagger _i}{I_i}\right)\frac{I_q}{I^\dagger_q}  -  \frac{S^\dagger_i}{S_i}\frac{R_i}{R^\dagger_i} - \frac{R^\dagger_i}{R_i} \frac{I_i}{I^\dagger_i}\right)  -\,{\dsum_{i=0}^{n}}\varpi_i\left(\nu_i - \widetilde{\nu}_i\right)H_i.
\end{array}
\end{equation*}
We choose $\widetilde\sigma_r=\sigma^{(1)}_r=1$ to make the initial terms vanish.
We also  choose $\varpi_i = \frac{a}{H}\frac{\phi_i}{\nu_i - \widetilde \nu_i}S^\dagger _q$, so that 
 $r_{suc}  S^\dagger _q = {\dsum_{i=0}^n}\varpi_i(\nu_i - \widetilde\nu_i)H_i$. Using the fact that $$r_{inf}  = \frac{a}{H}{\dsum_{i=0}^n}\phi_i(\xi_iI_i+\widetilde\xi_iv_iR_i) =  \frac{a}{H}\left({\dsum_{i\in\mc R}}\phi_i(\xi_iI_i) +  {\dsum_{i\in\overline{\mc R}}}\phi_i(\xi_iI_i+\widetilde\xi_iR_i)\right)$$ and $$\begin{array}{rcl}r_{suc}  &= &\frac{a}{H}{\dsum_{i=0}^n}\phi_i\left(S_i+u_iE_i+I_i+v_iR_i\right)\\& =&  \frac{a}{H}\left({\dsum_{i\in\mc E\cap\mc R}}\phi_i\left(S_i+E_i+I_i+R_i\right) +  {\dsum_{i\in\mc E\cap\overline{\mc R}}}\phi_i\left(S_i+E_i+I_i\right) +  {\dsum_{i\in\overline{\mc E}\cap\mc R}}\phi_i\left(S_i+I_i+R_i\right) + {\dsum_{i\in\overline{\mc E}\cap\overline{\mc R}}}\phi_i\left(S_i+I_i\right)\right)\end{array}$$, the above becomes: 

\begin{equation*}
\begin{array}{rcl}
F(\mathbf x)  &=& \delta S^\dagger _r  \left( \frac{S^\dagger _q}{S_q} + \frac{S_r}{S_r^\dagger }-\frac{S^\dagger _q}{S_q}\frac{S_r}{S_r^\dagger }-1\right) +\, \frac{\varpiVar  }{(f_qf_r)^{\ell}}I_q^\dagger  \left(2 - \frac{I_r}{I_r^\dagger }\frac{I_q^\dagger }{I_q}- \frac{I_q}{I_q^\dagger }\frac{I^\dagger _r}{I_r}\right) +\frac{a}{H}{\dsum_{i=0}^n}\phi_i \left(S^\dagger_i+u_iE^\dagger_i+I^\dagger_i+v_iR^\dagger_i\right)  S_q^\dagger \left(1-\frac{S^\dagger _q}{S_q}\right) \\&& +\,\frac{a}{H}{\dsum_{i=0}^n}\phi_i\left(S^\dagger_i+E^\dagger_i+\overline{\xi_i} I^\dagger_i+\overline{\widetilde\xi_i}v_i R^\dagger_i\right)S^\dagger_q  \left(1-\frac{S_r}{S_r^\dagger }\right)-\;\frac{a}{H}{\dsum_{i=0}^n}\phi_i \left(S_i+E_i+\overline{\xi_i} I_i+\overline{\widetilde\xi_i} v_iR_i\right)S_q \frac{S^\dagger _r}{S_r}  \\&& - \, \frac{a}{H}{\dsum_{i=0}^n}\phi_i(\xi_iI_i+\widetilde\xi_iv_iR_i)  S_q \frac{E^{(1)\dagger }_r}{E^{(1)}_r}+\, \frac{a}{H}{\dsum_{i=0}^n}\phi_i(\xi_iI^\dagger_i+\widetilde\xi_iv_iR^\dagger_i)  S^\dagger _q\left(2l+2- {\dsum_{j=1}^{\ell}}\frac{E^{(j)}_q}{E^{(j)\dagger }_q}\frac{E^{(j+1)\dagger }_r}{E^{(j+1)}_r}-\, {\dsum_{j=1}^{\ell}}\frac{ E^{(j)}_r}{ E^{(j)\dagger }_r}\frac{E^{(j)\dagger }_q}{E^{(j)}_q}- \frac{E^{(l+1)}_r}{E^{(\ell+1)\dagger }_r}\frac{I^\dagger _q}{I_q} - \frac{I_q}{I^\dagger _q}\right) \\&& +\,\frac{a}{H}{\dsum_{i=0}^{n}}\phi_iS^\dagger_qS^\dagger _i \left(2-\frac{S^\dagger _i}{S_i} \right)  + \,\frac{a}{H}{\dsum_{i\in\mc E}}\phi_iS^\dagger_qE^\dagger _i\left(2 + \frac{I_q}{I^\dagger_q} -\frac{S^\dagger _i}{S_i}   -  \frac{S _i}{S^\dagger_i}\frac{E^\dagger_i}{E_i} \frac{I_q}{I^\dagger_q}\right)   + \,\frac{a}{H}{\dsum_{i=0}^{n}}\phi_iS^\dagger_q I^\dagger_i\left(2 + u_i  + \frac{I_q}{I^\dagger_q} -\frac{S^\dagger _i}{S_i}  - u_i\frac{I^\dagger_i}{I_i} \frac{E_i}{E^\dagger_i}  \right. \\&& \left. -\,  \frac{S _i}{S^\dagger_i}\left(u_i \frac{E^\dagger_i}{E_i} + \bar u_i\frac{I^\dagger _i}{I_i}\right)\frac{I_q}{I^\dagger_q}\right) + \,\frac{a}{H}{\dsum_{i\in\mc R}}\phi_iS^\dagger_q R^\dagger _i\left(3 + u_i + \frac{I_q}{I^\dagger_q} - \frac{S^\dagger _i}{S_i}  -\, u_i\frac{I^\dagger_i}{I_i}\frac{E_i}{E^\dagger_i} -  \frac{S _i}{S^\dagger_i}\left(u_i \frac{E^\dagger_i}{E_i} + \bar u_i\frac{I^\dagger _i}{I_i}\right)\frac{I_q}{I^\dagger_q}  - \frac{R^\dagger_i}{R_i} \frac{I_i}{I^\dagger_i}\right) \\&&  +\,{\dsum_{i\in\mc E}} \varpi_i \widetilde\nu_iE^\dagger_i\left(1 + \frac{I_q}{I^\dagger_q} -\frac{S^\dagger _i}{S_i}\frac{E_i}{E^\dagger_i} -  \frac{S _i}{S^\dagger_i}\frac{E^\dagger_i}{E_i}\frac{I_q}{I^\dagger_q} \right)  +\,{\dsum_{i=0}^n} \varpi_i \widetilde\nu_i I^\dagger _i\left(1 + u_i + \frac{I_q}{I^\dagger_q} -\frac{S^\dagger _i}{S_i}\frac{I_i}{I^\dagger_i}  - u_i\frac{I^\dagger_i}{I_i}\frac{E_i}{E^\dagger_i} -  \frac{S _i}{S^\dagger_i}\left(u_i \frac{E^\dagger_i}{E_i} + \bar u_i\frac{I^\dagger _i}{I_i}\right)\frac{I_q}{I^\dagger_q} \right)  \\&& + \,{\dsum_{i\in\overline{\mc R}}} \varpi_i\gamma_i I^\dagger_i\left(1 + u_i + \frac{I_q}{I^\dagger_q} - \frac{S^\dagger _i}{S_i}\frac{I_i}{I^\dagger_i} -\, u_i\frac{I^\dagger_i}{I_i}\frac{E_i}{E^\dagger_i} -  \frac{S _i}{S^\dagger_i}\left(u_i \frac{E^\dagger_i}{E_i} + \bar u_i\frac{I^\dagger _i}{I_i}\right)\frac{I_q}{I^\dagger_q} \right) \\&&  + \,{\dsum_{i\in\mc R}} \varpi_i(\zeta_i+\widetilde\nu_i)R^\dagger_i\left(2 + u_i + \frac{I_q}{I^\dagger_q}   -\, u_i\frac{I^\dagger_i}{I_i}\frac{E_i}{E^\dagger_i} -  \frac{S _i}{S^\dagger_i}\left(u_i \frac{E^\dagger_i}{E_i} + \bar u_i\frac{I^\dagger _i}{I_i}\right)\frac{I_q}{I^\dagger_q}  -  \frac{S^\dagger_i}{S_i}\frac{R_i}{R^\dagger_i} - \frac{R^\dagger_i}{R_i} \frac{I_i}{I^\dagger_i}\right).              
\end{array}
\end{equation*}

\begin{equation*}
\begin{array}{rcl}
F(\mathbf x) 
&=& \delta S^\dagger _r  \left( \frac{S^\dagger _q}{S_q}+\frac{S_r}{S_r^\dagger }-\frac{S^\dagger _q}{S_q}\frac{S_r}{S_r^\dagger }-1\right)  +\, \frac{\varpiVar }{(f_qf_r)^{\ell}} I_q^\dagger \left(2 - \frac{I_r}{I_r^\dagger }\frac{I_q^\dagger }{I_q}- \frac{I_q}{I_q^\dagger }\frac{I^\dagger _r}{I_r}\right) +\,\frac{ a}{H}{\dsum_{i=0}^n}\phi_iS^\dagger _iS^\dagger _q \left(4 - \frac{S^\dagger _q}{S_q} -\frac{S^\dagger _i}{S_i} -\frac{S_i}{S^\dagger _i} \frac{S_q}{S^\dagger _q}\frac{S^\dagger _r}{S_r}-\frac{S_r}{S^\dagger _r}\right)\\&& +\,\frac{ a}{H}{\dsum_{i\in\mc E}}\phi_iE^\dagger _iS^\dagger _q \left(4 + \frac{I_q}{I^\dagger_q} - \frac{S^\dagger _q}{S_q} -\frac{S^\dagger _i}{S_i} -\frac{E_i}{E^\dagger _i} \frac{S_q}{S^\dagger _q}\frac{S^\dagger _r}{S_r}-\frac{S_r}{S^\dagger _r}-\frac{I_q}{I^\dagger_q}\frac{S_i}{S^\dagger_i}\frac{E^\dagger_i}{E_i}\right) +\,\frac{ a}{H}{\dsum_{i=0}^n}\phi_i\overline{\xi_i}I^\dagger _iS^\dagger _q\left(u_i\left(5 + \frac{I_q}{I^\dagger_q} - \frac{S^\dagger _i}{S_i}-\frac{I^\dagger_i}{I_i} \frac{E_i}{E^\dagger_i}-\,\frac{S^\dagger _q}{S_q}\right.\right.\\&& \left.\left. -\,\frac{S_r}{S^\dagger _r}  -\,\frac{I_i}{I^\dagger _i}\frac{S_q}{S^\dagger _q}\frac{S^\dagger _r}{S_r} - \frac{I_q}{I^\dagger _q}\frac{S_i}{S^\dagger_i}\frac{E^\dagger_i}{E_i}\right) + \bar u_i\left(4 + \frac{I_q}{I_q^\dagger} - \frac{S^\dagger _i}{S_i}  -\frac{S^\dagger _q}{S_q}  -\frac{I_i}{I^\dagger _i}\frac{S_q}{S^\dagger _q}\frac{S^\dagger _r}{S_r}-\,\frac{S_r}{S^\dagger _r} - \frac{I_q}{I_q^\dagger} \frac{S_i}{S_i^\dagger} \frac{I^\dagger_i}{I_i}\right)\right) +\,\frac{ a}{H}{\dsum_{i\in\mc R}}\phi_i\overline{\widetilde\xi_i}R^\dagger _iS^\dagger _q\left(u_i\left(6 + \frac{I_q}{I^\dagger_q}\right.\right.\\&& \left.\left. - \frac{S^\dagger _i}{S_i} -\frac{I^\dagger_i}{I_i} \frac{E_i}{E^\dagger_i} -\frac{S^\dagger _q}{S_q} -\,\frac{S_r}{S^\dagger _r}-\,\frac{R_i}{R^\dagger _i}\frac{S_q}{S^\dagger _q}\frac{S^\dagger _r}{S_r}  -\frac{I_i}{I^\dagger_i} \frac{R^\dagger_i}{R_i} - \frac{I_q}{I^\dagger _q}\frac{S_i}{S^\dagger_i} \frac{E^\dagger_i}{E_i} \right) + \bar u_i\left(5 + \frac{I_q}{I_q^\dagger} - \frac{S^\dagger _i}{S_i}  -\frac{S^\dagger _q}{S_q}  -\,\frac{S_r}{S^\dagger_r} -\frac{R_i}{R^\dagger _i}\frac{S_q}{S^\dagger _q}\frac{S^\dagger _r}{S_r} - \frac{R^\dagger_i}{R_i} \frac{I_i}{I^\dagger_i} - \frac{I_q}{I_q^\dagger} \frac{S_i}{S_i^\dagger} \frac{I^\dagger_i}{I_i} \right)\right) \\&& +\,\frac{a}{H}{\dsum_{i=0}^n}\phi_i\xi_iI^\dagger _iS^\dagger _q\left(u_i\left(2\ell+6  - \frac{S^\dagger _i}{S_i}-\frac{I^\dagger_i}{I_i} \frac{E_i}{E^\dagger_i} -\frac{S^\dagger _q}{S_q} -\; \frac{I_i}{I^\dagger _i}\frac{S_q}{S^\dagger _q}\frac{E^{(1)\dagger }_r}{E^{(1)}_r}- {\dsum_{j=1}^{\ell}}\frac{E^{(j)}_q}{E^{(j)\dagger }_q}\frac{E^{(j+1)\dagger }_r}{E^{(j+1)}_r}-\, {\dsum_{j=1}^{\ell}}\frac{ E^{(j)}_r}{ E^{(j)\dagger }_r}\frac{E^{(j)\dagger }_q}{E^{(j)}_q} -\frac{E^{(l+1)}_r}{E^{(l+1)\dagger }_r}\frac{I^\dagger _q}{I_q}- \frac{I_q}{I^\dagger _q}\frac{S_i}{S^\dagger_i} \frac{E^\dagger_i}{E_i} \right)\right. \\&& \left.+\, \bar u_i\left(2\ell+5  - \frac{S^\dagger _i}{S_i} -\frac{S^\dagger _q}{S_q} -\; \frac{I_i}{I^\dagger _i}\frac{S_q}{S^\dagger _q}\frac{E^{(1)\dagger }_r}{E^{(1)}_r}- {\dsum_{j=1}^{l}}\frac{E^{(j)}_q}{E^{(j)\dagger }_q}\frac{E^{(j+1)\dagger }_r}{E^{(j+1)}_r}-\, {\dsum_{j=1}^{l}}\frac{ E^{(j)}_r}{ E^{(j)\dagger }_r}\frac{E^{(j)\dagger }_q}{E^{(j)}_q} -\frac{E^{(l+1)}_r}{E^{(l+1)\dagger }_r}\frac{I^\dagger _q}{I_q}- \frac{I_q}{I^\dagger _q}\frac{S_i}{S^\dagger_i}\frac{I^\dagger_i}{I_i}\right)\right)  \\&& +\,\frac{ a}{H}{\dsum_{i\in\mc R}}\phi_i\widetilde\xi_iR^\dagger _iS^\dagger _q\left(u_i\left(2\ell+7 - \frac{S^\dagger _i}{S_i} -\frac{I^\dagger_i}{I_i} \frac{E_i}{E^\dagger_i} -\frac{I_i}{I^\dagger_i} \frac{R^\dagger_i}{R_i} -\frac{S^\dagger _q}{S_q} -\; \frac{R_i}{R^\dagger _i}\frac{S_q}{S^\dagger _q}\frac{E^{(1)\dagger }_r}{E^{(1)}_r}- {\dsum_{j=1}^{l}}\frac{E^{(j)}_q}{E^{(j)\dagger }_q}\frac{E^{(j+1)\dagger }_r}{E^{(j+1)}_r}-\, {\dsum_{j=1}^{\ell}}\frac{ E^{(j)}_r}{ E^{(j)\dagger }_r}\frac{E^{(j)\dagger }_q}{E^{(j)}_q} -\frac{E^{(l+1)}_r}{E^{(\ell+1)\dagger }_r}\frac{I^\dagger _q}{I_q} \right.\right.\\&& \left.\left.-\, \frac{I_q}{I^\dagger _q}\frac{S_i}{S^\dagger_i}\frac{E^\dagger_i}{E_i}\right) +\bar u_i\left(2\ell+6 - \frac{S^\dagger _i}{S_i} -\frac{I_i}{I^\dagger_i} \frac{R^\dagger_i}{R_i} -\frac{S^\dagger _q}{S_q} -\; \frac{R_i}{R^\dagger _i}\frac{S_q}{S^\dagger _q}\frac{E^{(1)\dagger }_r}{E^{(1)}_r}- {\dsum_{j=1}^{l}}\frac{E^{(j)}_q}{E^{(j)\dagger }_q}\frac{E^{(j+1)\dagger }_r}{E^{(j+1)}_r}-\, {\dsum_{j=1}^{l}}\frac{ E^{(j)}_r}{ E^{(j)\dagger }_r}\frac{E^{(j)\dagger }_q}{E^{(j)}_q} -\frac{E^{(l+1)}_r}{E^{(\ell+1)\dagger }_r}\frac{I^\dagger _q}{I_q}- \frac{I_q}{I^\dagger _q}\frac{S_i}{S^\dagger_i}\frac{I^\dagger_i}{I_i}\right)\right) \\&&   +\,{\dsum_{i\in\mc E}} \varpi_i \widetilde\nu_iE^\dagger_i\left(1 + \frac{I_q}{I^\dagger_q} - \frac{S^\dagger_i}{S_i} \frac{E_i}{E^\dagger_i} - \frac{I_q}{I^\dagger_q}\frac{S_i}{S^\dagger_i}\frac{E^\dagger_i}{E_i}\right)  + \,{\dsum_{i\in\overline{\mc R}}}\varpi_i (\gamma_i + \widetilde\nu_i)I^\dagger_i\left(u_i\left(2 + \frac{I_q}{I^\dagger_q} -\frac{S^\dagger _i}{S_i} \frac{I_i}{I^\dagger_i} -\frac{I^\dagger_i}{I_i}\frac{E_i}{E^\dagger_i} - \frac{I _q}{I_q^\dagger}\frac{S_i}{S^\dagger_i}\frac{E^\dagger _i}{E_i}\right) + \bar u_i\left(1 + \frac{I_q}{I^\dagger_q} \right.\right. \\&&\left.\left.-\frac{S^\dagger _i}{S_i} \frac{I_i}{I^\dagger_i}  - \,\frac{I _q}{I_q^\dagger}\frac{S_i}{S^\dagger_i}\frac{I^\dagger _i}{I_i}\right)\right) + \,{\dsum_{i\in\mc R}}\varpi_i \widetilde\nu_iI^\dagger_i\left(u_i\left(2 + \frac{I_q}{I^\dagger_q} -\frac{S^\dagger _i}{S_i} \frac{I_i}{I^\dagger_i} -\frac{I^\dagger_i}{I_i}\frac{E_i}{E^\dagger_i} - \frac{I _q}{I_q^\dagger}\frac{S_i}{S^\dagger_i}\frac{E^\dagger _i}{E_i}\right) + \bar u_i\left(1 + \frac{I_q}{I^\dagger_q} -\frac{S^\dagger _i}{S_i} \frac{I_i}{I^\dagger_i}  - \,\frac{I _q}{I_q^\dagger}\frac{S_i}{S^\dagger_i}\frac{I^\dagger _i}{I_i}\right)\right) \\&&   +\,{\dsum_{i\in\mc R}} \varpi_i\left(\zeta_i+\widetilde \nu_i\right)R^\dagger_i\left(u_i\left(3 + \frac{I_q}{I^\dagger_q} -\frac{I^\dagger_i}{I_i}\frac{E_i}{E^\dagger_i} - \frac{I_i}{I^\dagger_i}\frac{R^\dagger_i}{R_i}  - \frac{R_i}{R^\dagger_i}\frac{S^\dagger _i}{S_i} - \frac{I _q}{I_q^\dagger}\frac{S_i}{S^\dagger_i}\frac{E^\dagger _i}{E_i}\right) +\bar u_i\left(2 + \frac{I_q}{I^\dagger_q}  - \frac{I_i}{I^\dagger_i}\frac{R^\dagger_i}{R_i}  - \frac{R_i}{R^\dagger_i}\frac{S^\dagger _i}{S_i} - \frac{I _q}{I_q^\dagger}\frac{S_i}{S^\dagger_i}\frac{I^\dagger _i}{I_i}\right)\right)  
\end{array}
\end{equation*}
where $\overline{\xi_i}=1-\xi_i$ and $\overline{\widetilde\xi_i}=1-\widetilde\xi_i$, $i=0,\;\cdots,\;n$. Setting $\hat \nu_i = \nu_i - \widetilde \nu_i$ and replacing $H_i$ by $S^\dagger _i+I^\dagger _i$ for $i=0,\;\cdots,\;n$ gives finally
\begin{equation*}
\begin{array}{rcl}
F(\mathbf x) 
&=& \delta S^\dagger _r  \left( \frac{S^\dagger _q}{S_q}+\frac{S_r}{S_r^\dagger }-\frac{S^\dagger _q}{S_q}\frac{S_r}{S_r^\dagger }-1\right)  +\, \frac{r_{suc} }{(f_qf_r)^{\ell}} I_q^\dagger \left(2 - \frac{I_r}{I_r^\dagger }\frac{I_q^\dagger }{I_q}- \frac{I_q}{I_q^\dagger }\frac{I^\dagger _r}{I_r}\right) +\,{\dsum_{i=0}^n}\varpi_i\hat\nu_iS^\dagger _i \left(4 - \frac{S^\dagger _q}{S_q} -\frac{S^\dagger _i}{S_i} -\frac{S_i}{S^\dagger _i} \frac{S_q}{S^\dagger _q}\frac{S^\dagger _r}{S_r}-\frac{S_r}{S^\dagger _r}\right)\\&& +\,{\dsum_{i=0}^n}\varpi_i\hat\nu_iE^\dagger _i \left(4 + \frac{I_q}{I^\dagger_q} - \frac{S^\dagger _q}{S_q} -\frac{S^\dagger _i}{S_i} -\frac{E_i}{E^\dagger _i} \frac{S_q}{S^\dagger _q}\frac{S^\dagger _r}{S_r}-\frac{S_r}{S^\dagger _r}-\frac{I_q}{I^\dagger_q}\frac{S_i}{S^\dagger_i}\frac{E^\dagger_i}{E_i}\right) +\,{\dsum_{i=0}^n}\varpi_i\hat\nu_i\overline{\xi_i}I^\dagger _i\left(u_i\left(5 + \frac{I_q}{I^\dagger_q} - \frac{S^\dagger _i}{S_i}-\frac{I^\dagger_i}{I_i} \frac{E_i}{E^\dagger_i}-\,\frac{S^\dagger _q}{S_q}\right.\right.\\&& \left.\left. -\,\frac{S_r}{S^\dagger _r}  -\,\frac{I_i}{I^\dagger _i}\frac{S_q}{S^\dagger _q}\frac{S^\dagger _r}{S_r} - \frac{I_q}{I^\dagger _q}\frac{S_i}{S^\dagger_i}\frac{E^\dagger_i}{E_i}\right) + \bar u_i\left(4 + \frac{I_q}{I_q^\dagger} - \frac{S^\dagger _i}{S_i}  -\frac{S^\dagger _q}{S_q}  -\frac{I_i}{I^\dagger _i}\frac{S_q}{S^\dagger _q}\frac{S^\dagger _r}{S_r}-\,\frac{S_r}{S^\dagger _r} - \frac{I_q}{I_q^\dagger} \frac{S_i}{S_i^\dagger} \frac{I^\dagger_i}{I_i}\right)\right) +\,{\dsum_{i\in\mc R}}\varpi_i\hat\nu_i\overline{\widetilde\xi_i}R^\dagger _i\left(u_i\left(6 + \frac{I_q}{I^\dagger_q}\right.\right.\\&& \left.\left. - \frac{S^\dagger _i}{S_i} -\frac{I^\dagger_i}{I_i} \frac{E_i}{E^\dagger_i} -\frac{S^\dagger _q}{S_q} -\,\frac{S_r}{S^\dagger _r}-\,\frac{R_i}{R^\dagger _i}\frac{S_q}{S^\dagger _q}\frac{S^\dagger _r}{S_r}  -\frac{I_i}{I^\dagger_i} \frac{R^\dagger_i}{R_i} - \frac{I_q}{I^\dagger _q}\frac{S_i}{S^\dagger_i} \frac{E^\dagger_i}{E_i} \right) + \bar u_i\left(5 + \frac{I_q}{I_q^\dagger} - \frac{S^\dagger _i}{S_i}  -\frac{S^\dagger _q}{S_q}  -\,\frac{S_r}{S^\dagger_r} -\frac{R_i}{R^\dagger _i}\frac{S_q}{S^\dagger _q}\frac{S^\dagger _r}{S_r} - \frac{R^\dagger_i}{R_i} \frac{I_i}{I^\dagger_i} - \frac{I_q}{I_q^\dagger} \frac{S_i}{S_i^\dagger} \frac{I^\dagger_i}{I_i} \right)\right) \\&& +\,{\dsum_{i=0}^n}\varpi_i\hat\nu_i\xi_iI^\dagger _i\left(u_i\left(2\ell+6  - \frac{S^\dagger _i}{S_i}-\frac{I^\dagger_i}{I_i} \frac{E_i}{E^\dagger_i} -\frac{S^\dagger _q}{S_q} -\; \frac{I_i}{I^\dagger _i}\frac{S_q}{S^\dagger _q}\frac{E^{(1)\dagger }_r}{E^{(1)}_r}- {\dsum_{j=1}^{\ell}}\frac{E^{(j)}_q}{E^{(j)\dagger }_q}\frac{E^{(j+1)\dagger }_r}{E^{(j+1)}_r}-\, {\dsum_{j=1}^{\ell}}\frac{ E^{(j)}_r}{ E^{(j)\dagger }_r}\frac{E^{(j)\dagger }_q}{E^{(j)}_q} -\frac{E^{(l+1)}_r}{E^{(l+1)\dagger }_r}\frac{I^\dagger _q}{I_q}- \frac{I_q}{I^\dagger _q}\frac{S_i}{S^\dagger_i} \frac{E^\dagger_i}{E_i} \right)\right. \\&& \left.+\, \bar u_i\left(2\ell+5  - \frac{S^\dagger _i}{S_i} -\frac{S^\dagger _q}{S_q} -\; \frac{I_i}{I^\dagger _i}\frac{S_q}{S^\dagger _q}\frac{E^{(1)\dagger }_r}{E^{(1)}_r}- {\dsum_{j=1}^{l}}\frac{E^{(j)}_q}{E^{(j)\dagger }_q}\frac{E^{(j+1)\dagger }_r}{E^{(j+1)}_r}-\, {\dsum_{j=1}^{l}}\frac{ E^{(j)}_r}{ E^{(j)\dagger }_r}\frac{E^{(j)\dagger }_q}{E^{(j)}_q} -\frac{E^{(l+1)}_r}{E^{(l+1)\dagger }_r}\frac{I^\dagger _q}{I_q}- \frac{I_q}{I^\dagger _q}\frac{S_i}{S^\dagger_i}\frac{I^\dagger_i}{I_i}\right)\right)  \\&& +\,{\dsum_{i\in\mc R}}\varpi_i\hat\nu_i\widetilde\xi_iR^\dagger _i \left(u_i\left(2\ell+7 - \frac{S^\dagger _i}{S_i} -\frac{I^\dagger_i}{I_i} \frac{E_i}{E^\dagger_i} -\frac{I_i}{I^\dagger_i} \frac{R^\dagger_i}{R_i} -\frac{S^\dagger _q}{S_q} -\; \frac{R_i}{R^\dagger _i}\frac{S_q}{S^\dagger _q}\frac{E^{(1)\dagger }_r}{E^{(1)}_r}- {\dsum_{j=1}^{l}}\frac{E^{(j)}_q}{E^{(j)\dagger }_q}\frac{E^{(j+1)\dagger }_r}{E^{(j+1)}_r}-\, {\dsum_{j=1}^{\ell}}\frac{ E^{(j)}_r}{ E^{(j)\dagger }_r}\frac{E^{(j)\dagger }_q}{E^{(j)}_q} -\frac{E^{(l+1)}_r}{E^{(\ell+1)\dagger }_r}\frac{I^\dagger _q}{I_q} \right.\right.\\&& \left.\left.-\, \frac{I_q}{I^\dagger _q}\frac{S_i}{S^\dagger_i}\frac{E^\dagger_i}{E_i}\right) +\bar u_i\left(2\ell+6 - \frac{S^\dagger _i}{S_i} -\frac{I_i}{I^\dagger_i} \frac{R^\dagger_i}{R_i} -\frac{S^\dagger _q}{S_q} -\; \frac{R_i}{R^\dagger _i}\frac{S_q}{S^\dagger _q}\frac{E^{(1)\dagger }_r}{E^{(1)}_r}- {\dsum_{j=1}^{l}}\frac{E^{(j)}_q}{E^{(j)\dagger }_q}\frac{E^{(j+1)\dagger }_r}{E^{(j+1)}_r}-\, {\dsum_{j=1}^{l}}\frac{ E^{(j)}_r}{ E^{(j)\dagger }_r}\frac{E^{(j)\dagger }_q}{E^{(j)}_q} -\frac{E^{(l+1)}_r}{E^{(\ell+1)\dagger }_r}\frac{I^\dagger _q}{I_q}- \frac{I_q}{I^\dagger _q}\frac{S_i}{S^\dagger_i}\frac{I^\dagger_i}{I_i}\right)\right) \\&&   +\,{\dsum_{i\in\mc E}} \varpi_i \widetilde\nu_iE^\dagger_i\left(1 + \frac{I_q}{I^\dagger_q} - \frac{S^\dagger_i}{S_i} \frac{E_i}{E^\dagger_i} - \frac{I_q}{I^\dagger_q}\frac{S_i}{S^\dagger_i}\frac{E^\dagger_i}{E_i}\right)  \\&& + \,{\dsum_{i=0}^n}\varpi_i(\bar v_i\gamma_i + \widetilde\nu_i)I^\dagger_i\left(u_i\left(2 + \frac{I_q}{I^\dagger_q} -\frac{S^\dagger _i}{S_i} \frac{I_i}{I^\dagger_i} -\frac{I^\dagger_i}{I_i}\frac{E_i}{E^\dagger_i} - \frac{I _q}{I_q^\dagger}\frac{S_i}{S^\dagger_i}\frac{E^\dagger _i}{E_i}\right) + \bar u_i\left(1 + \frac{I_q}{I^\dagger_q} -\frac{S^\dagger _i}{S_i} \frac{I_i}{I^\dagger_i}  - \,\frac{I _q}{I_q^\dagger}\frac{S_i}{S^\dagger_i}\frac{I^\dagger _i}{I_i}\right)\right) \\&&   +\,{\dsum_{i\in\mc R}} \varpi_i\left(\zeta_i+\widetilde \nu_i\right)R^\dagger_i\left(u_i\left(3 + \frac{I_q}{I^\dagger_q} -\frac{I^\dagger_i}{I_i}\frac{E_i}{E^\dagger_i} - \frac{I_i}{I^\dagger_i}\frac{R^\dagger_i}{R_i}  - \frac{R_i}{R^\dagger_i}\frac{S^\dagger _i}{S_i} - \frac{I _q}{I_q^\dagger}\frac{S_i}{S^\dagger_i}\frac{E^\dagger _i}{E_i}\right) +\bar u_i\left(2 + \frac{I_q}{I^\dagger_q}  - \frac{I_i}{I^\dagger_i}\frac{R^\dagger_i}{R_i}  - \frac{R_i}{R^\dagger_i}\frac{S^\dagger _i}{S_i} - \frac{I _q}{I_q^\dagger}\frac{S_i}{S^\dagger_i}\frac{I^\dagger _i}{I_i}\right)\right). 
\end{array}
\end{equation*}
which together with \eqref{eq:Fexp} yields expression \eqref{eq:eqliapeeder2}.

\section{Nonstandard finite difference scheme}\label{subsec.nsfd}
The Nonstandard finite difference scheme uses for the simulations is:

\begin{equation}
   \left\{
    \begin{aligned}
    	\frac{S^{p+1}_i - S^{p}_i}{\phi(\delta t)} &~~=~~\Lambda_i+\widetilde{\nu_i}H^{p}_i - \nu_iS^{p}_i- a\,\miVar \frac{I^{p}_q}{H^{p}}S^{p+1}_i +\bar v_i\gamma_i I^{p}_i  + v_i\zeta_i R^{p}_i & i = 0,\;1,\;\cdots,\; n \\
        \frac{S^{p+1}_q - S^{p}_q}{\phi(\delta t)}&~~=~~\Gamma -\mu S^{p}_q- (d^{p}+\varpiVar ^{p}) S^{p+1}_q  + \delta S^{p}_r  & \,\\
        \frac{S^{p+1}_r - S^{p}_r}{\phi(\delta t)}&~~=~~(\varpiVar ^{p}-\varphiVar ^p) S^{p+1}_q    -  (\mu+\delta) S^{p}_r & \,\\
        \frac{E^{(1)p+1}_r - E^{(1)p}_r}{\phi(\delta t)}&~~=~~\varphiVar ^p  S^{p+1}_q-(\mu+\delta) E^{(1)p}_r& \,\\
       \frac{E^{p+1}_i - E^{p}_i}{\phi(\delta t)}&~~=~~u_ia\,\miVar \frac{I^{p}_q}{H^{p}}S^{p+1}_i -u_i\left(\varepsilon_i+\nu_i\right) E^{p}_i & 
i = 0,\;1,\;\cdots,\; n \,\\
       \frac{E^{(j)p+1}_q - E^{(j)p}_q}{\phi(\delta t)}&~~=~~\delta E^{(j)p}_r - \mu E^{(j)p}_q - (d^{p}+\varpiVar ^{p})E^{(j)p+1}_q &
j = 1,\;2,\;\cdots,\; l  \,\\
       \frac{E^{(j+1)p+1}_r - E^{(j+1)p}_r}{\phi(\delta t)}&~~=~~\varpiVar ^{p}  E^{(j)p+1}_q-(\mu+\delta)
E^{(i+1)p}_r& j = 1,\;2,\;\cdots,\; l  \,\\
       \frac{I^{p+1}_i - I^{p}_i}{\phi(\delta t)}&~~=~~\bar u_ia\,\miVar \frac{I^{p}_q}{H^{p}}S^{p+1}_i +u_i\varepsilon_iE^{p}_i -\left(\gamma_i+\nu_i\right) I^{p}_i & 
 i = 0,\;1,\;\cdots,\; n \,\\
  \frac{R^{p+1}_i - R^{p}_i}{\phi(\delta t)}&~~=~~v_i\gamma_iI^{p}_i  -v_i\left(\zeta_i+\nu_i\right) R^{p}_i & 
 i = 0,\;1,\;\cdots,\; n \,\\
       \frac{I^{p+1}_q - I^{p}_q}{\phi(\delta t)}&~~=~~\delta E^{(l+1)p}_r -\mu I^{p}_q- (d^{p}+\varpiVar ^{p}) I^{p+1}_q +\delta
I^{p}_r& \,\\
       \frac{I^{p+1}_r - I^{p}_r}{\phi(\delta t)}&~~=~~\varpiVar ^{p} I^{p+1}_q - (\delta+\mu)I^{p}_r
    \end{aligned}
    \right. 
     \label{eq:eqbednetnfds}
\end{equation} 
where 
\[
\varphiVar ^p = \frac{a}{H^{p}}{\dsum_{i=1}^n}f_ic_i\bar k_iI^p_i;\quad \varpiVar ^p = \frac{a}{H^{p}}{\dsum_{i=1}^n}f_ic_i\bar k_iH^p_i; \quad d^p = \frac{a}{H^{p}}{\dsum_{i=1}^n} k_iH^p_i.
\] 
The time step function $\phi(t)$  is defined as 
\[
\phi(t) = \dfrac{1-\mathrm e^{-th}}{h}, \textrm{~~with~} h = \max(\delta+\mu,\;\nu_1+\gamma_1,\;\cdots,\;\nu_n+\gamma_n,\;u_1(\varepsilon_1+\nu_1),\;\cdots,\; u_n(\varepsilon_n+\nu_n) ,\; v_1(\zeta_1+\nu_1),\;\cdots,\; v_n(\zeta_n+\nu_n)).
\]
Solving~\eqref{eq:eqbednetnfds} for  the $p+1^{th}$ terms give the 
following semi-implicit system of  difference equations:
\begin{equation}
   \left\{
    \begin{aligned}
S^{p+1}_i  &~=~\frac{\phi(\delta t)\Lambda_i}{1+ \phi(\delta t)a \miVar  \frac{I^{p}_q}{H^{p}}}+\frac{1-\phi(\delta t)\left(\nu_i-\widetilde\nu_i\right)}{1+ \phi(\delta t)a \miVar  \frac{I^{p}_q}{H^{p}}}S^{p}_i+\frac{
\phi(\delta t)\widetilde \nu_i}{1+ \phi(\delta t)a \miVar  \frac{I^{p}_q}{H^{p}}}E^{p}_i \\ &~~~+~ \frac{
\phi(\delta t)\left(\widetilde \nu_i + \bar v_i\gamma_i \right) }{1+ \phi(\delta t)a \miVar  \frac{I^{p}_q}{H^{p}}}I^{p}_i + \frac{
\phi(\delta t) \left(\widetilde \nu_i + v_i\zeta_i \right)}{1+ \phi(\delta t)a \miVar  \frac{I^{p}_q}{H^{p}}}R^{p}_i 
 & i=0,\;1,\;\cdots,\; n \\
 S^{p+1}_q&~=~\frac{\phi(\delta t) \Gamma}{1+\phi(\delta t)(d^{p}+\varpiVar ^{p})}+\frac{1-\phi(\delta t) \mu}{1+\phi(\delta t)(d^{p}+\varpiVar ^{p})} S^{p}_q +\frac{\phi(\delta t)\delta}{1+\phi(\delta t)(d^{p}+\varpiVar ^{p})} S_r^p & \,\\
 S^{p+1}_r&~=~\frac{{\phi(\delta t)}^2(\varpiVar ^p-\varphiVar ^p) \Gamma}{1+\phi(\delta t)(d^{p}+\varpiVar ^{p})}+\frac{\phi(\delta t)(\varpiVar ^p-\varphiVar ^p)\left(1-\phi(\delta t) \mu\right)}{1+\phi(\delta t)(d^{p}+\varpiVar ^{p})} S^{p}_q \\ &~~~+~\frac{1+ \phi(\delta t)(d^p+\varpiVar ^p-\delta-\mu) -  {\phi(\delta t)}^2((\mu+\delta) d^p+\mu\varpiVar ^{p}+\delta\varphiVar ^p)}{1+\phi(\delta t)(d^{p}+\varpiVar ^{p})} S_r^p  & \,\\
E^{p+1}_i &~~=~~u_ia\,\miVar \frac{I^{p}_q}{H^{p}}\phi(\delta t)S^{p+1}_i +(1-u_i\phi(\delta t))\left(\varepsilon_i+\nu_i\right) E^{p}_i & 
i = 0,\;1,\;\cdots,\; n \,\\
E^{(1)p+1}_r&~=~\phi(\delta t)\varphiVar ^p  S^{p+1}_q + \left(1 - \phi(\delta t)(\mu+\delta)\right) E^{(1)p}_r & \,\\
      E^{(j)p+1}_q&~=~\frac{\phi(\delta t) \delta}{1+\phi(\delta t)(d^{p}+\varpiVar ^{p})} E^{(j)p}_r +\frac{1-\phi(\delta t) \mu}{1+\phi(\delta t)(d^{p}+\varpiVar ^{p})}E^{(j)p}_q &
j = 1,\;2,\;\cdots,\; l \,\\
    E^{(j+1)p+1}_r&~=~ \frac{{\phi(\delta t)}^2\varpiVar ^{p} \delta}{1+\phi(\delta t)(d^{p}+\varpiVar ^{p})} E^{(j)p}_r +\frac{\phi(\delta t)\varpiVar ^{p}(1-\phi(\delta t) \mu)}{1+\phi(\delta t)(d^{p}+\varpiVar ^{p})}E^{(j)p}_q \\ &~~~+~ \left(1-\phi(\delta t)(\mu+\delta)\right)
E^{(j+1)p}_r& \;j = 1,\;2,\;\cdots,\; l \,\\
    I^{p+1}_i&~=~\bar u_ia\miVar\frac{I^p_q}{H^p}\phi(\delta t)S^{p+1}_i  +u_i\varepsilon_i\phi(\delta t)E^p_i + \left(1-\phi(\delta t)(\nu_i+\gamma_i)\right)I^p_i  &
 i = 0,\;1,\;\cdots,\; n \,\\
R^{p+1}_i&~~=~~v_i\gamma_i\phi(\delta t)I^{p}_i + (1 -v_i\left(\zeta_i+\nu_i\right)\phi(\delta t)) R^{p}_i & 
i = 0,\;1,\;\cdots,\; n \,\\
     I^{p+1}_q&~=~\frac{\phi(\delta t) \delta}{1+\phi(\delta t)(d^{p}+\varpiVar ^{p})} E^{(l+1)p}_r +\frac{1 - \phi(\delta t) \mu}{1+\phi(\delta t)(d^{p}+\varpiVar ^{p})} I^{p}_q +\frac{\phi(\delta t) \delta}{1+\phi(\delta t)(d^{p}+\varpiVar ^{p})}
I^{p}_r & \,\\
    I^{p+1}_r&~=~ \frac{{\phi(\delta t)}^2 \varpiVar ^{p}\delta}{1+\phi(\delta t)(d^{p}+\varpiVar ^{p})} E^{(l+1)p}_r +\frac{\phi(\delta t)\varpiVar ^{p}(1 - \phi(\delta t) \mu)}{1+\phi(\delta t)(d^{p}+\varpiVar ^{p})} I^{p}_q \\ &~~~+~\frac{1+ \phi(\delta t)(d^p+\varpiVar ^p-\delta-\mu) -  {\phi(\delta t)}^2((\mu+\delta) d^p+\mu\varpiVar ^{p}) }{1+\phi(\delta t)(d^{p}+\varpiVar ^{p})}
    I^{p}_r 
    \end{aligned}
    \right. 
     \label{eq:eqbednetnfds1}
\end{equation} 

\noindent The system  \eqref{eq:eqbednetnfds1} can be written in shortened form as 
\begin{equation}\label{eq:eqbednetnfds1_short}
\left\{
 \begin{array}{lcr}
 \mathbf x_1^{p+1}-\mathbf x_1^*& = & \mathbf A_{S,\delta t}\left(\mathbf x^{p}\right)\,.\,\left(\mathbf x_1^p - \mathbf x_1^*\right) +\mathbf A_{S,\,I,\delta t}(\mathbf x^p)\,.\,\mathbf x_2^p\\\mathbf x^{p+1}_2 & = & \mathbf A_{I,\delta t} \left(\mathbf x^{p},\;\mathbf x^{p+1}_1\right)\,.\,\mathbf x_2^p
 \end{array}
\right.  ,
\end{equation} 
where the sub-matrices in \eqref{eq:eqbednetnfds1_short} are defined as follows: First we have the $(n+2)\times (n+2)$ diagonal matrix
$$\mathbf A_{S,h}\left(\mathbf x^{p}\right) = \mathrm{diag}\left(\mathbf A_{S_{H},h}\left(\mathbf x^{p} \right) ,\;\mathbf A_{S_{V},h}\left(\mathbf x^{p} \right)\right),$$
where
\begin{align*}
\mathbf A_{S_{H},h}\left(\mathbf x^{p} \right) &= \mathrm{diag}\left(\frac{1-(\nu_i-\widetilde \nu_i)\phi(h)}{1+a\phi(h)\miVar\frac{I^p_q}{H^p}}\right)_{1\leqslant i \leqslant n}; \qquad \mathbf A_{S_{V},h}\left(\mathbf x^{p} \right) =\begin{pmatrix}
\frac{1-\phi(\delta t)\mu}{1+\phi(h)(d^p+\varpiVar ^p)} & \frac{\phi(h)\delta}{1+\phi(h)(d^p+\varpiVar ^p)}\cr \frac{\phi(h)(\varpiVar ^p-\varphiVar ^p)(1-\phi(h)\mu)}{1+\phi(h)(d^p+\varpiVar ^p)} &\xi(\mathbf x^p,\; h) \end{pmatrix};\\ 
\xi(\mathbf x^p,\;h) &=\dfrac{1+ \phi(h)(d^p+\varpiVar ^p-\delta-\mu) -  {\phi(h)}^2((\mu+\delta) d^p+\mu\varpiVar ^{p}+\delta\varphiVar ^p)}{1+\phi(h)(d^p+\varpiVar ^p)}.
\end{align*}

\noindent  The $(n+2)\times (3n+2l+3)$ matrix $\mathbf A_{S,\,I,h}(\mathbf x^p)$ may be written in block form as: $$\mathbf A_{S,\,I,h}(\mathbf x^p) = \begin{pmatrix}\mathbf A_{S,\,E_H,h}(\mathbf x^p) &\mathbf 0_{n,\,2l+1} & \mathbf A_{S,\,I_H,h}(\mathbf x^p) & \mathbf A_{S,\,R_H,h}(\mathbf x^p) & \mathbf 0_{n,\,2}\cr \mathbf 0_{2,\,n} & \mathbf 0_{2,\,2l+1} & \mathbf 0_{2,\,n} &  \mathbf 0_{2,\,n} & \mathbf 0_{2,\,2}  \end{pmatrix},$$ 
where $\mathbf A_{S,\,E_H,h}(\mathbf x^p)$, $\mathbf A_{S,\,I_H,h}(\mathbf x^p)$ and $\mathbf A_{S,\,R_H,h}(\mathbf x^p)$ are respectively
$$\ \diag\begin{pmatrix}
	\dfrac{
		u_i\phi(h)\widetilde \nu_i  }{1+ \phi(h)a \miVar  \frac{I^{p}_q}{H^{p}}}
\end{pmatrix}_{1\leqslant i\leqslant n},\  \diag\begin{pmatrix}
\dfrac{
    \phi(h)\left(\widetilde \nu_i + \bar v_i\gamma_i \right) }{1+ \phi(h)a \miVar  \frac{I^{p}_q}{H^{p}}}
\end{pmatrix}_{1\leqslant i\leqslant n}, \hbox{  and  }  \diag\begin{pmatrix}
\dfrac{
	v_i\phi(h) \left(\widetilde \nu_i + \zeta_i \right) }{1+ \phi(h)a \miVar  \frac{I^{p}_q}{H^{p}}}
\end{pmatrix}_{1\leqslant i\leqslant n}.$$
 
 
\noindent The  matrix $\mathbf A_{I,\,h}(\mathbf x^p,\;\mathbf x_1^{p+1})$ may be written in block form  as 
\begin{equation}
\mathbf A_{I,\,h}(\mathbf x^p,\;\mathbf x_1^{p+1}) =    \left(\begin{array}{cc} \mathbf A_{I_E,\,h}(\mathbf x^p)
&   \mathbf A_{I_{I,\,E},\,h}(\mathbf x^p,\;\mathbf x_1^{p+1}) \\ \mathbf A_{I_{E,\,I},\,h}(\mathbf x^p) &  \mathbf
A_{I_I,\,h}(\mathbf x^p,\;\mathbf x_1^{p+1}) \end{array}\right),
\label{eq:matia}
\end{equation}
where the four  block matrices may be described as follows: 
\medskip

\noindent
The  $(n+2l+1)\times (n+2l+1)$ matrix $\mb A_{I_E,\,h}(\mathbf x^p)$ may be written as $\mb A_{I_E,\,h}(\mathbf x^p)=\diag\left(\mb A_{I_{E_H},\,h},~ \mb A_{I_{E_V},\,h}(\mathbf x^p)\right)$  with $$\mb A_{I_{E_H},\,h}=\diag((1-u_i\phi(h))\left(\varepsilon_i+\nu_i\right))_{1\leqslant i\leqslant n}$$ and $\mb A_{I_{E_V},\,h}(\mathbf x^p)$ is a 3-banded sub-diagonal matrix whose diagonal and sub-diagonal elements are given by  the vectors $\mathbf d_{0,\,h}(\mathbf x^p)$, $\mathbf d_{-1,\,h}(\mathbf x^p)$ and  $\mathbf d_{-2,\,h}(\mathbf x^p)$ respectively, defined by
\begin{align*}
\mathbf d_{0,\,h}(\mathbf x^p) &= \Big(\underbrace{1-\phi(h)(\mu+\delta),\; \frac{1-\phi(h)\mu}{1+\phi(h)(d^p+\varpiVar ^p)},\;\cdots,1-\phi(h)(\mu+\delta),\;
\frac{1-\phi(h)\mu}{1+\phi(h)(d^p+\varpiVar ^p)}}_{2l\;\; components},\;1-\phi(h)(\mu+\delta) \Big),\\
\mathbf d_{-1,\,h}(\mathbf x^p) &= \frac{\phi(h)}{1+\phi(h)(d^p+\varpiVar ^p)}{\left(\right.} \underbrace{ \delta , \;\varpiVar ^p(1-\phi(h)\mu), \;\cdots,
\;\delta,\;\varpiVar ^p(1-\phi(h)\mu)}_{2l\;\;components}{\left.\right.)},\\
\mathbf d_{-2,\,h}(\mathbf x^p) &= \frac{\phi(h)^2\varpiVar ^p}{1+\phi(h)(d^p+\varpiVar ^p)}{\left(\right.} \underbrace{ \delta , \;0, \;\cdots,
\;\delta,\;0}_{2(l-1)\;\;components},\;\delta).
\end{align*}

\noindent The  $(n+2l+1)\times 2(n+1)$ matrix  $ \mathbf A_{I_{I,\,E},\,h}(\mathbf x^p,\;\mathbf x_1^{p+1})$ that may be written in block form as $$ \mb A_{I_{I,\,E},\,h}(\mathbf x^p,\;\mathbf x_1^{p+1}) = \begin{pmatrix}\mb 0 & \mb A_{I_{I_V,\,E_H},\,h}(\mathbf x^p,\;\mathbf x_1^{p+1})\cr \mb A_{I_{I_H,\,E_V},\,h}(\mathbf x^p,\;\mathbf x_1^{p+1}) & \mb 0\end{pmatrix}$$ with blocks $\mb A_{I_{I_V,\,E_H},\,h}(\mathbf x^p,\;\mathbf x_1^{p+1})$and $\mb A_{I_{I_H,\,E_V},\,h}(\mathbf x^p,\;\mathbf x_1^{p+1})$  given respectively by
$$
\mathbf A_{I_{I_V,\,E_H},\,h}(\mathbf x^p,\;\mathbf x_1^{p+1})= a\frac{\phi(h)}{H^p}\left(\begin{array}{cc}m_1S_1^{p+1}&0\\  m_2S_2^{p+1}&0\\\vdots&\vdots\\  m_nS_n^{p+1}&0\end{array}\right), \qquad \mathbf A_{I_{I_H,\,E_V},\,h}(\mathbf x^p,\;\mathbf x_1^{p+1})= a\phi(h)\frac{S^{p+1}_q}{H^p}\begin{pmatrix}\xi_1\phi_1&\cdots & \xi_n\phi_n & \widetilde\xi_1\phi_1&\cdots & \widetilde\xi_n\phi_n\\0 & \cdots&0 & 0 & \cdots&0\\\vdots&\cdots&\vdots & \vdots&\cdots&\vdots\\0 & \cdots&0 & 0 & \cdots&0\end{pmatrix}.
$$

\noindent
The $2n+2)\times (n+2l+1)$ matrix  $ \mathbf A_{I_{E,\,I},\,h}(\mathbf x^p,\;\mathbf x_1^{p+1})$ may be written in block form as $$ \mathbf A_{I_{E,\,I},\,h}(\mathbf x^p,\;\mathbf x_1^{p+1}) = \begin{pmatrix} \mathbf  A_{I_{{E_H},\,{I_H}},\,h}
	&   \mathbf 0 \\ \mathbf 0 &  \mathbf
	A_{I_{{E_V},\,{I_V}},\,h}(\mathbf x^p) \end{pmatrix} $$ where the block matrix $\mathbf  A_{I_{{E_H},\,{I_H}},\,h}$ is the $n\times n$ matrix $$\mathbf  A_{I_{{E_H},\,{I_H}},\,h}=\diag(u_0\phi(h)\varepsilon_{0},~\cdots,~ u_n\phi(h)\varepsilon_{n})$$ and the block matrix $\mathbf
A_{I_{{E_V},\,{I_V}},\,h}(\mathbf x^p)$ is the $(n+2)\times (2l+1)$ matrix that has only two non-zero entries; the $(1,2l+1)$ entry is equal to \\$\frac{\phi(h) \delta}{1+\phi(\delta t)(d^{p}+\varpiVar ^{p})}$, and the $(2,2l+1)$ entry is equal to $\frac{{\phi(\delta t)}^2 \varpiVar ^{p}\delta}{1+\phi(\delta t)(d^{p}+\varpiVar ^{p})}$. 

\noindent 
The $(2n+2)\times (2n+2)$  matrix
$\mathbf
A_{I_I,\,h}(\mathbf x^p,\;\mathbf x_1^{p+1})$ may be written in block  form as: $$\mathbf
A_{I_I,\,h}(\mathbf x^p,\;\mathbf x_1^{p+1})=\begin{pmatrix}\mathbf A_{I_{I_H},\,h}&\mathbf A_{I_{I_H, I_V},\,h}(\mathbf x^p,\;\mathbf x_1^{p+1}) \\\mathbf 0 &\mathbf A_{I_{I_V},\,h}(\mathbf x^p)\end{pmatrix}$$  where
the block matrix $\mathbf A_{I_{I_H},\,h}$ is the two band diagonal matrix of order $2n\times 2n$ with the principal diagonal constituted with the terms on the vectors $$\mathbf a_0 = \left((1-\phi(h)(\nu_1+\gamma_1))
,\;\cdots,\;(1-\phi(h)(\nu_n+\gamma_n)),\;(1-\phi(h)v_1(\nu_1+\zeta_1)),\;\cdots,\;(1-\phi(h)v_n(\nu_n+\zeta_n))\right)$$  and the $n^{th}$ sub-diagonal constituted with terms of the vector $\mathbf a_n = \phi(h)\left(v_1\gamma_1,\; \cdots,\; v_n\gamma_n\right)$ and $\mathbf A_{I_{I_H, I_V},\,h}(\mathbf x^p,\;\mathbf x_1^{p+1})$, that is $2n\times2$ is $\mathbf A_{I_{I_H, I_V},\,h}(\mathbf x^p,\;\mathbf x_1^{p+1}) = \dfrac{a\phi(h)}{H^p}\begin{pmatrix}
	\bar u_1\varphi_1 S_1^{p+1}&0\cr\vdots&\vdots\cr\bar u_n\varphi_n S_n^{p+1}&0\cr0&0\cr\vdots&\vdots\cr0&0
\end{pmatrix}$

\noindent The last $2\times 2$ square block matrix $A_{I_{I_V},\,h}(\mathbf x^p)$ is $$A_{I_{I_V},\,h}(\mathbf x^p) = \begin{pmatrix}\frac{1-\phi(h)\mu}{1+ \phi(h)(d^p+\varpiVar ^p)} & \frac{\phi(h)\delta}{1+ \phi(h)(d^p+\varpiVar ^p)}\\\frac{\phi(h)\varpiVar ^p(1-\phi(h)\mu)}{1+ \phi(h)(d^p+\varpiVar ^p)} & \frac{1+ \phi(\delta t)(d^p+\varpiVar ^p-\delta-\mu) -  {\phi(\delta t)}^2((\mu+\delta) d^p+\mu\varpiVar ^{p}) }{1+\phi(\delta t)(d^{p}+\varpiVar ^{p})}\end{pmatrix}.$$ 


}

\bibliographystyle{plain}
\bibliography{BiblioBiomath}
\end{document}